%% file: ANA-STDM-2017-32-PAPER.tex
\pdfoutput=1
 
\newcommand*{\ATLASLATEXPATH}{}
\documentclass[cernpreprint,texlive=2016,txfonts,UKenglish]{\ATLASLATEXPATH atlasdoc}
 
\usepackage{\ATLASLATEXPATH atlaspackage}
\usepackage{\ATLASLATEXPATH atlasphysics}
 
\newcommand{\AtlasCoordFootnote}{ATLAS uses a right-handed coordinate
system with its origin at the nominal interaction point (IP) in the
centre of the detector and the $z$-axis along the beam pipe. The
$x$-axis points from the IP to the centre of the LHC ring, and the
$y$-axis points upwards. Cylindrical coordinates $(r,\phi)$ are used
in the transverse plane, $\phi$ being the azimuthal angle around the
$z$-axis. The transverse energy is defined as
$E_{\mathrm{T}}=E\sin\theta$, where $E$ is the energy and $\theta$
is the polar angle. The pseudorapidity is defined as
$\eta=-\ln\tan(\theta/2)$ and the angular distance is measured in
units of $\Delta R\equiv\sqrt{(\Delta\eta)^{2}+(\Delta\phi)^{2}}$.}
 
\graphicspath{{logos/}}
 
\usepackage[backref=false]{\ATLASLATEXPATH atlasbiblatex}
 
\def\pb1{pb$^{-1}$}
\def\fb1{fb$^{-1}$}
 
\def\kt{k_{\mathrm{t}}}
\def\akt{k_{t}}

\def\qq{q\bar q}

\def\dsetg{{\mathrm{d}}\sigma/{\mathrm{d}}\etg}
\def\dspt{{\mathrm{d}}\sigma/{\mathrm{d}}\ptjet}
\def\dsyj{{\mathrm{d}}\sigma/{\mathrm{d}}|\rapjet|}
\def\dsrapgj{{\mathrm{d}}\sigma/{\mathrm{d}}|\deltarapgj|}
\def\dsphigj{{\mathrm{d}}\sigma/{\mathrm{d}}\deltaphigj}
\def\dsrapjj{{\mathrm{d}}\sigma/{\mathrm{d}}|\deltarapjj|}
\def\dsphijj{{\mathrm{d}}\sigma/{\mathrm{d}}\deltaphijj}
\def\dsmjj{{\mathrm{d}}\sigma/{\mathrm{d}}\mjj}
\def\dsmgjj{{\mathrm{d}}\sigma/{\mathrm{d}}\mgjj}

\def\rbg{R^{\mathrm{bg}}}

\def\ppgjj{pp\rightarrow\gamma + {\mathrm{jet}} + {\mathrm{jet}} + {\mathrm{X}}}

\def\z0{Z^0}
\def\mz{m_Z}

\def\as{\alpha_{\mathrm{s}}}
\def\asz{\alpha_{\mathrm{s}}(\mz)}
\def\oalphas2{{\cal O}(\alpha\as^2)}

\def\muR{\mu_{\mathrm{R}}}
\def\muF{\mu_{\mathrm{F}}}

\def\etg{E_{\mathrm{T}}^{\gamma}}
\def\ptjet{p_{\mathrm{T}}^{\mathrm{jet}}}
\def\ptjetl{p_{\mathrm{T}}^{\mathrm{jet1}}}
\def\ptjetsl{p_{\mathrm{T}}^{\mathrm{jet2}}}
\def\rapjet{y^{\mathrm{jet}}}
\def\etag{\eta^{\gamma}}

\def\mgjj{m^{\gamma-{\mathrm{jet}}-{\mathrm{jet}}}}
\def\mjj{m^{{\mathrm{jet}}-{\mathrm{jet}}}}

\def\dr{\Delta R^{\gamma-{\mathrm{jet}}}}
\def\deltaphigj{\Delta\phi^{\gamma-{\mathrm{jet}}}}
\def\deltarapgj{\Delta y^{\gamma-{\mathrm{jet}}}}
\def\deltaphijj{\Delta\phi^{{\mathrm{jet}}-{\mathrm{jet}}}}
\def\deltarapjj{\Delta y^{{\mathrm{jet}}-{\mathrm{jet}}}}
 
\def\pt{p_{\mathrm{T}}}

\def\etiso{E_{\mathrm{T}}^{\mathrm{iso}}}
\def\etisoc{E_{\mathrm{T,cut}}^{\mathrm{iso}}}

\def\etisocut{$0.0042\cdot\etg+4.8\ {\mathrm{\GeV}}$}

\def\etisocutp{$0.0042\cdot\etg+10\ {\mathrm{\GeV}}$}
 
\def\sher{{\textsc{Sherpa}}}
\def\pyt{{\textsc{Pythia}}}

\mathchardef\mhyphen="2D
 
\def\rbg{R_{\mathrm{bg}}}
 
\AtlasTitle{Measurement of isolated-photon plus two-jet production
in {\boldmath $pp$} collisions at {\boldmath $\sqrt s=13$}~\TeV\
with the ATLAS detector}
 
\author{The ATLAS Collaboration}
 
\AtlasJournalRef{JHEP 03 (2020) 179}
\AtlasDOI{10.1007/JHEP03(2020)179}
\PreprintIdNumber{CERN-EP-2019-210}
 
\AtlasAbstract{
The dynamics of isolated-photon plus two-jet production in $pp$
collisions at a centre-of-mass energy of $13$~\TeV\ are studied with
the ATLAS detector at the LHC using a dataset corresponding to an
integrated luminosity of $36.1$~\fb1. Cross sections are measured as
functions of a variety of observables, including angular correlations
and invariant masses of the objects in the final state,
$\gamma+{\mathrm{jet}}+{\mathrm{jet}}$. Measurements are also
performed in phase-space regions enriched in each of the two
underlying physical mechanisms, namely direct and fragmentation
processes. The measurements cover the range of photon (jet) transverse
momenta from $150$~\GeV\ ($100$~\GeV) to $2$~\TeV. The tree-level plus
parton-shower predictions from \sher\ and \pyt\ as well as the
next-to-leading-order QCD predictions from \sher\ are compared with the
measurements. The next-to-leading-order
QCD predictions describe the data adequately in shape and
normalisation except for regions of phase space such as those with high values of
the invariant mass or rapidity separation of the two jets, where the
predictions overestimate the data.
}
 
\hypersetup{pdftitle={ATLAS document},pdfauthor={The ATLAS Collaboration}}
 
\begin{document}
 
\maketitle
 
\tableofcontents
 
\section{Introduction}
\label{intro}
 
The production of prompt photons\footnote{All photons that are not
produced in hadron decays are considered to be `prompt'.} in
association with two jets in proton--proton collisions, $\ppgjj$,
provides a means of testing perturbative QCD (pQCD) predictions.
Measurements of the angular correlations between the photon and each
of the jets as well as between the two jets can be used to probe the
dynamics of the hard-scattering process. In addition, measurements of the
invariant-mass distributions of the dijet system and the
photon plus dijet system are sensitive to the dynamics of the
hard interaction. A comprehensive study of the observables describing
this final state is of relevance for the development of pQCD
calculations as well as for the tuning of Monte Carlo (MC)
models.
 
\begin{figure}[h]
\vfill
\setlength{\unitlength}{1.0cm}
\begin{picture} (18.0,4.9)
\put (3.0,-0.5){\includegraphics[bb = 0 0 567 567,width=3.8cm]{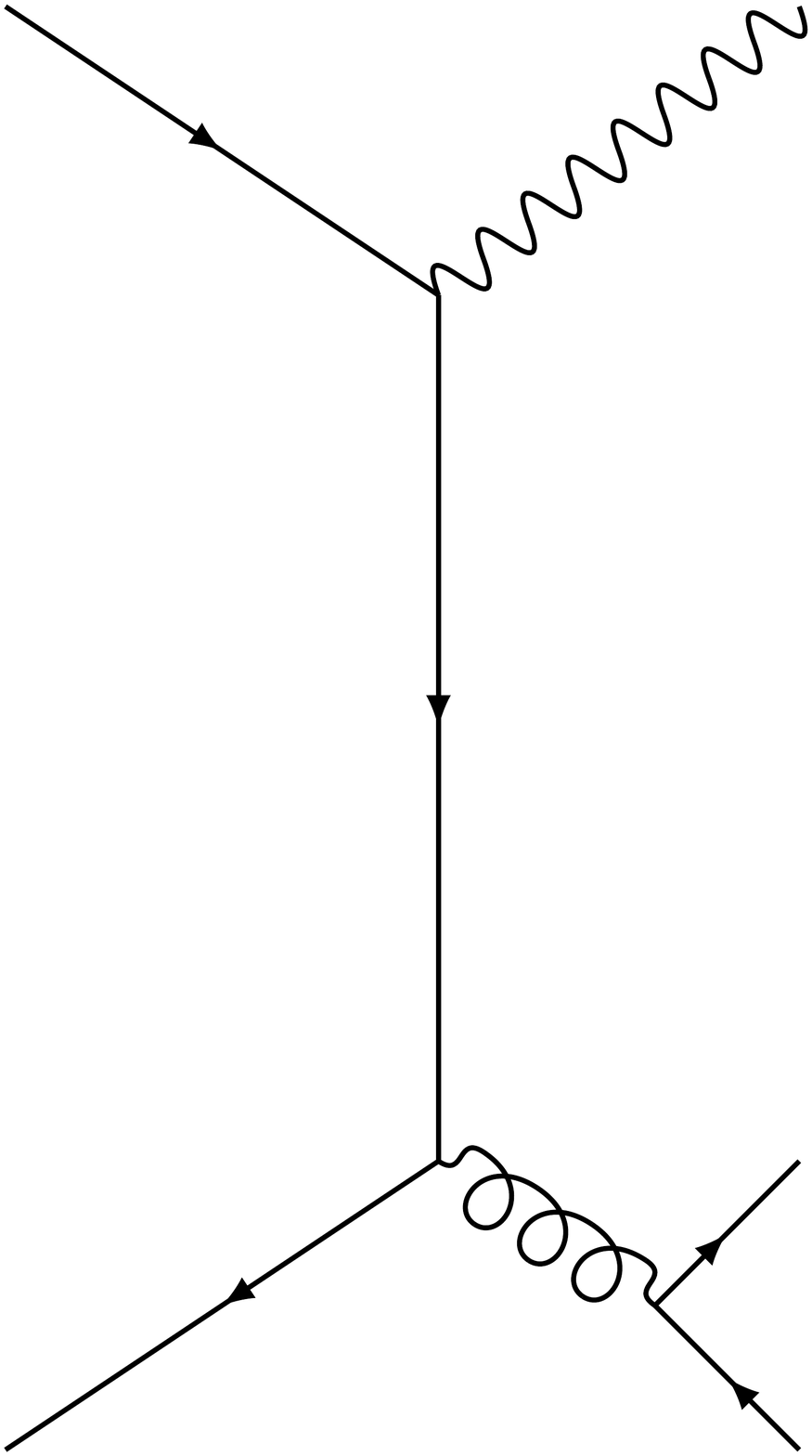}}
\put (9.0,-0.5){\includegraphics[bb = 0 0 567 567,width=6.8cm]{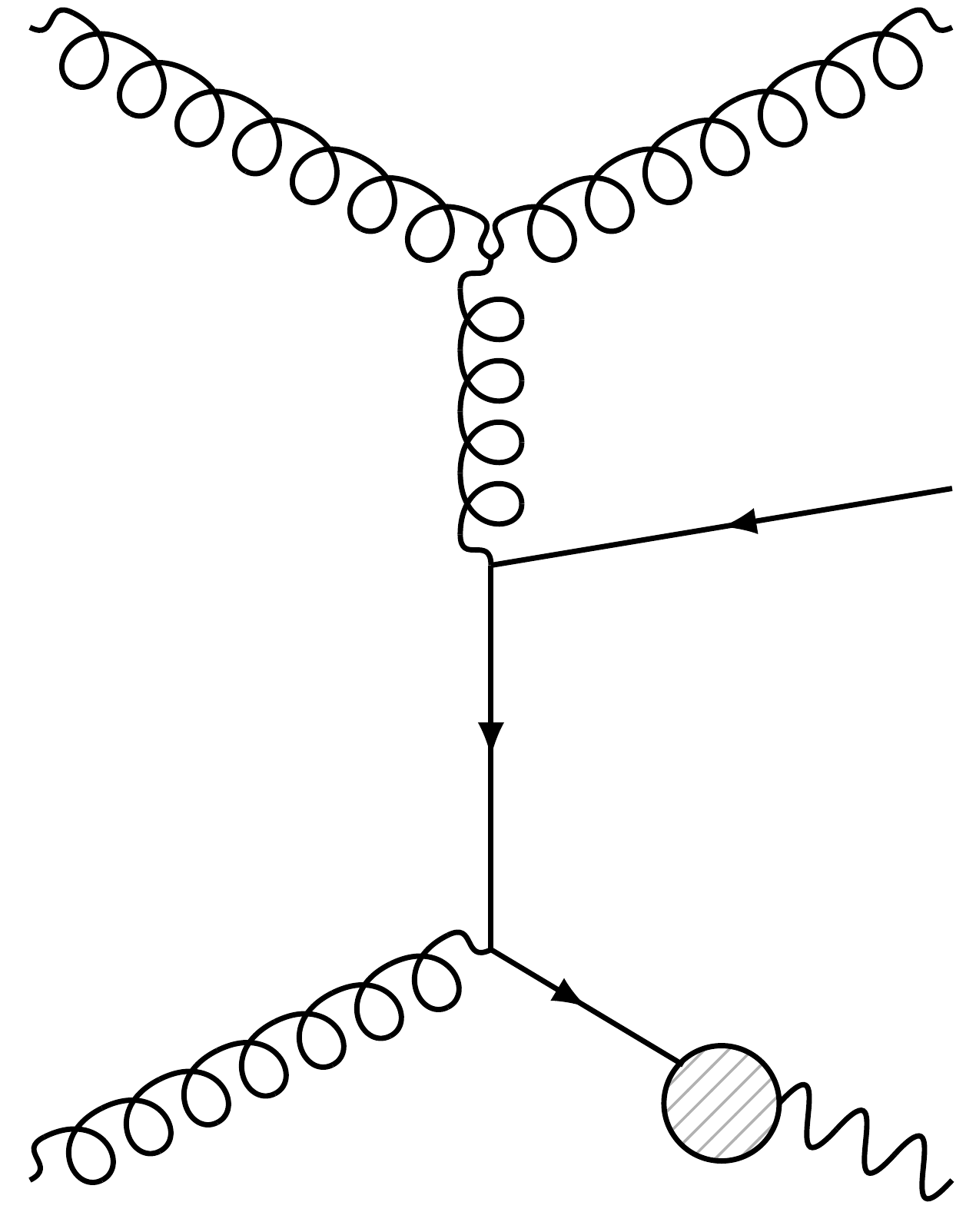}}
\put (4.0,3.7){\small $q$}
\put (4.0,0.3){\small $\bar q$}
\put (6.0,3.7){\small $\gamma$}
\put (6.7,0.6){\small $q$}
\put (6.7,-0.6){\small $\bar q$}
\put (10.0,3.7){\small $g$}
\put (10.0,0.7){\small $g$}
\put (13.0,0.2){\small $\gamma$}
\put (12.0,3.7){\small $g$}
\put (12.0,2.2){\small $\bar{q}$}
\put (4.9,-0.5){\small (a)}
\put (10.8,-0.5){\small (b)}
\end{picture}
\vspace{0.1cm}
\caption
{
Examples of diagrams for $\gamma+{\mathrm{jet}}+{\mathrm{jet}}$
production through (a) direct-photon and (b) fragmentation-photon processes.
}
\label{fig01}
\end{figure}

Prompt-photon plus two-jet production proceeds via two mechanisms (see
Figure~\ref{fig01}): direct processes, in which the photon originates
from the hard interaction, and fragmentation processes, in which the
photon arises from the fragmentation of a high transverse
momentum\footnote{\AtlasCoordFootnote} ($\pt$)
parton~\cite{pr:d76:034003,pr:d79:114024}. The direct and
fragmentation contributions are well defined only at leading order
(LO) in QCD; at higher orders this distinction is no longer possible.
Furthermore, pure electroweak processes or electroweak virtual
corrections are expected to play an important role for photon transverse
energies at the \TeV\ scale~\cite{jhep:0603:059,pr:d75:013005,pr:d88:013009,pr:d92:073011}
and dijet configurations with large invariant mass and large separation in rapidity.
 
Measurements of prompt-photon production in a final state with
accompanying hadrons necessitate an isolation requirement on the
photon to minimise the large multi-jet background where hadrons in
jets decay into photons. The production of isolated photons in association with jets
in $pp$ collisions at $\sqrt s=7$, $8$ and $13$~\TeV\ was
studied by
ATLAS~\cite{pr:d85:092014,np:b875:483,np:b918:257,pl:b780:578} and
CMS~\cite{pr:d88:112009,jhep:1406:009,jhep:1510:128,epj:c79:20,epj:c79:969}.
Previous measurements of the production cross sections of photons
accompanied by two jets were done for angular correlations and the
transverse momenta of the photon and the jets at
$\sqrt s=8$~TeV~\cite{np:b918:257}. With the increase in centre-of-mass
energy to $13$~\TeV, new observables have been introduced to explore the dynamics
where electroweak contributions could be relevant, in particular, the invariant
mass of the two jets and the invariant mass of the photon--jet--jet system.
 
Measurements of photon plus two-jet production provide a deeper understanding of
the dynamics of the underlying processes and complement the recent
measurements of photon plus one-jet production at $13$~TeV~\cite{pl:b780:578}.
The analysis presented here includes the study of the kinematics of the photon plus
two-jet system via the measurement of the cross section as a function
of the leading-photon transverse energy ($\etg$), and of the transverse
momentum ($\ptjet$) and rapidity ($\rapjet$) of each of the two
jets. The dynamics of the photon plus two-jet system is studied by
measuring the azimuthal angular separation between the photon and each
of the jets ($\deltaphigj$), the difference in rapidity between the
photon and each of the jets ($\deltarapgj$), the invariant mass of the
dijet system ($\mjj$), the azimuthal angular separation between the
two jets ($\deltaphijj$), the difference in rapidity between the two
jets ($\deltarapjj$) and the invariant mass of the photon--jet--jet
system ($\mgjj$).
Measurements of the final state $\gamma+{\mathrm{jet}}+{\mathrm{jet}}$
have the advantage that the second jet is explicitly identified and the
interplay between the two underlying production mechanisms can be tested more
directly than in the $\gamma+{\mathrm{jet}}$ final state.
The two underlying production mechanisms exhibit distinct
features in kinematic variables as well as in angular correlations
or invariant masses of the objects in the final state. For example,
the spectrum in $\etg$ ($\ptjet$, $\mjj$) is expected to be harder
(softer) for direct processes than for fragmentation processes; as
another example, angular correlations such as $\deltaphijj$ are
expected to be different for the two processes, with the two jets
being closer to each other in direct processes than in fragmentation
processes. These distinct features arise from the matrix elements of
the two processes and the fraction of the centre-of-mass energy of the
colliding partons carried away by the final-state photon~\cite{pl:b269:445}.
In order to enhance the sensitivity to the two underlying
mechanisms,  measurements are performed in two complementary regions
of phase space, in addition to the inclusive phase space, by selecting
those events in which $\etg$ is larger (smaller) than the leading
(sub-leading) jet $\pt$.\footnote{The leading (sub-leading) jet is the one
with (next-to) highest $\ptjet$.} This separation allows a more
in-depth study of the direct and fragmentation contributions since it
is performed in terms of kinematic variables
and, therefore, does not rely on distinguishing between direct and
fragmentation processes using a MC generator.
Furthermore, the photon is required to be isolated and the distance in the
$\eta$--$\phi$ plane between the
photon and each of the jets is required to be larger than $0.8$;
the region of phase space in which a jet is close to the photon is removed to
avoid a bias in the photon isolation energy
as well as to suppress the dependence on the modelling
of the fragmentation component in that region.
The analysis is performed using $36.1$~\fb1\ of ATLAS data at $\sqrt s=13$~\TeV\
taken during 2015 and 2016. The predictions of the tree-level plus parton shower models
of \pyt~\cite{cpc:178:852} and \sher~\cite{jhep:0902:007}, as well as
the next-to-leading-order (NLO) QCD predictions from
\sher~\cite{scipost:7:034,jhep:0202:044,prl:108:111601,jhep:0803:038,jhep:1304:027},
are compared with the measurements.
 
\section{ATLAS detector}
\label{detector}
The ATLAS experiment uses a multipurpose particle detector~\cite{PERF-2007-01}
with a forward--backward symmetric cylindrical geometry. It consists of
an inner tracking detector surrounded by a thin superconducting
solenoid, electromagnetic and hadronic calorimeters, and a muon
spectrometer incorporating three large superconducting toroidal
magnets. The inner-detector system is immersed in a $2$~T axial
magnetic field and provides charged-particle tracking in the range
$|\eta|<2.5$. The high-granularity silicon pixel detector is closest
to the interaction region and provides four measurements per track;
the innermost layer, known as the insertable
B-layer~\cite{ATLAS-TDR-19,Abbott:2018ikt}, provides high-resolution
hits at small radius to improve the tracking performance. The pixel detector
is followed by a silicon microstrip tracker, which typically provides
four three-dimensional space-point measurements per track. These
silicon detectors are complemented by a transition radiation
tracker, which enables radially extended track reconstruction up to
$|\eta|=2.0$. The calorimeter system covers the range
$|\eta|<4.9$. Within the region $|\eta|<3.2$, electromagnetic (EM)
calorimetry is provided by barrel and endcap high-granularity
lead/liquid-argon (LAr) EM calorimeters, with an additional thin LAr
presampler covering $|\eta|<1.8$ to correct for energy loss in
material upstream of the calorimeters; for $|\eta|<2.5$, the EM
calorimeter is divided into three layers in depth. Hadronic
calorimetry is provided by a steel/scintillator-tile calorimeter,
segmented into three barrel structures within $|\eta|<1.7$, and two
copper/LAr hadronic endcap calorimeters, which cover the region
$1.5<|\eta|<3.2$. The solid-angle coverage is completed out to
$|\eta|=4.9$ with forward copper/LAr and tungsten/LAr calorimeter
modules, which are optimised for EM and hadronic measurements,
respectively. Events are initially selected using a first-level trigger
implemented in custom electronics, which reduces the
event-acceptance rate from the maximum bunch crossing
rate of $40$~MHz to a design value of $100$~kHz using a subset of
detector information. Software algorithms with access to the full
detector information are then used in the high-level trigger to yield
a recorded event rate of about $1$~kHz~\cite{epj:c77:317}.
 
\section{Data selection}
\label{datsel}
The data used in this analysis were collected with the ATLAS detector
during the $pp$ collision running periods of 2015 and 2016, when the
LHC operated at a centre-of-mass energy of $\sqrt s=13$~\TeV. Only
events taken during stable beam conditions and satisfying detector and
data-quality requirements~\cite{1911.04632}, which include the calorimeters
and inner tracking detectors being in nominal operation, are considered. The
total integrated luminosity of the collected sample amounts to
{\mbox{$36.1\pm 0.8$~\fb1}}.
The uncertainty in the combined 2015--2016 integrated luminosity of
$2.1\%$~\cite{ATLAS-CONF-2019-021} is obtained using the LUCID-2
detector~\cite{LUCID2} for the primary luminosity measurements.
 
\subsection{Event selection}
\label{evesel}
Events were recorded using a single-photon trigger, with a transverse
energy threshold of $140$~\GeV; `loose' identification criteria for the
photon, which are based on the shower shapes in the second layer of
the electromagnetic calorimeter and on the fraction of energy deposited
in the hadronic calorimeter, were also applied~\cite{1909.00761}. Events
are required to have a reconstructed primary vertex~\cite{ATL-PHYS-PUB-2015-026}.
If multiple primary vertices are reconstructed, the one with the highest
sum of the $\pt^2$ of the associated tracks is selected as the primary vertex.
 
\subsection{Photon selection}
\label{phosel}
The photon-candidate reconstruction, selection and calibration closely
follow those reported in Ref.~\cite{jhep:1910:203} and are summarised
in the following. Photon candidates are reconstructed from clusters of
energy deposited in the electromagnetic calorimeter and
classified~\cite{epj:c79:205} as unconverted photons (those candidates
without a matching track or reconstructed conversion vertex in the
inner detector) or converted photons (those candidates with a matching
reconstructed conversion vertex or a matching track consistent with
originating from a photon conversion).
 
The photon identification is primarily based on shower shapes in the
calorimeter~\cite{epj:c79:205}. The information from the hadronic
calorimeter and the lateral shower shape in the second layer of the EM
calorimeter are used in an initial selection. The final `tight'
selection comprises more stringent criteria for these variables as well
as requirements on the shower shapes in the first layer of the EM
calorimeter. Small differences are observed in the
average values of the shower-shape variables between data and
simulation and are corrected for in simulated events before applying
the photon identification criteria. The photon identification efficiencies are
measured to be above $90\%$ ($95\%$) for unconverted (converted)
photon candidates with $\etg>150$~\GeV.
 
The measurement of the photon energy is based on the energy collected
in calorimeter cells in an area of size
$\Delta\eta\times\Delta\phi=0.075\times 0.175$ in the barrel and
$0.125\times 0.125$ in the endcaps.
The photon energy calibration procedure is described in Ref.~\cite{PERF-2017-03}.
The uncertainty in the photon
energy scale at $\etg = 150$~\GeV\ is in the range $0.4\%$--$3.1\%$
($0.4\%$--$2.8\%$) for unconverted (converted) candidates depending on
$|\etag|$, being largest in the region $1.81<|\etag|<2.37$.
 
The selection of isolated photons is based on the amount of transverse energy
inside a cone of size $\Delta R=0.4$ in the $\eta$--$\phi$ plane
around the photon candidate, excluding an area of size $\Delta\eta
\times \Delta\phi = 0.125 \times 0.175$ centred on the photon. The
isolation transverse energy ($\etiso$), which is computed from
topological clusters of calorimeter cells~\cite{epj:c77:490}, is
corrected for the leakage of the photon energy into the isolation cone
and the estimated contributions from the underlying event (UE) and
additional inelastic $pp$ interactions
(pile-up)~\cite{pr:d83:052005}.  The combined correction
to $\etiso$ for the last two effects is computed on an event-by-event basis using
the jet-area method~\cite{jhep:0804:005} and typically amounts
to $3.5$~\GeV\ ($1.3$~\GeV) for $|\etag|<1.37$
($1.56<|\etag|<2.37$). {\textit{In situ}} corrections to $\etiso$ are applied
in simulated events such that the peak position in the $\etiso$
distribution coincides with that in the data. After corrections,
$\etiso$ is required to be less than
$\etisoc\equiv$~\etisocut.
 
Events with at least one isolated photon candidate with
$\etg>150$~\GeV\ and $|\etag|<2.37$ are selected. Candidates in the
region $1.37<|\etag|<1.56$, which includes the transition region
between the barrel and endcap calorimeters, are not considered. In
events with more than one photon candidate satisfying the selection
criteria, only the highest-$\etg$ (leading) photon is considered for
further study.
 
\subsection{Jet selection}
\label{jetsel}
The anti-$\akt$ algorithm~\cite{jhep:0804:063,epj:c72:1896} with radius
parameter $R=0.4$ is used to reconstruct jets. Topological clusters of
calorimeter cells are used as input (jet constituents). The
calorimeter cell energies are measured at the EM scale, which
corresponds to the energy deposited by electromagnetically interacting
particles. The jet four-momenta are computed from the sum of the
jet-constituent four-momenta, treating each as a four-vector with zero
mass. At this stage, the jet four-momentum values refer to the EM energy scale.
 
The calibration of the jets is performed following the methods described
in Ref.~\cite{pr:d96:072002} and is discussed below. The four-momentum
of jets is recalculated to point to the selected primary vertex of the event
rather than the centre of the detector. The jet-area method is then used on a
jet-by-jet basis~\cite{epj:c76:581} to subtract the contributions from the UE and pile-up.
Subsequently, a calibration of the jet energy and direction is applied that corrects the
reconstructed jet four-momentum to the particle level. For this
purpose, simulated events are employed and the same jet algorithm used
on data is applied to the generated stable particles. Stable particles
are defined as those with a decay length $c\tau > 10$~mm. Muons,
neutrinos and particles from pile-up interactions are
excluded. The resulting jets are referred to as particle-level
jets. Following the previous calibration, residual dependencies of the
reconstructed jet energy on the longitudinal and transverse
characteristics of the jet are observed. Corrections to reduce these
dependencies are derived from simulated events, using global properties
of the jet obtained from tracking information, calorimeter energy deposits
and muon spectrometer information. These corrections are applied
sequentially to the jet four-momentum while conserving the average jet
response in the sample of dijet simulated events used to derive the
corrections. These corrections improve the jet energy resolution and
account for differences in energy response between quark- and
gluon-initiated jets. Since the jet flavour composition of the final
state $\gamma+{\mathrm{jet}}+{\mathrm{jet}}$ is, in principle,
different from that used for the calibration (dijet events), the
corrections help to reduce the dependence on the jet flavour
composition assumed. In addition, systematic uncertainties due to the
jet flavour response and composition are considered.
A final correction is applied to account for differences
in the jet response between data and simulations. The final correction
is derived {\textit{in situ}} from a combination of dijet,
$\gamma+{\mathrm{jet}}$, $Z+{\mathrm{jet}}$ and multi-jet $\pt$-balance
methods. Dijet events are used to correct the
average response for forward jets to that for well-measured central
jets. The other three {\textit{in situ}} calibrations correct for differences between
the average responses for central jets and
well-measured reference objects. The uncertainty in the jet energy
scale at $\ptjet=100$~\GeV\ varies in the range $1.5\%$--$2\%$ depending
on the jet pseudorapidity.
 
Quality criteria are applied to reject events with jets reconstructed
from calorimeter signals not originating from a $pp$
collision~\cite{ATLAS-CONF-2015-029}. These
criteria suppress events with jets from beam-induced background due to
proton losses upstream of the interaction point, cosmic-ray air
showers overlapping with collision events and calorimeter noise from
large-scale coherent noise or isolated pathological cells. Jets are
required to have calibrated transverse momenta greater than $100$~\GeV\
and rapidity $|\rapjet|<2.5$. Jets overlapping with the photon
candidate are not considered if the jet axis lies within a cone of
size $\Delta R=0.8$ around the photon candidate; this requirement
prevents any overlap between the photon isolation cone ($\Delta
R=0.4$) and the jet cone ($\Delta R=0.4$). Finally, the event is selected if
there are at least two jets (leading and sub-leading jets) satisfying the
requirements above. The threshold $\ptjet = 100$~\GeV\ is chosen to avoid
the increasingly large uncertainties in the jet energy scale for lower values of $\ptjet$.
 
\subsection{Data samples}
The number of data events selected by the requirements listed in the
previous subsections is  $755\,270$. Three samples of events are
selected according to the following criteria: a `total' sample in
which no further requirement is applied; a `fragmentation-enriched'
sample in which the photon is required to have $\etg<\ptjetsl$, where
$\ptjetsl$ is the $p_{\mathrm{T}}$ of the sub-leading jet; and a
`direct-enriched' sample in which the photon is required to have
$\etg>\ptjetl$, where $\ptjetl$ is the $p_{\mathrm{T}}$ of the leading
jet. A summary of the requirements on the photon and the jets is given
in Table~\ref{tableone}, together with the number of selected events
in each data sample.
 
\begin{table}[ht!]
\renewcommand{\arraystretch}{1.2}
\caption{
Phase-space regions for the measurements and predictions. The number of
data events selected in each phase space is also shown.
}
\setlength{\unitlength}{1.0cm}
\begin{picture} (18.0,4.0)
\put (0.0,1.5){\scalebox{1.0}{\centerline{
\begin{tabular}{|l|c|c|c|} \hline
{\textbf{Requirements on photon}}
& \multicolumn{3}{c|}{$\etg>150$~\GeV, $|\etag|<2.37$ (excluding
$1.37<|\etag|<1.56$)} \\
& \multicolumn{3}{c|}{$\etiso<$~\etisocut\ (reconstruction level)} \\
& \multicolumn{3}{c|}{$\etiso<$~\etisocutp\ (particle level)} \\ \hline
{\textbf{Requirements on jets}}
& \multicolumn{3}{c|}{at least two jets using anti-$\akt$ algorithm with $R=0.4$}  \\
& \multicolumn{3}{c|}{$\ptjet>100$~\GeV, $|\rapjet|<2.5$, $\dr>0.8$} \\ \hline
\textbf{Phase space} & \textbf{total} & \textbf{fragmentation enriched} & \textbf{direct enriched}  \\
&  & $\etg<\ptjetsl$ & $\etg>\ptjetl$ \\ \hline
\textbf{Number of events} & $755\,270$ & $111\,666$ & $386\,846$ \\ \hline
\end{tabular}
}}}
\end{picture}
\label{tableone}
\end{table}
 
\section{Monte Carlo simulations and Standard Model theory predictions}
\label{mcnlo}
\subsection{Signal Monte Carlo simulations}
\label{mc}
Samples of events were generated to study the characteristics of the
$\ppgjj$ signal events. The programs \pyt~8.186~\cite{cpc:178:852} and
\sher~2.1.1~\cite{jhep:0902:007} were used for the predictions.
In both cases, the event generation is performed
using tree-level matrix elements, with the inclusion of initial- and
final-state parton showers. The Lund string model~\cite{prep:97:31}
in the case of \pyt\ and a modified version of the cluster
model~\cite{epj:c36:381} in \sher\ are used to describe the
fragmentation into hadrons. The LO NNPDF2.3~\cite{np:b867:244} parton
distribution functions (PDFs) were used by \pyt\ while the NLO
CT10~\cite{pr:d82:074024} PDFs were used by \sher\ to parameterise the
proton structure. Both samples include a simulation of the UE. The
event-generator parameter values were set according to the A14 tune for
\pyt~\cite{ATL-PHYS-PUB-2014-021} and the CT10 tune for \sher.
 
The \pyt\ simulation of the signal includes LO photon-plus-jet events
from both the direct processes (the subprocesses $qg\rightarrow
q\gamma$ and $\qq\rightarrow g\gamma$, called the `hard' component)
and the photon bremsstrahlung in LO QCD dijet events (called the
`bremsstrahlung' component). The bremsstrahlung component was
modelled by final-state QED radiation arising from calculations of all
$2\rightarrow 2$ QCD processes. The \sher\ samples were generated with
LO matrix elements for photon-plus-jet final states with up to three
additional partons ($2\rightarrow n$ processes with $n$ from $2$ to
$5$); the matrix elements were merged with the \sher\ parton
shower~\cite{jhep:0905:053} using the ME+PS@LO prescription.
The bremsstrahlung component is accounted for in \sher\ through
the matrix elements of $2\rightarrow n$ processes with $n\geq 3$. In
the generation of the \sher\ samples, a requirement on the photon
isolation at the matrix-element level was imposed using the criterion
defined in Ref.~\cite{pl:b429:369}. This criterion, commonly called
Frixione's criterion, requires the total transverse energy inside a
cone of size ${\cal V}$ around the generated final-state photon,
excluding the photon itself, to be below a certain threshold,
$E_{\mathrm{T}}^{\mathrm{max}}({\cal V})=\epsilon\etg ((1-\cos{{\cal V}})/(1-\cos{{\cal R}}))^{n}$,
for all ${\cal V} < {\cal R}$, and is applied to avoid singularities in the
matrix-elements calculation. The parameter values for the threshold
were chosen to be ${\cal R}=0.3$, $n=2$ and $\epsilon=0.025$.
These values guarantee a selection loose enough to
minimise possible biases from the application of the photon isolation
requirement at reconstruction and particle levels (see
Table~\ref{tableone}).\footnote{Since Frixione's criterion requires
the upper limit on the transverse energy isolation to be exactly
zero at ${\cal V}=0$, the criterion cannot be strictly looser than
any non-zero photon isolation requirement at reconstruction or
particle level for all ${\cal V}<{\cal R}$.} A possible bias is
accounted for by a systematic uncertainty (see Section~\ref{md}) estimated
from a comparison with \pyt\ samples, for which no photon isolation
requirement is applied during the event generation. The \sher\ predictions
from this computation are referred to as LO \sher.
 
Pile-up from additional $pp$ collisions in the same and neighbouring
bunch crossings was simulated by overlaying each MC event with a
variable number of simulated inelastic $pp$ collisions generated using
\pyt~8.186 with the ATLAS set of tuned parameters for minimum-bias
events (A2 tune)~\cite{ATL-PHYS-PUB-2012-003} and the MSTW2008LO
PDF set~\cite{epj:c63:189}. The MC events are
weighted to reproduce the distribution of the average number of
interactions per bunch crossing observed in the data,
referred to as `pile-up reweighting'.
 
All the samples of generated events were passed through the
{\textsc{Geant4}}-based~\cite{nim:a506:250} ATLAS detector- and
trigger-simulation programs~\cite{epj:c70:823}. They are reconstructed
and analysed with the same program chain as the data.
 
\subsection{Background Monte Carlo simulations}
The main background to isolated-photon events arises from jets
misidentified as photons. This background, which is typically
below $5\%$, is subtracted using an {\textit{in situ}} technique, as described in Section~\ref{bacsub}.
The background from electrons or positrons misidentified as photons is
estimated using MC samples generated with the program
\sher~2.2.1~\cite{scipost:7:034,jhep:0202:044,prl:108:111601}. The
$Z^{(*)}/\gamma^*\rightarrow e^+e^-$ and $W^{(*)}\rightarrow e\nu$
processes were generated with matrix elements calculated for up to
two additional partons at NLO and up to four partons at LO.
 
\subsection{Next-to-leading-order pQCD predictions}
\label{nlo}
The NLO pQCD predictions presented in this paper are computed using
the program \sher~2.2.2~\cite{scipost:7:034}. This program consistently
combines parton-level calculations of $\gamma+ {\textrm{(1,\ 2)-jet}}$
events at NLO and  $\gamma+{\textrm{(3,\ 4)-jet}}$ events
at LO~\cite{jhep:0202:044,prl:108:111601} supplemented with a parton
shower~\cite{jhep:0803:038} while avoiding double-counting
effects~\cite{jhep:1304:027}. A requirement on the photon isolation at
the matrix-element level is imposed using Frixione's criterion with
${\cal R}=0.1$, $n=2$ and $\epsilon=0.1$; for the \sher\ NLO
calculations the parameters are fixed to looser values than for the LO
calculations to further minimise any possible bias from the application
of the photon isolation requirement at particle level. The NNLO
NNPDF3.0 PDFs~\cite{jhep:1504:040} are used in conjunction with the
corresponding \sher\ default tuning. Dynamic factorisation and
renormalisation scales are adopted and set equal to $\etg$, as well as a dynamical merging
scale with $\bar Q_{\mathrm{cut}}=20$~\GeV~\cite{jp:g44:044007}. The
strong coupling constant is set to $\asz=0.118$. Fragmentation into
hadrons and simulation of the UE are made using the same models as for
the LO \sher\ samples of simulated events (see Section~\ref{mc}).
All the \sher\ NLO predictions are based on the particle-level
observables from this computation after applying the requirements
listed in Table~\ref{tableone}. In particular, the photon isolation
requirement is applied at particle level using the procedure described
in Section~\ref{cor}.
 
\section{Signal extraction}
\label{bacsub}
 
\subsection{Multi-jet background subtraction}
The main background to prompt-photon production consists of multi-jet
events where one jet typically contains a $\pi^0$ or $\eta$ meson that
carries most of the jet energy and is misidentified as a photon
because it decays into an almost collinear photon pair. The multi-jet
background estimation relies on an {\textit{in situ}} method based on
signal-suppressed control regions. The method uses the same
two-dimensional sideband technique as in previous
analyses~\cite{pr:d83:052005,pl:b706:150,pr:d85:092014,jhep:1608:005,np:b875:483,pl:b770:473}
and it is described briefly in the following. The background subtraction
procedure is performed bin by bin for each distribution. Two
variables, the photon identification quality and $\etiso$, are used to
define the control regions for the background estimation. The photon
candidate is classified as `isolated' if $\etiso<\etisoc$ and as
`non-isolated' if $\etisoc +2~{\mathrm{\GeV}}<\etiso< 50$~\GeV.
The non-isolated region is separated by $2$~\GeV\ from the isolated
region to reduce the contamination from signal events. The upper bound in $\etiso$
is applied to avoid highly non-isolated photons; in this way, the determination of the
signal yield does not depend on the description by the MC generators of
the distribution of $\etiso$ for prompt photons with high values of $\etiso$.
A photon candidate is classified as `tight' if tight identification criteria are satisfied.
It is classified as `non-tight' if it fails at least one of four
tight requirements on the shower-shape variables computed from the
energy deposits in the first layer of the electromagnetic calorimeter,
but satisfies the tight requirements listed below: the tight requirement on
the total lateral shower width in the first layer, all the other tight
identification criteria in other layers and the tight requirement on
the leakage into the hadronic calorimeter~\cite{epj:c79:205}.
 
The two-dimensional plane defined by $\etiso$ and the tight/non-tight
photon-identification variable is divided into four non-overlapping
regions: the `signal' region of tight isolated candidates (region
$A$) and three background control regions of tight non-isolated
(region $B$), isolated non-tight (region $C$) and non-isolated
non-tight (region $D$) photon candidates. The signal yield
$N_A^{\mathrm{sig}}$ in region $A$ is determined by using the relation
 
\begin{equation}
N_A^{\mathrm{sig}}=N_A-\rbg\cdot(N_B-f_B N_A^{\mathrm{sig}})\cdot
\frac{(N_C-f_C N_A^{\mathrm{sig}})}{(N_D-f_D N_A^{\mathrm{sig}})},
\label{eqone}
\end{equation}
where $N_K$, with $K=A,B,C,D$, is the number of events in region $K$ and
$\rbg=N_A^{\mathrm{bg}}\cdot N_D^{\mathrm{bg}}/(N_B^{\mathrm{bg}}\cdot N_C^{\mathrm{bg}})$
is the so-called background correlation and is taken as $\rbg=1$ for
the nominal results; $N_K^{\mathrm{bg}}$ with $K=A,B,C,D$ is the
unknown number of background events in each
region. Equation~(\ref{eqone}) takes into account the expected number
of signal events in the three background control regions
($N_K^{\mathrm{sig}}$) via the signal leakage fractions,
$f_K=N_K^{\mathrm{sig}}/N_A^{\mathrm{sig}}$ with $K=B,C,D$, which are
estimated using the MC simulations of the signal.
 
The signal purity, defined as $N_A^{\mathrm{sig}}/N_A$, is typically
above $95\%$. The use of \pyt\ or LO \sher\ samples to extract the
signal leakage fractions leads to similar signal purities; the
difference in the signal purity is taken as a systematic uncertainty
(see Section~\ref{md}). The photon isolation and identification
variables are assumed to be uncorrelated ($\rbg=1$) for background events. The
dependence of the signal yield on this assumption is investigated in
validation regions, which are defined within the background-enriched
regions $B$ and $D$. A study of $\rbg$ in the validation regions,
accounting for signal leakage using either the \pyt\ or LO \sher\
simulations, shows deviations from unity, ranging from $10\%$ to $50\%$,
which are then propagated through Eq.~(\ref{eqone}) and taken as systematic
uncertainties (see Section~\ref{idisocorrana}).
 
\subsection{Other backgrounds}
The background from electrons or positrons misidentified as photons,
mainly produced in Drell--Yan $Z^{(*)}/\gamma^*\rightarrow e^+e^-$ and
$W^{(*)}\rightarrow e\nu$ processes, is also studied. Such events are
largely suppressed by the photon selection. The remaining background
contribution is estimated using MC samples of fully simulated events
and found to be sub-percent for most of the phase space considered. No
subtraction is performed and a systematic uncertainty equal to
the size of the estimated background is assigned (see Section~\ref{noeb}).
 
\section{Unfolding kinematic distributions to particle level}
\label{cor}
Cross sections for isolated-photon production in association with at
least two jets are measured in the fiducial phase-space regions outlined in
Table~\ref{tableone}. The differential cross sections are obtained through
the formula
\begin{equation}
\frac{{\mathrm{d}}\sigma}{{\mathrm{d}}O}(i)=\frac{N^{\mathrm{unf,sig}}(i)}{{\cal L}\ \Delta O(i)}, \nonumber
\end{equation}
where ${\mathrm{d}}\sigma/{\mathrm{d}}O(i)$ is the cross section as a function of observable
$O$ in bin $i$, $N^{\mathrm{unf,sig}}(i)$ is the unfolded number of background-subtracted
data events in bin $i$ and is obtained as described below,  ${\cal L}$ is the integrated luminosity
and $\Delta O(i)$ is the width of bin $i$.
 
The particle-level isolation transverse energy
of the photon is calculated by summing the transverse energy of all
stable particles, except for muons and neutrinos, in a cone of size
$\Delta R = 0.4$ around the photon direction after the contribution
from the underlying event is subtracted. The same subtraction
procedure used on data is applied at the particle level; the particles
associated with the overlaid $pp$ collisions are not considered. The
particle-level requirement on $\etiso$ is determined using the \pyt\
and LO \sher\ MC samples by comparing the calorimeter isolation
transverse energy with the particle-level isolation on an
event-by-event basis. The effect of the experimental isolation
requirement used in the data is found to be close to a particle-level
requirement of $\etiso({\mathrm{particle}})<$~\etisocutp; the same
requirement is obtained whether \pyt\ or LO \sher\ is
used. Particle-level jets are reconstructed with the anti-$k_t$
algorithm with $R=0.4$ using all stable final-state particles as
input; muons, neutrinos and particles from pile-up activity are
excluded.
 
The efficiency of the selection, defined as the fraction of the
events generated in the fiducial region that are also reconstructed in
that region, is evaluated using MC simulations. The selection
efficiency using LO \sher\ (\pyt)  samples is $76\%$ ($76\%$), $74\%$
($77\%$), $71\%$ ($70\%$) for the total, fragmentation- and direct-enriched
phase-space regions, respectively.
 
The data distributions, after background subtraction, are unfolded to
the particle level using an unfolding method~\cite{nim:a362:487} based
on the iterative application of Bayes' theorem, as implemented in
RooUnfold~\cite{1105.1160}. In this method, an unfolding matrix is
constructed using the samples of simulated events. The unfolding matrix
encapsulates the migrations across bin boundaries of the observable when
switching between particle and reconstruction levels. The binning of the
distributions is chosen according to the resolution in those variables and in
some cases, such as at high $\etg$ or $\ptjet$, larger bin sizes are used to
reduce the statistical uncertainties in the data. There are observables that
are filled once per event, such as $\etg$, $\mjj$,  $\deltaphijj$,
$|\deltarapjj|$ and $\mgjj$, and observables which are filled twice
per event, such as $\ptjet$, $|\rapjet|$, $\deltaphigj$ and
$|\deltarapgj|$. The unfolding matrix is filled with those events that
pass the selection requirements at both the reconstruction and
particle levels. In addition, the reconstructed jet (photon) is
required to be matched to a particle-level jet (photon) within $\Delta
R =0.4$ ($0.2$). Corrections are also applied to take into account the
fractions of events for which there is no matching either at particle or
reconstruction level; the fraction of
unmatched particle-level (reconstruction-level) events is typically in
the range $20\%$--$40\%$ ($5\%$--$25\%$). In the evaluation of the
systematic uncertainties (see Section~\ref{syst}), modifications of
the unfolding matrix and of the corrections are applied simultaneously
in a consistent way. Four iterations of the unfolding procedure are
performed. The nominal cross sections are measured using the signal
leakage fractions and unfolding matrices
from LO \sher\ since these simulations describe the shape of the
distributions of the signal yield better than those of \pyt. The
results obtained by unfolding the data with \pyt\ are used to estimate
the model dependence and their deviations from the nominal
results are taken as systematic uncertainties, as explained in
Section~\ref{md}. The results of the Bayesian unfolding
procedure are checked with a bin-by-bin method based on LO \sher\
simulations, giving consistent results; the systematic uncertainty due
to the unfolding procedure is negligible.
 
\section{Experimental and theoretical uncertainties}
\label{systandthe}
 
\subsection{Uncertainties in the cross-section measurements}
\label{syst}
The sources of systematic uncertainty that affect the measurements are
the photon energy scale and resolution,  the photon-identification efficiency,
the photon-isolation modelling, the jet energy scale and resolution, the
parton-shower and hadronisation model dependence, the definition of the
background control regions, the identification and isolation correlation in the
background, the background from electrons or positrons faking photons,
the modelling of pile-up, the trigger efficiency and the luminosity measurement.
These sources of uncertainty and their effects are described below.
 
\subsubsection{Photon energy scale and resolution}
\label{geser}
The uncertainties in the photon energy scale and resolution are
described in Ref.~\cite{PERF-2017-03}. The sources
of uncertainty include: the uncertainty in the overall energy scale
adjustment using $Z\rightarrow e^+e^-$ decays; the uncertainty in the
non-linearity of the energy measurement at the calorimeter cell level;
the uncertainty in the relative calibration of the different
calorimeter layers; the uncertainty in the amount of material in front
of the calorimeter; the uncertainty in the modelling of the
reconstruction of photon conversions; the uncertainty in the modelling
of the lateral shower shape; the uncertainty in the modelling of the
sampling term;\footnote{The relative energy resolution is
parameterised as $\sigma(E)/E = a/\sqrt{E} \oplus b/E \oplus c$,
where $a$ is the sampling term, $b$ is the noise term and $c$ is
the constant term.} and the uncertainty
in the measurement of the constant term in $Z$-boson decays. The
uncertainties are split into independent components to account for
correlations and their $\eta$ dependence. The individual sources of
uncertainty are varied by $\pm 1\sigma$ in the MC simulations and
propagated through the analysis separately, to maintain the full
information about the correlations. The uncertainties are propagated
to the cross sections by modifying the unfolding matrix, the unmatched
fractions and the signal yields. The impact of the photon energy
resolution uncertainty is much smaller than that of the photon energy
scale uncertainty except for high $\etg$ ($\etg>500$~\GeV), low $\mjj$
($\mjj < 400$~\GeV) and low $\deltaphijj$ ($\deltaphijj < 0.6$~rad)
in the fragmentation-enriched region.
The resulting uncertainty in the measured cross
sections is obtained by adding in quadrature all the individual
components. For $\dsetg$ this uncertainty increases from $0.5\%$ at
$\etg=150$~\GeV\ to $6\%$ at $\etg=2$~\TeV.
 
\subsubsection{Photon-identification efficiency}
Scale factors are applied to simulated events to match the efficiency
for tight photon identification measured in data~\cite{epj:c79:205}.
The uncertainties in the scale factors are propagated to the measured
cross sections by modifying the unfolding matrix, the unmatched
fractions and the signal yields. The resulting uncertainty in the measured cross sections is
in the range $1.5\%$--$2\%$.
 
\subsubsection{Photon-isolation modelling}
\label{isocorsyst}
{\textit{In situ}} corrections to $\etiso$ are applied in simulated events
for the nominal results. The differences between the nominal results
and those obtained without applying the aforementioned corrections are
taken as systematic uncertainties due to the modelling of $\etiso$ in
the MC simulation. The uncertainties are propagated to the cross sections
by modifying the unfolding matrix, the unmatched fractions and the signal
yields; the resulting uncertainties are not symmetrised. The resulting
uncertainty in the measured cross sections is typically smaller than $1\%$.
\nopagebreak
\subsubsection{Jet energy scale and resolution}
The uncertainties in the jet energy scale and resolution are
considered using the method reported in Ref.~\cite{pr:d96:072002}. The
sources of uncertainty include: the uncertainty in the overall energy
scale adjustment using $Z+{\mathrm{jet}}$,
$\gamma+{\mathrm{jet}}$\footnote{The correlation between the
uncertainty in the photon energy scale and the uncertainty in the
jet energy scale arising from the {\textit{in situ}} method based on
$\gamma+{\mathrm{jet}}$ events has negligible impact.} and
multi-jet $\pt$-balance methods; the uncertainty due to modelling,
statistics and calibration non-closure in the $\eta$ intercalibration;
the uncertainty from single-particle response studies of high-$\pt$
jets;  the uncertainty due to the modelling of the pile-up
contribution; the uncertainty in the jet response and simulated jet
composition of light-quark, $b$-quark, and gluon-initiated jets; and the
uncertainty due to punch-through jets.\footnote{High-$\pt$ jets that
are not fully contained in the calorimeter are referred to as
punch-through jets.} The individual sources of uncertainty are
varied by $\pm 1\sigma$ in the MC simulations and propagated through
the analysis separately, to maintain the full information about the
correlations. The uncertainties are propagated to the cross sections
by modifying the unfolding matrix and the unmatched fractions.
The resulting uncertainty in the measured cross sections
is obtained by adding in quadrature all the individual components.
For $\dspt$ this uncertainty is in the range $3\%$--$8\%$.
 
\subsubsection{Parton-shower and hadronisation model dependence}
\label{md}
The dependence of the signal purity and unfolding corrections on the
parton-shower and hadronisation models used in the simulations
are studied separately. The differences observed in the signal purity
between the nominal results and those obtained using the \pyt\ MC
samples for the determination of the signal leakage fractions are taken
as systematic uncertainties. The uncertainties are propagated
to the cross sections by modifying the signal yields. The resulting
uncertainties in the measured cross sections are typically smaller
than $1\%$. The differences between the nominal results and those
obtained using the \pyt\ MC samples for the unfolding are taken as
systematic uncertainties and are typically smaller than $2\%$. In both
cases, for the signal leakage fractions and for the unfolding, the
uncertainties also account for a possible bias due to the
application of the Frixione's criterion at particle level in the LO
\sher\ MC samples.
 
\subsubsection{Definition of the background control regions}
The signal yield depends on the definition of the background control
regions. The dependence is investigated by (i) varying the lower limit
on $\etiso$ of regions $B$ and $D$ by $\pm 1$~\GeV, (ii) removing the
upper limit on $\etiso$ of regions $B$ and $D$, and (iii) changing the
choice of inverted photon identification variables used in the
selection of non-tight photons. The signal yields
obtained after each variation are compared with the nominal results and
the differences are taken as systematic uncertainties. The uncertainties
are propagated to the cross sections by modifying the signal yields.
The uncertainty in the measured cross sections is typically below $0.2\%$, $0.2\%$,
and $1\%$ for the (i), (ii) and (iii) variations, respectively.
 
\subsubsection{Identification and isolation correlation in the
background}
\label{idisocorrana}
In the {\textit{in situ}} background subtraction (Eq.~(\ref{eqone})),
$\rbg$ is assumed to be unity. A range in $\rbg$ is set to cover the
deviations from unity observed in the validation regions after
subtracting the signal leakage with either \pyt\ or LO \sher\ MC
samples. The maximum deviations of $\rbg$ from unity in the validation regions
vary in the range $10\%$--$50\%$ depending on the observable.
The uncertainties are propagated to the cross sections by
modifying the signal yields. The resulting uncertainty in the measured
cross sections is typically smaller than $1\%$.
 
\subsubsection{Background from electrons or positrons faking photons}
\label{noeb}
A systematic uncertainty is assigned by taking the full size of the
estimated contribution; the $W+{\mathrm{jets}}$ and
$Z+{\mathrm{jets}}$ contributions are added linearly.
The uncertainties are propagated to the cross sections by
modifying the signal yields. The resulting
uncertainty in the measured cross sections is typically smaller than
$1\%$.
 
\subsubsection{Pile-up, trigger efficiency and integrated luminosity}
The pile-up reweighting of simulated events is varied to cover the
uncertainty in the ratio of the predicted and measured inelastic cross
sections~\cite{prl:117:182002}. The uncertainties are propagated
to the cross sections by modifying the unfolding matrix, the unmatched
fractions and the signal yields. The resulting uncertainty in the
measured cross sections is typically smaller than $1\%$. There might
be double counting between this source of uncertainty and that in the
modelling of the photon isolation. However, given the smallness of the
uncertainties, the impact is limited.
The uncertainty in the trigger efficiency is estimated using the same
methodology as in Ref.~\cite{1909.00761} and amounts to $0.22\%$ for
$150<\etg<175$~\GeV, $0.08\%$ for $175<\etg<200$~\GeV,
$0.04\%$ for $200<\etg<470$~\GeV, and $0.40\%$ for
$470<\etg<2500$~\GeV. It is propagated linearly to the measured cross
sections.
The uncertainty in the integrated luminosity is $2.1\%$ and is fully
correlated between all bins of all the measured cross sections.
 
\subsubsection{Total systematic uncertainty}
The total systematic uncertainty is computed by adding in quadrature
the uncertainties from the sources listed above and the statistical
uncertainty from the MC samples. The latter is computed using the bootstrap
resampling technique~\cite{ISBN:978-3-935702-41-6}.
There are large correlations in the systematic uncertainties across bins of one
observable, particularly in the uncertainties due to the photon and jet energy
scales, which are dominant. The total systematic uncertainty in the measured cross
sections, excluding the uncertainty of the luminosity determination, for the
total phase space varies in the range shown in Table~\ref{tabletwo} depending on the
variable.
 
\begin{table}[ht!]
\renewcommand{\arraystretch}{1.2}
\caption{
The total systematic uncertainty in the measured cross
sections, excluding the uncertainty of the luminosity determination, for the
total phase space: the range for each variable is shown in percentage.
}
\setlength{\unitlength}{1.0cm}
\begin{picture} (18.0,3.0)
\put (0.0,1.5){\scalebox{1.0}{\centerline{
\begin{tabular}{|c|c|c|c|c|c|c|c|c|} \hline
\multicolumn{9}{|c|}{\textbf{Range of the relative uncertainty (in $\%$) for each variable}} \\ \hline
$\etg$ & $\ptjet$ & $|\rapjet|$ & $|\deltarapgj|$ & $\deltaphigj$ & $|\deltarapjj|$ &
$\deltaphijj$ & $\mjj$ & $\mgjj$  \\ \hline
$3.5\%$--$6.5\%$  &   $4\%$--$15\%$ & $4\%$--$6\%$ & $4\%$--$9\%$ &
$3.5\%$--$6\%$  & $4\%$--$8\%$ & $3\%$--$7\%$ & $4\%$--$9\%$ & $4\%$--$11\%$ \\ \hline
\end{tabular}
}}}
\end{picture}
\label{tabletwo}
\end{table}
 
\begin{figure}[p]
\setlength{\unitlength}{1.0cm}
\begin{picture} (18.0,21.0)
\put (1.0,14.0){\includegraphics[width=7cm]{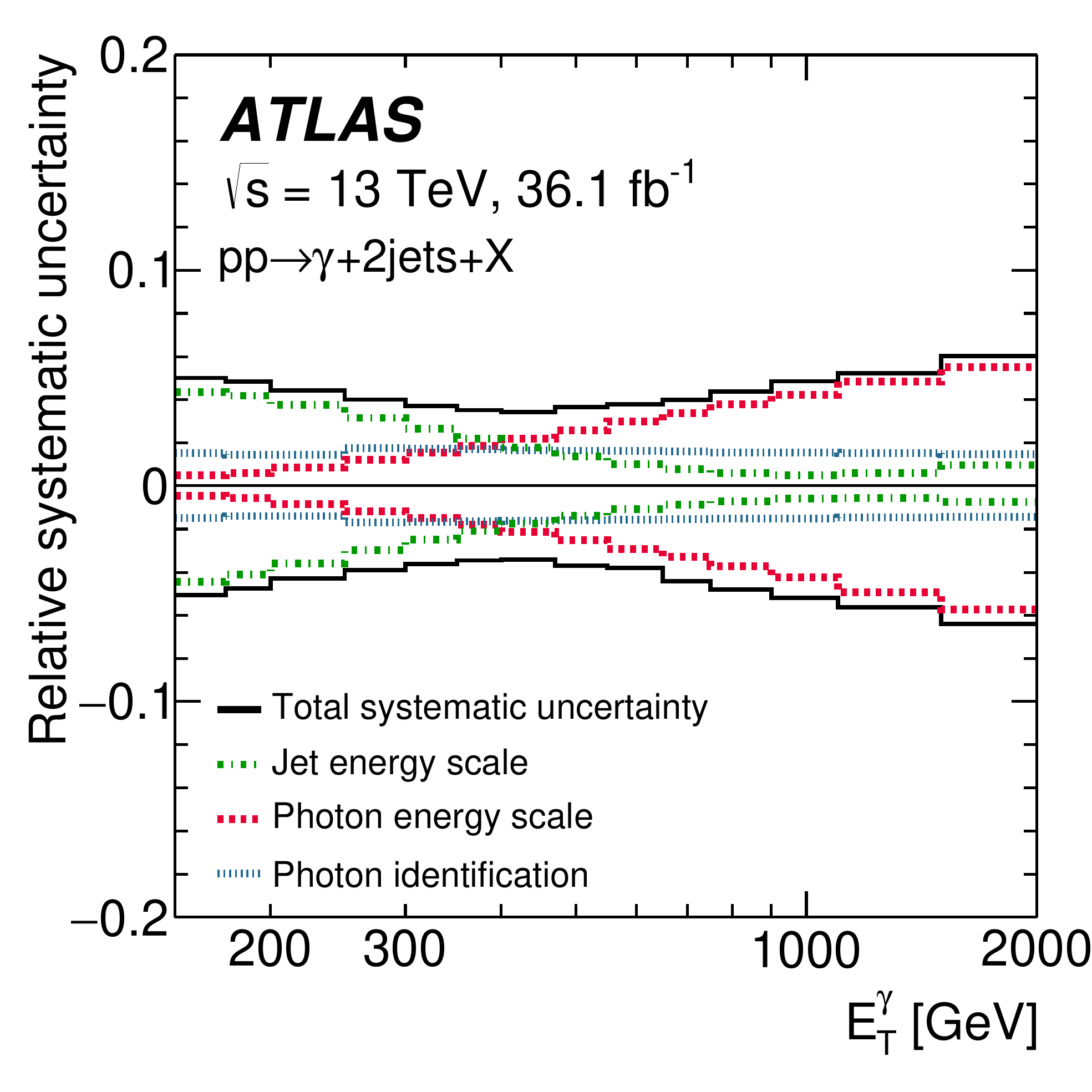}}
\put (8.0,14.0){\includegraphics[width=7cm]{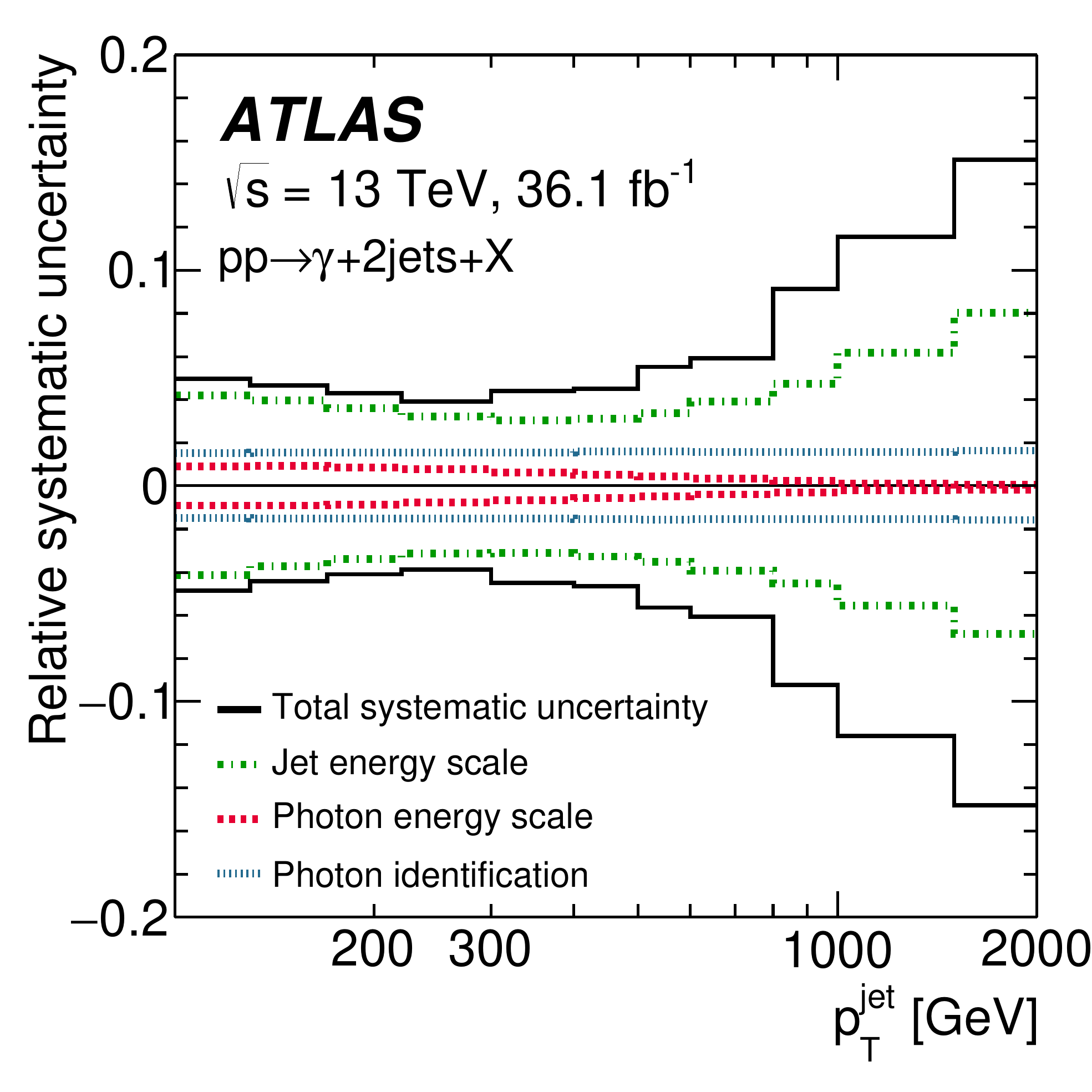}}
\put (1.0,7.0){\includegraphics[width=7cm]{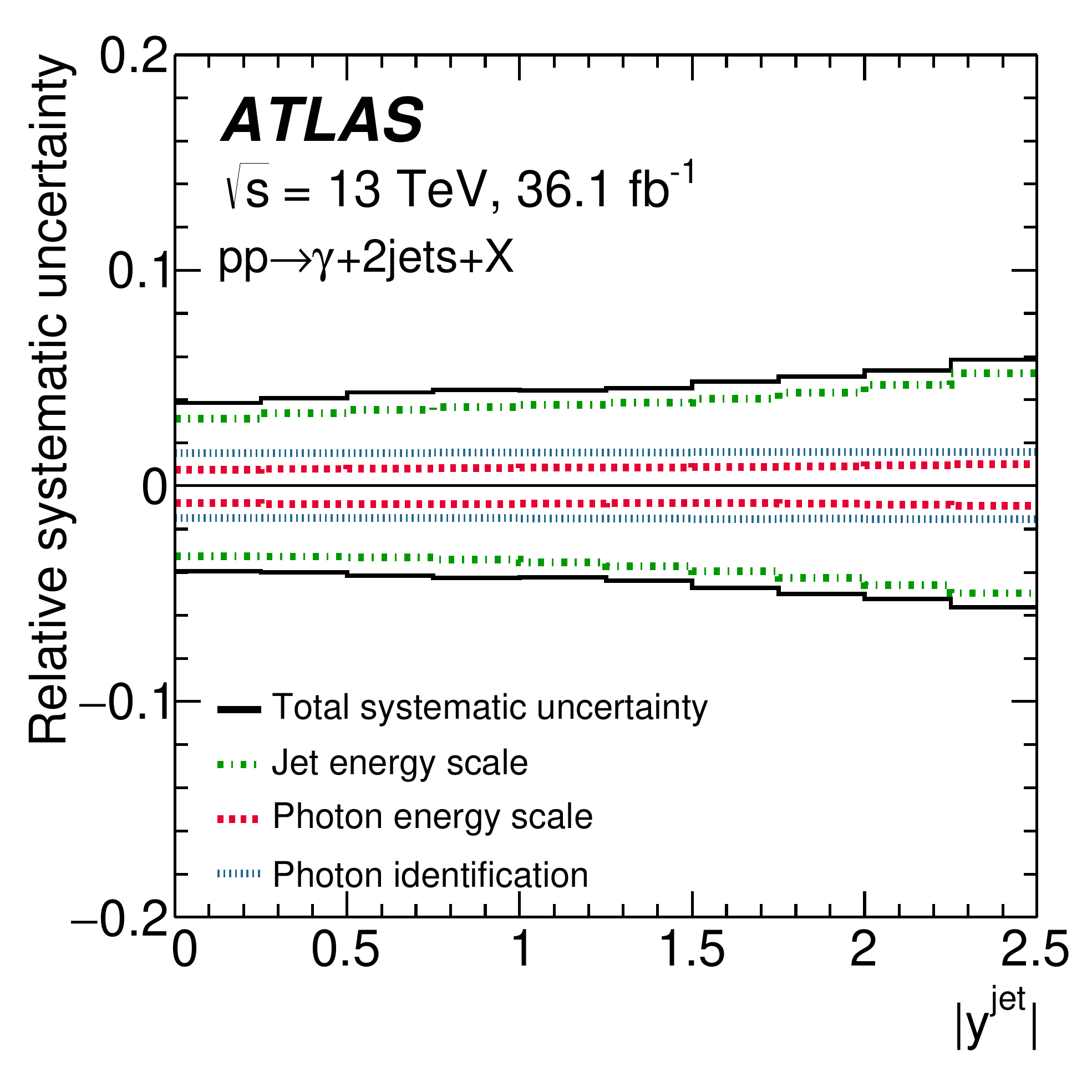}}
\put (1.0,0.0){\includegraphics[width=7cm]{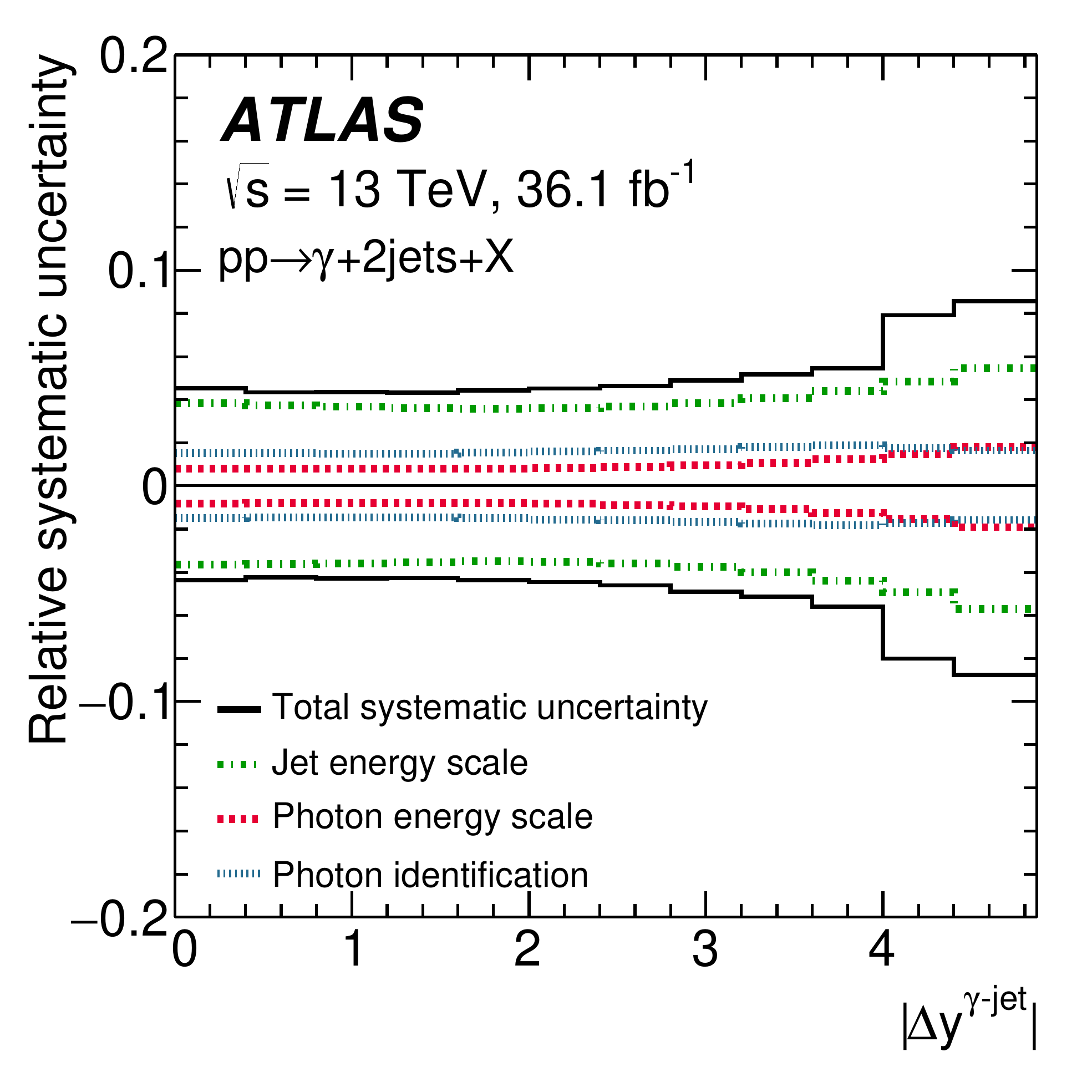}}
\put (8.0,0.0){\includegraphics[width=7cm]{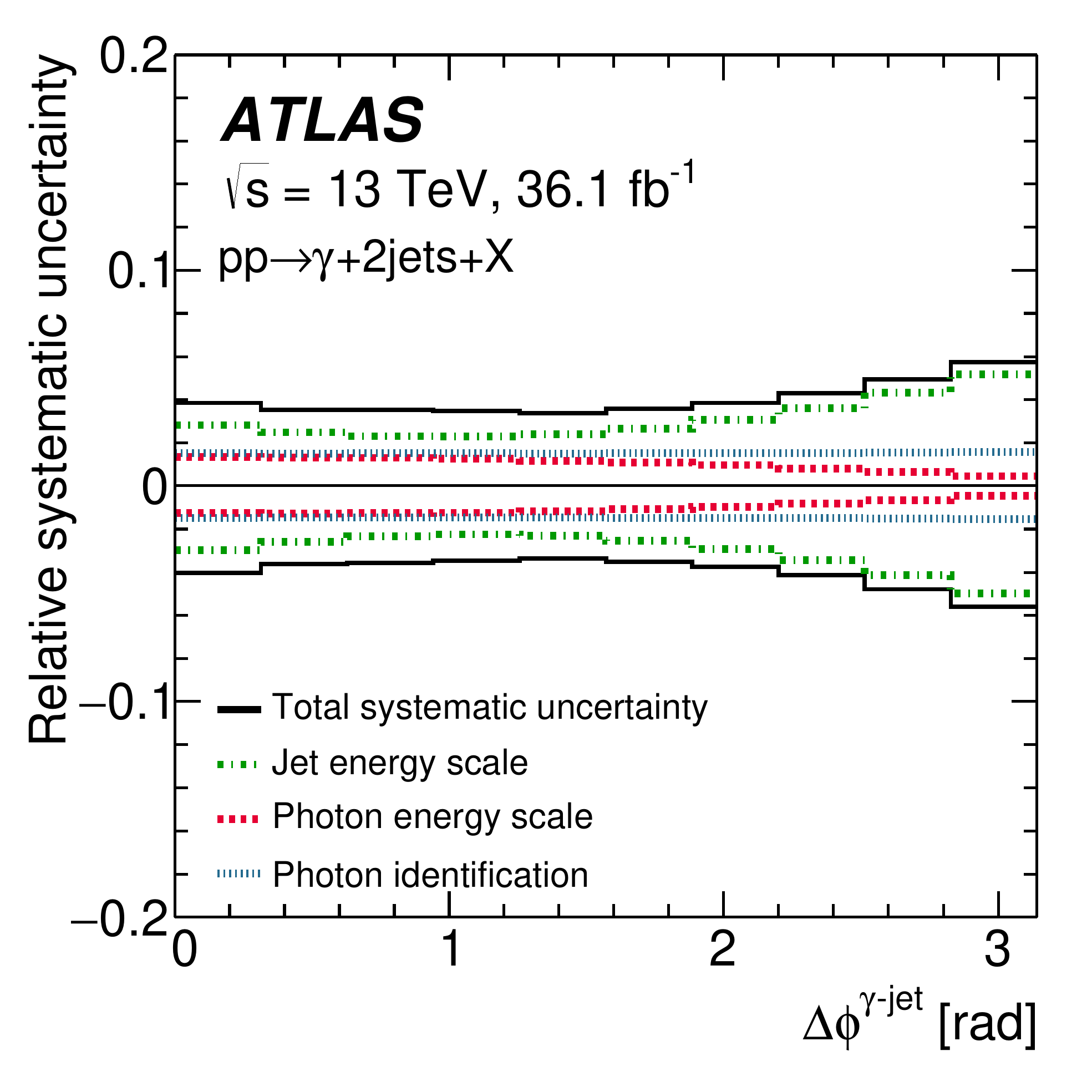}}
\put (4.0,14.5){{\small (a)}}
\put (11.0,14.5){{\small (b)}}
\put (4.0,7.5){{\small (c)}}
\put (4.0,0.5){{\small (d)}}
\put (11.0,0.5){{\small (e)}}
\end{picture}
\caption
{
Total relative systematic uncertainty in the measured cross-sections (black lines), excluding the
uncertainty of the luminosity determination, together with the three main
contributions, namely jet energy scale (dot-dashed lines), photon
energy scale (dashed lines) and photon identification (dotted
lines), as a function of (a) $\etg$, (b) $\ptjet$, (c) $|\rapjet|$,
(d) $|\deltarapgj|$ and (e) $\deltaphigj$ for the total phase
space.
}
\label{fig02}
\end{figure}

\begin{figure}[p]
\setlength{\unitlength}{1.0cm}
\begin{picture} (18.0,17.0)
\put (1.0,7.0){\includegraphics[width=7cm]{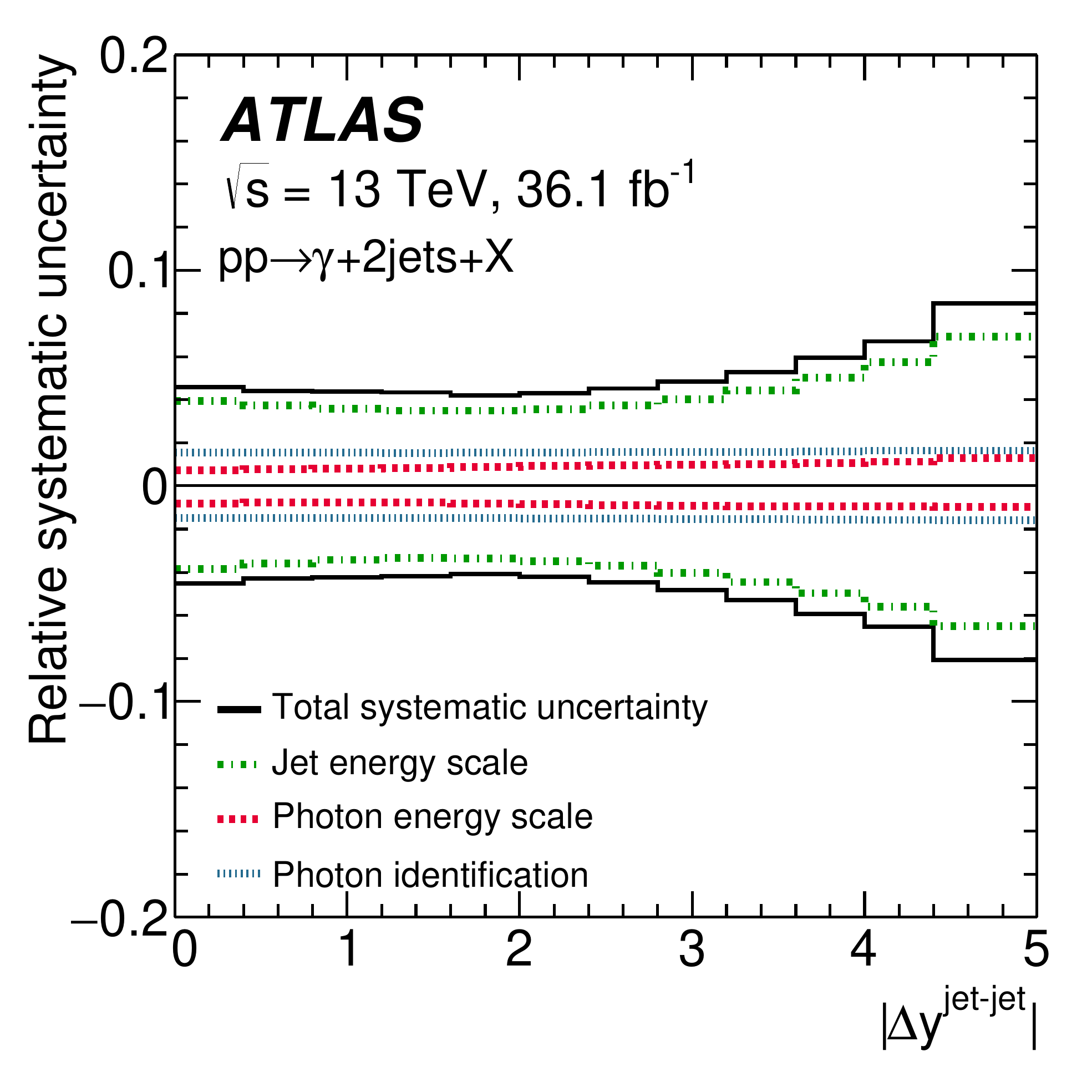}}
\put (8.0,7.0){\includegraphics[width=7cm]{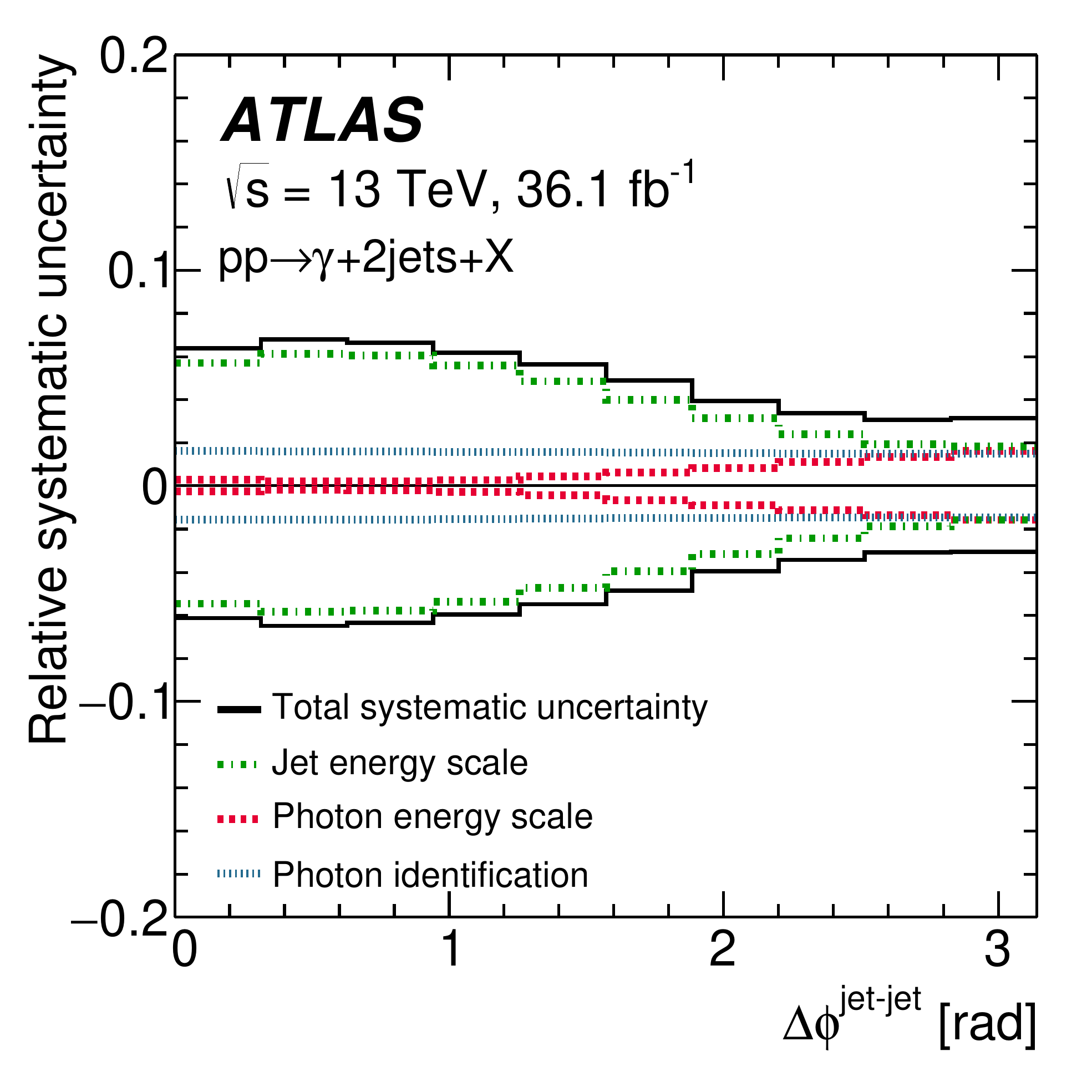}}
\put (1.0,0.0){\includegraphics[width=7cm]{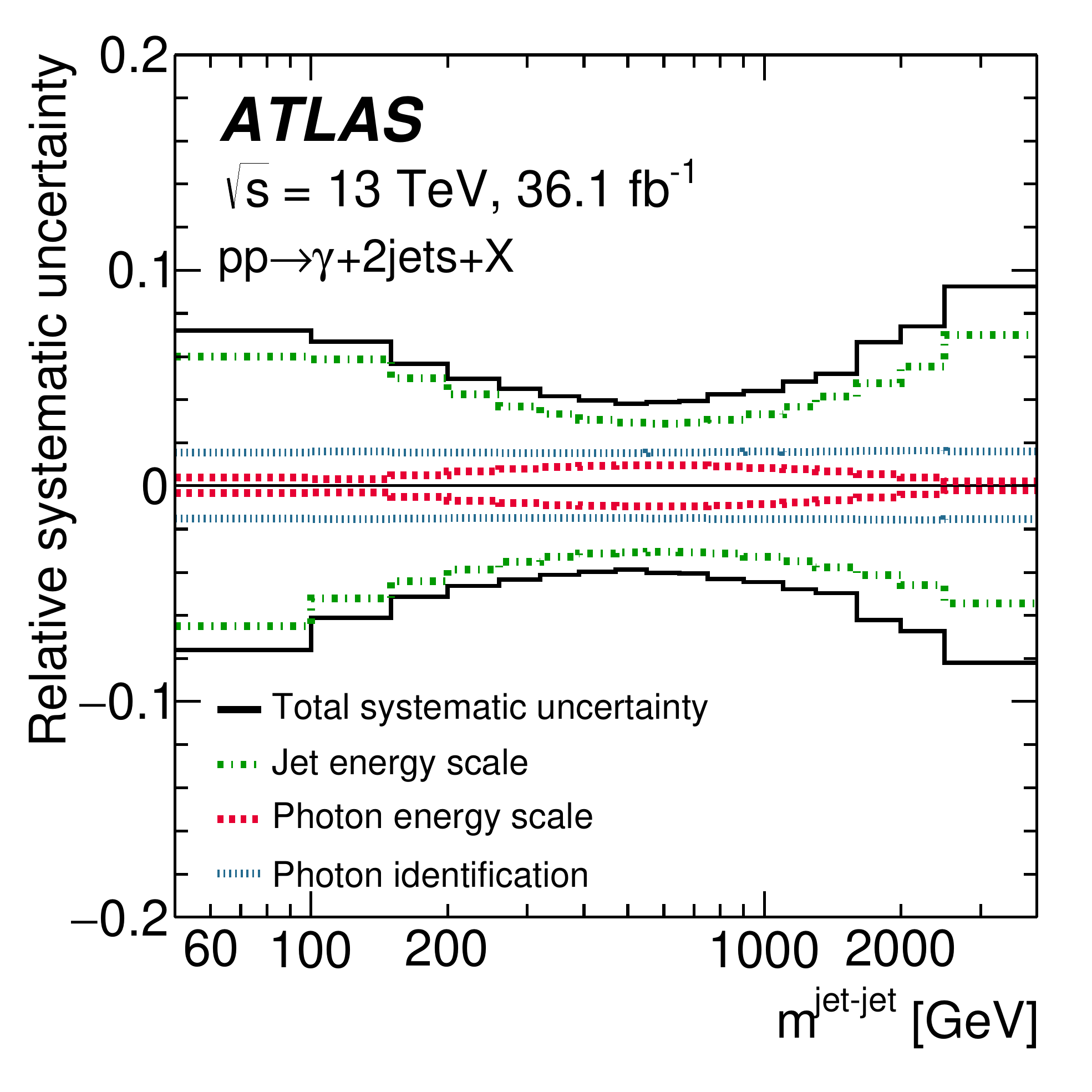}}
\put (8.0,0.0){\includegraphics[width=7cm]{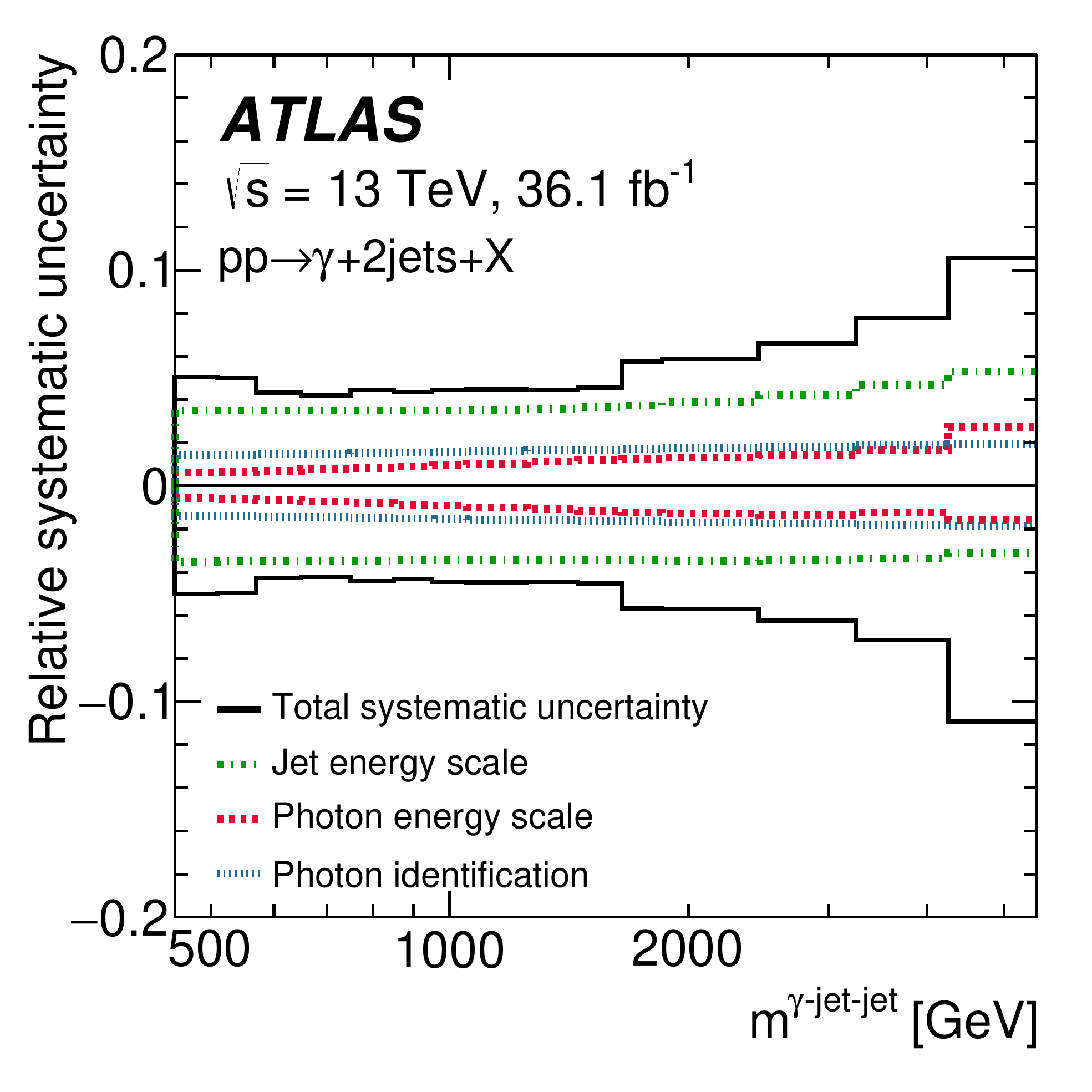}}
\put (4.0,7.5){{\small (a)}}
\put (11.0,7.5){{\small (b)}}
\put (4.0,0.5){{\small (c)}}
\put (11.0,0.5){{\small (d)}}
\end{picture}
\caption
{
Total relative systematic uncertainty in the measured cross-sections (black lines), excluding the
uncertainty of the luminosity determination, together with the three main
contributions, namely jet energy scale (dot-dashed lines), photon
energy scale (dashed lines) and photon identification (dotted
lines), as a function of (a) $|\deltarapjj|$, (b) $\deltaphijj$, (c)
$\mjj$ and (d) $\mgjj$ for the total phase space.
}
\label{fig03}
\end{figure}

Figures~\ref{fig02} and \ref{fig03} show the total systematic uncertainty,
excluding the uncertainty of the luminosity determination, for each
measured cross section for the total phase space. The dominant
components, namely the jet energy scale, the photon energy scale and
the photon identification, are shown separately in these figures.
There are variables such as $\deltaphijj$ for which the systematic
uncertainties due to the jet and photon energy scales depend on
the variable itself. These dependencies arise from the event topology:
at low $\deltaphijj$ the two jets are close together and recoil against
the photon whereas at high $\deltaphijj$ the two jets recoil against each
other and the photon has lower $\pt$. Thus, the populations in jet and
photon $\pt$ vary with $\deltaphijj$.
In the total phase space, the systematic uncertainty dominates the total
experimental uncertainty for $\etg\lesssim 1$~\TeV, $\ptjet\lesssim
1.5$~\TeV\ and $\mgjj\lesssim 4$~\TeV. For higher $\etg$, $\ptjet$ and
$\mgjj$ values, the statistical uncertainty of the data limits
the precision of the measurements. For other observables, the
systematic uncertainty dominates in the whole measured range. In
particular, in the tails of the distributions of $\ptjet$, $|\deltarapgj|$
and $\mgjj$ the contributions from the statistical uncertainties of the
MC samples and the systematic uncertainty in the unfolding correction
due to the parton-shower and hadronisation model dependence
can be as large as the systematic uncertainty due to the jet energy scale.
 
\subsection{Theoretical uncertainties in the predictions}
\label{thunc}
The following sources of uncertainty in the theoretical predictions
are considered. The uncertainty in the NLO pQCD predictions from
\sher\ due to terms beyond NLO is estimated by repeating the
calculations using values of the renormalisation ($\muR$) and
factorisation ($\muF$) scales multiplied by the factors $0.5$ or $2$. The
two scales are varied either simultaneously or individually. In all
cases, the condition $0.5\leq\muR/\muF\leq 2$ is imposed.
The largest deviation from each nominal prediction
among the six possible variations is taken as the uncertainty. The
uncertainty in the predictions due to the proton PDFs is estimated by
using 100 replicas from the NNPDF3.0
analysis~\cite{jhep:1504:040}. The uncertainty in the predictions due
to the uncertainty in $\as$ is estimated by repeating the calculations
using two additional sets of proton PDFs from the NNPDF3.0 analysis,
for which different values of $\as(\mz)$ were assumed in the fits,
namely $0.117$ and $0.119$; in this way, the correlation between $\as$
and the PDFs is preserved.
 
The total theoretical uncertainty is obtained by adding in quadrature
the individual uncertainties mentioned above. The dominant theoretical
uncertainty is that arising from terms beyond NLO and typically varies
between about $20\%$ and about $40\%$, depending on the observable and region of
phase space. The uncertainty in the predictions arising from those in
the PDFs is below $6\%$ and that arising from the value of $\asz$ is
below $6\%$ for all observables and phase-space regions.
 
\section{Results}
\label{res}
The measured isolated-photon plus two-jets cross sections in the three
phase-space regions are shown in Figures~\ref{fig04} to \ref{fig09}.
 
The measured $\dsetg$ is shown in Figure~\ref{fig04}(a) for the total
phase space. The cross section decreases by approximately five orders
of magnitude in the $150$--$2000$~\GeV\ range. The measured $\dsetg$
for the fragmentation- and direct-enriched phase-space regions are shown in
Figures~\ref{fig06}(a) and \ref{fig08}(a), respectively; it is observed
that the $\etg$ spectrum for the direct-enriched phase space is harder
than that for the fragmentation-enriched one.
 
The measured $\dspt$ for the total phase space, shown in
Figure~\ref{fig04}(b), decreases by approximately five orders of
magnitude in the measured range. Values of $\ptjet$ up to $2$~\TeV\
are measured. The measured $\dspt$ for the fragmentation-enriched
phase space (see Figure~\ref{fig06}(b)) has a harder spectrum than that
for the direct-enriched one (see Figure~\ref{fig08}(b)).
 
Figure~\ref{fig04}(c) shows the measured $\dsyj$ for the total phase
space. The measured cross section decreases as $|\rapjet|$
increases. The measured $\dsyj$ for the fragmentation- and
direct-enriched phase-space regions are shown in Figures~\ref{fig06}(c) and
\ref{fig08}(c), respectively, and are similar for $|\rapjet|<1$, but for
$1<|\rapjet|<2.5$ the spectrum for the fragmentation-enriched phase space
decreases faster as $|\rapjet|$ increases.
 
The measured $\dsrapgj$ for the total, fragmentation-enriched and
direct-enriched phase-space regions are shown in Figures~\ref{fig04}(d),
\ref{fig06}(d) and \ref{fig08}(d), respectively. The measured cross
sections decrease as $|\deltarapgj|$ increases in all three phase-space regions.
 
The measured $\dsphigj$ for the total phase space is shown in
Figure~\ref{fig04}(e). The measured cross sections
for the total and direct-enriched (see Figure~\ref{fig08}(e)) phase-space
regions increase as $\deltaphigj$ increases and display a maximum at
$\deltaphigj\sim 2.8$~rad. In contrast,  $\dsphigj$ for the
fragmentation-enriched phase space (see Figure~\ref{fig06}(e)) exhibits
a peak at lower $\deltaphigj$ values, $\deltaphigj\sim 2.2$~rad. The
difference between the distributions of $\deltaphigj$ for the direct- and
fragmentation-enriched phase-space regions is partly due to the kinematical
constraints that define those regions, namely $\etg>\ptjetl$ and $\etg<\ptjetsl$,
respectively.
 
Figures~\ref{fig05}(a), \ref{fig07}(a) and \ref{fig09}(a) show the measured
$\dsrapjj$ for the total, fragmentation- and direct-enriched phase-space
regions, respectively. The measured cross sections have similar shapes
in the three phase-space regions. However, the measured $\dsphijj$ (see
Figures~\ref{fig05}(b), \ref{fig07}(b) and \ref{fig09}(b)) have very
different shapes in the three phase-space regions. For the
fragmentation-enriched phase space, the two jets primarily populate
the range $\deltaphijj>2.2$~rad whereas for the direct-enriched one
they mostly populate the range $0<\deltaphijj<2.2$~rad. The jet radius
parameter of $R=0.4$ has an effect for $\deltaphijj<0.4$ as can be seen
in Figures~\ref{fig05}(b) and \ref{fig09}(b). The kinematical constraints
that define the direct- and fragmentation-enriched phase-space regions
have also an effect on the distributions of $\deltaphijj$; for instance,
$\dsphijj$ for the fragmentation-enriched phase-space region is suppressed
at low $\deltaphijj$.
 
The dependence of the cross sections on $\mjj$ and $\mgjj$ is measured
up to values of $4$~\TeV\ and $5.5$~\TeV, respectively. For $\dsmjj$
(see Figures~\ref{fig05}(c), \ref{fig07}(c) and \ref{fig09}(c)), the
measured spectra have different shapes in the three phase-space regions. The
cross section for the fragmentation-enriched phase space is suppressed
at low $\mjj$ values, but it is harder than that for the
direct-enriched one at high $\mjj$ values. The kinematical constraints have
some effect on the differences between the distributions of $\mjj$ for the direct-
and fragmentation-enriched phase-space regions. For $\dsmgjj$ (see
Figures~\ref{fig05}(d), \ref{fig07}(d) and \ref{fig09}(d)), the shapes of
the spectra are more similar, but the distribution for the
fragmentation-enriched phase space is harder than for the
direct-enriched one.
 
\subsection{Comparison with tree-level plus parton-shower Monte Carlo
models}
\label{compalo}
The predictions of the tree-level plus parton-shower MC models of
\pyt\ and \sher\ are compared with the data in Figures~\ref{fig04} to
\ref{fig09}. These predictions are normalised to the measured
integrated cross section in each phase space to compare the shapes
of the predictions with the data. The normalisation factors relative to
the cross-section predictions of the generators are indicated in
Figures~\ref{fig04} to \ref{fig09}. Being tree-level predictions, the
theoretical uncertainties in these models can be as large as $50\%$
and are not included in the figures.
 
The predictions of LO \sher\ provide a good description of the shape
of the data, except at high $\etg$, $|\deltarapjj|$ and $\mgjj$. The
predictions of \pyt, in general, fail to describe the shape of the
data. There are two reasons for this disagreement. First, since \pyt\ uses
$2\rightarrow 2$ matrix elements, the second jet necessarily originates
from the parton shower; \sher\ incorporates matrix elements for the
processes $2\rightarrow n$ with $n=2$ to $5$ and thus the second jet
is better simulated. Second, the fragmentation component in
\pyt\ is overestimated for the parameter settings used here. This is
particularly visible in the cross
section for the total phase space as a function of $\ptjet$ in the
region $\ptjet > 300$~GeV, where the prediction overestimates the data
by $50\%$ or more. It is also visible in the cross section for the
total phase space as a function of $\mjj$ and $\mgjj$, where the
predictions of \pyt\ overestimate the data by up to an order of
magnitude. The overestimation of the fragmentation component is also
supported by the fact that the normalisation factor for the
predictions from \pyt\ is $1.2$ ($0.7$) for the total
(fragmentation-enriched) phase space. The inclusion of higher-order
tree-level matrix elements for the processes $2\rightarrow n$ with
$n=3$ to $5$ in \sher\ significantly improves the description of the
data for all observables and phase-space regions.
 
\subsection{Comparison with next-to-leading-order plus parton-shower QCD
predictions}
\begin{figure}[h]
\setlength{\unitlength}{1.0cm}
\begin{picture} (18.0,20.5)
\put (1.0,14.0){\includegraphics[width=7cm]{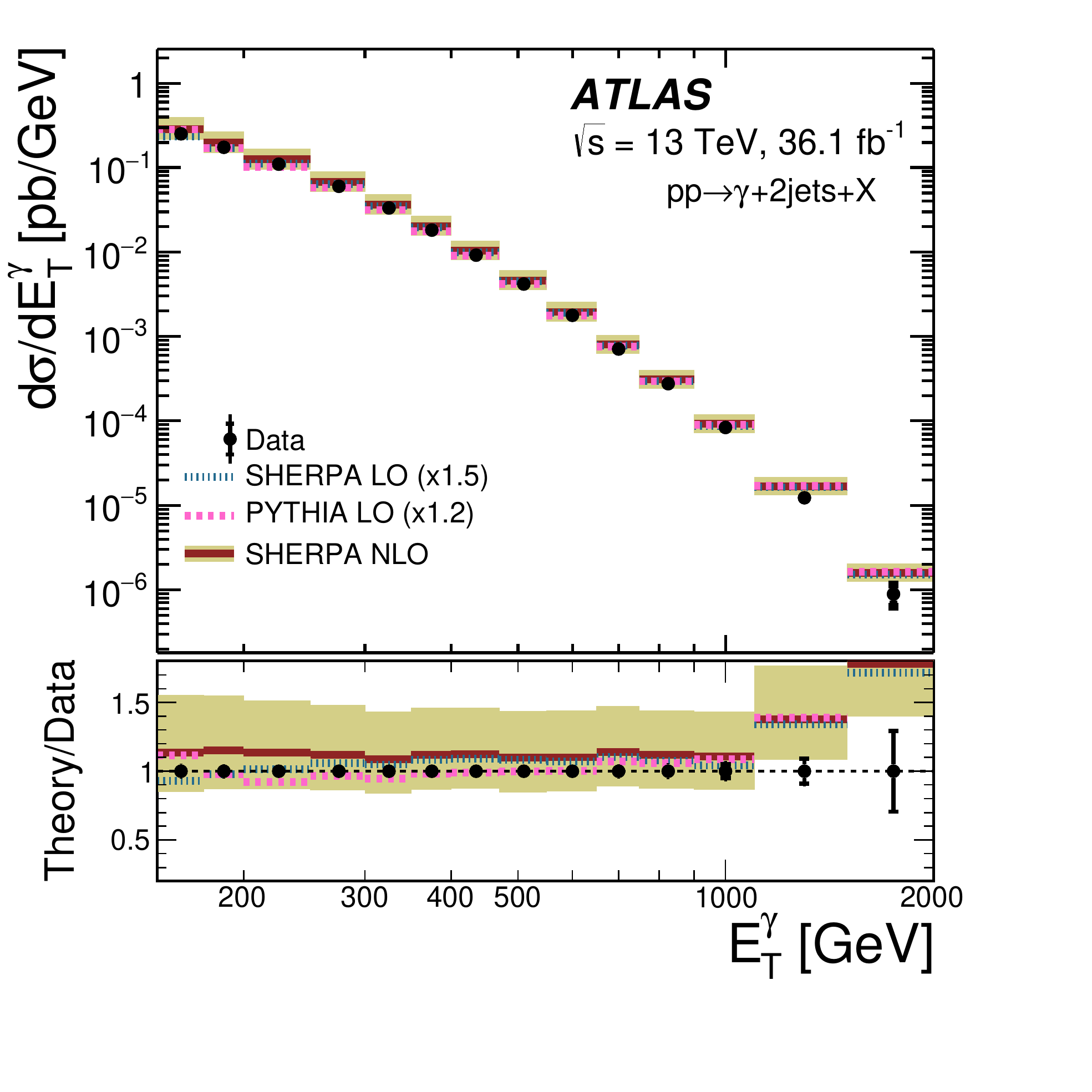}}
\put (8.0,14.0){\includegraphics[width=7cm]{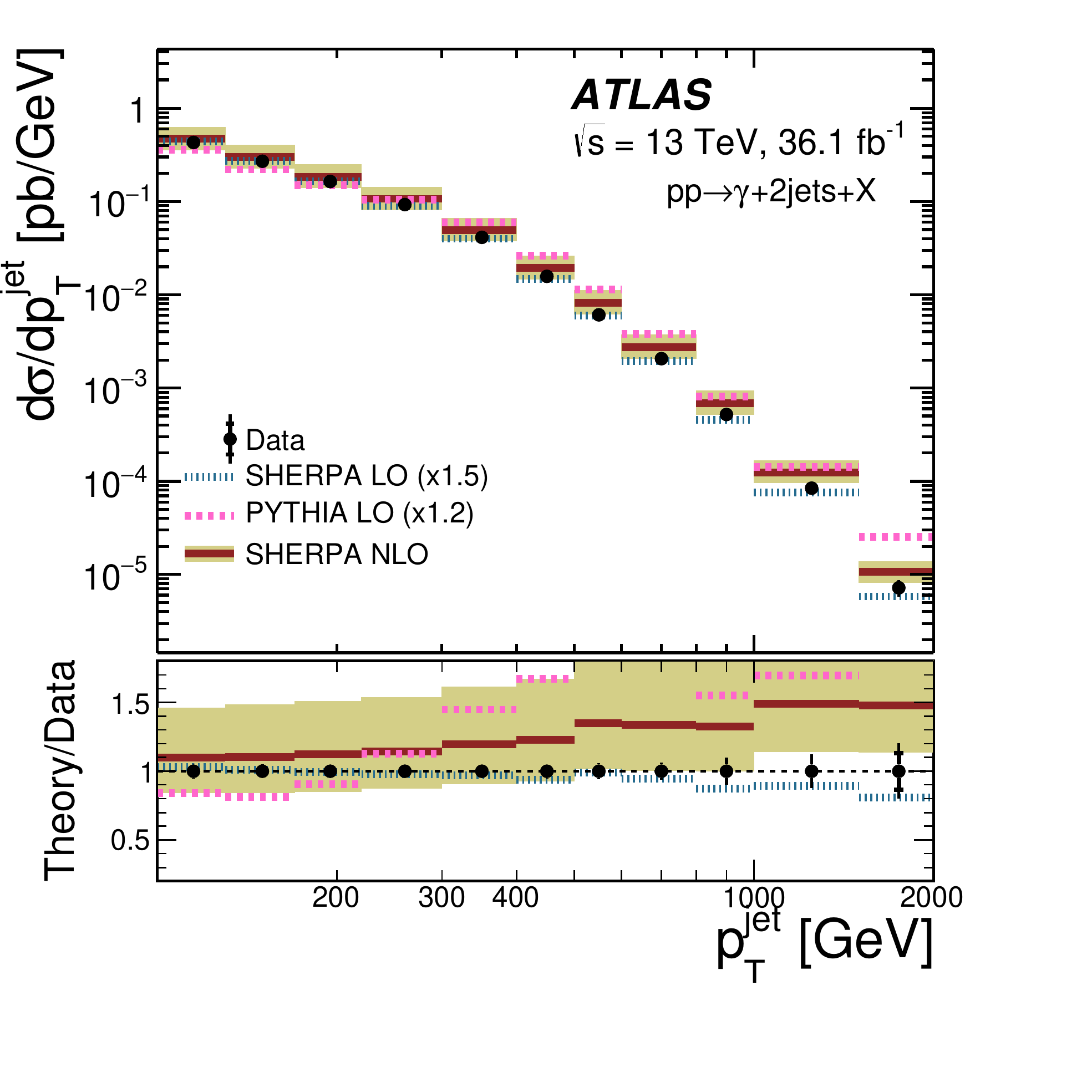}}
\put (1.0,7.5){\includegraphics[width=7cm]{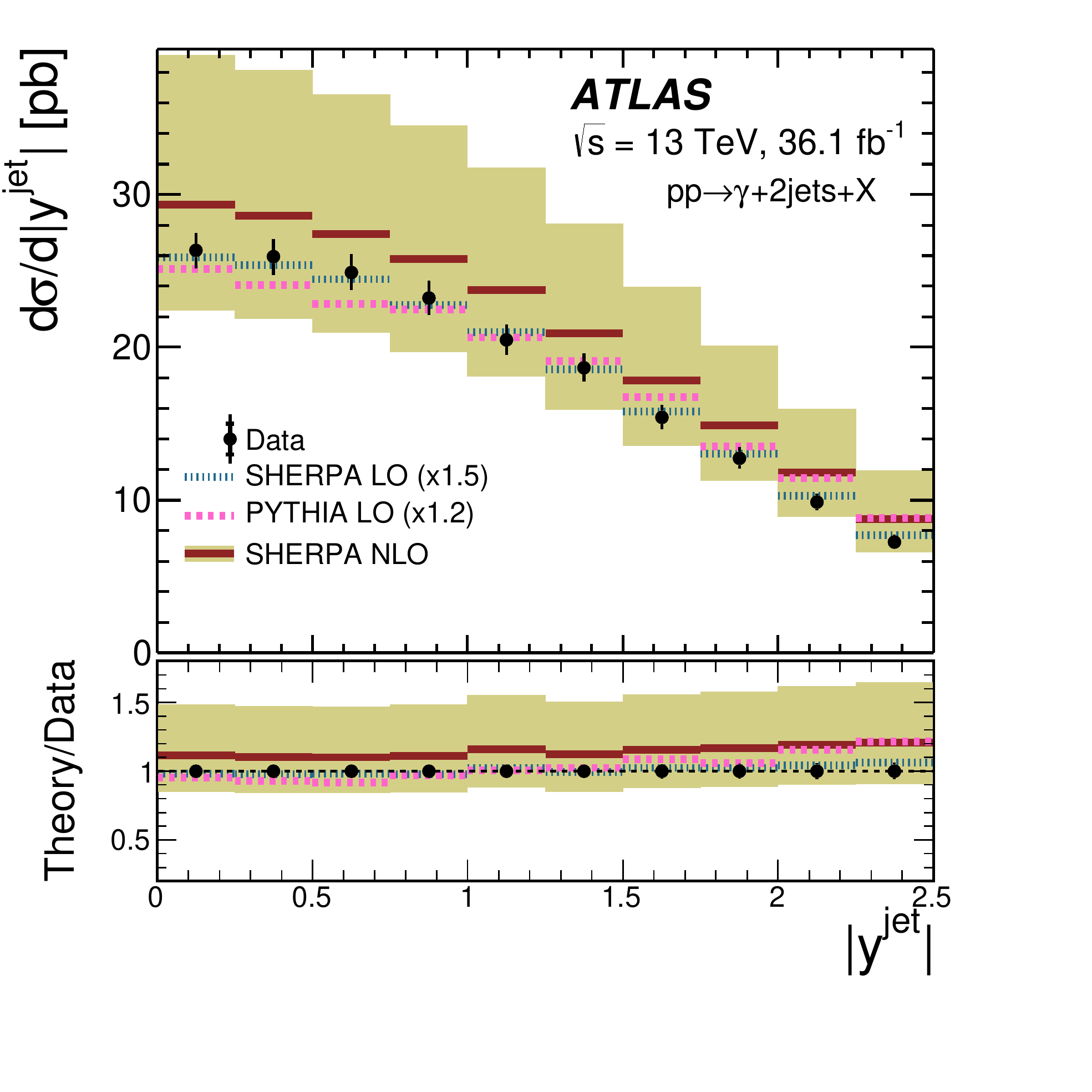}}
\put (1.0,1.0){\includegraphics[width=7cm]{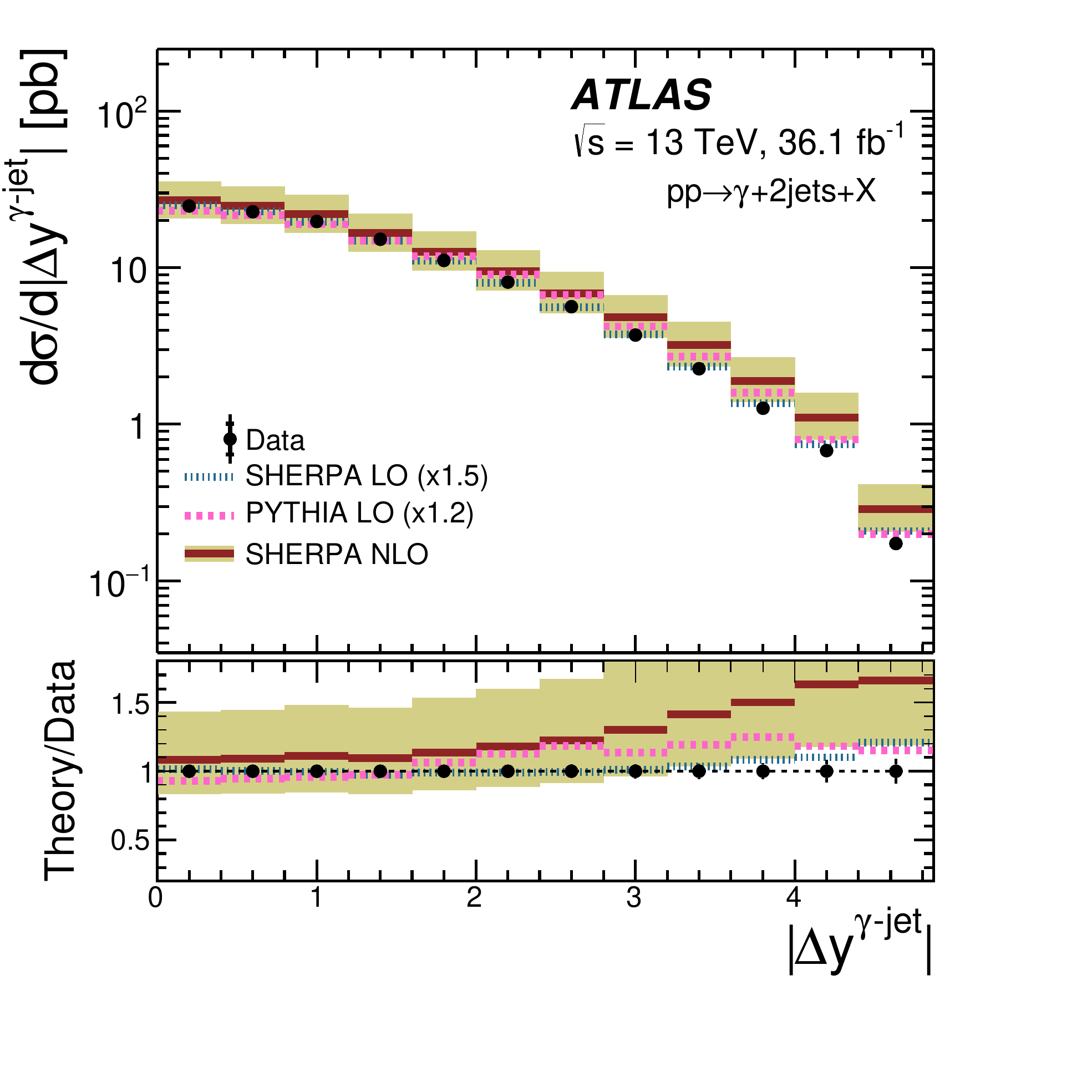}}
\put (8.0,1.0){\includegraphics[width=7cm]{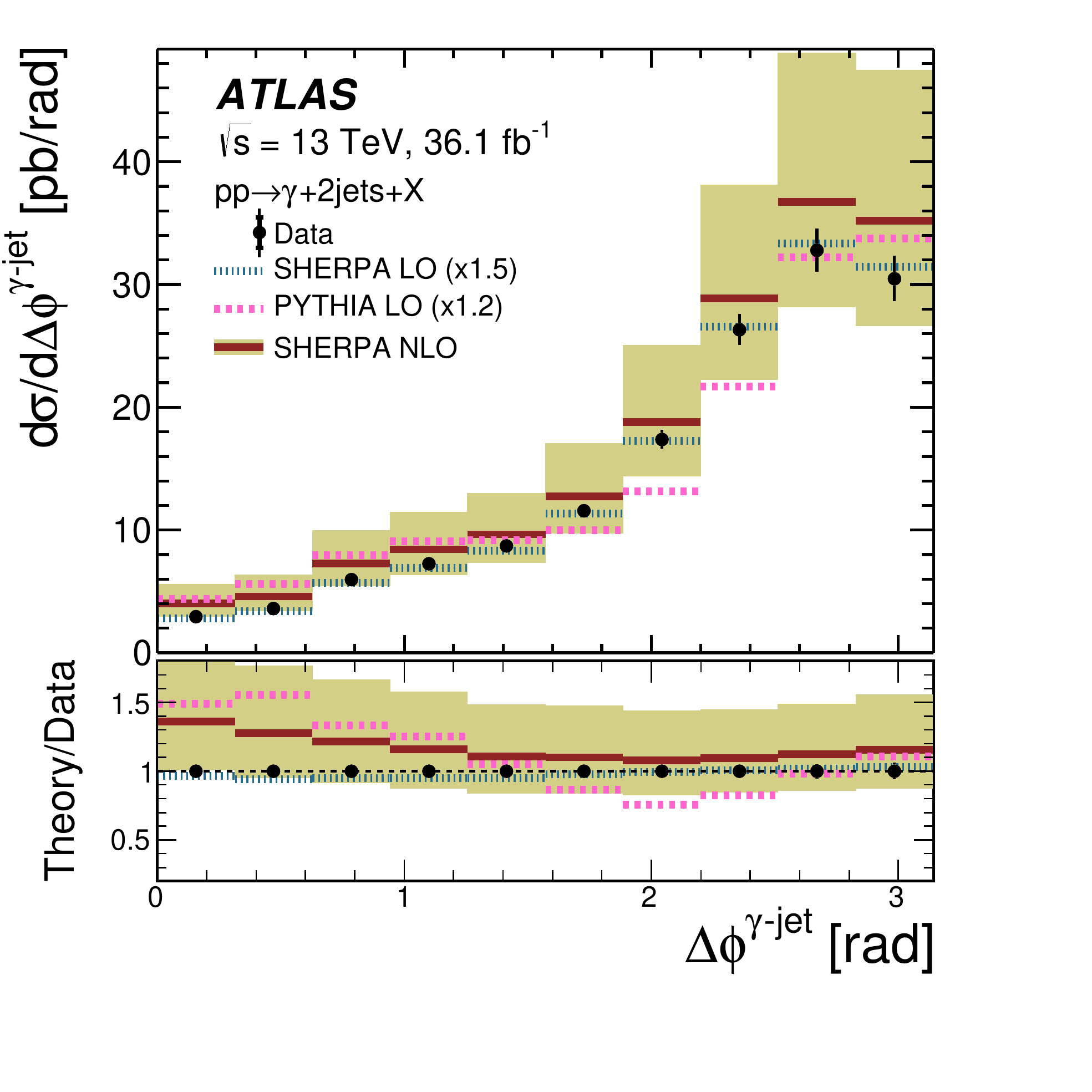}}
\put (4.0,14.5){{\small (a)}}
\put (11.0,14.5){{\small (b)}}
\put (4.0,8.0){{\small (c)}}
\put (4.0,1.5){{\small (d)}}
\put (11.0,1.5){{\small (e)}}
\end{picture}
\vspace{-2cm}
\caption
{
Measured cross sections for isolated-photon plus two-jet production
(dots) as functions of (a) $\etg$, (b) $\ptjet$, (c) $|\rapjet|$,
(d) $|\deltarapgj|$ and (e) $\deltaphigj$ for the total phase
space. The NLO QCD predictions from \sher\ (solid lines) based on
the NNPDF3.0 PDFs are also shown. The tree-level plus parton-shower
predictions from LO \sher\ (dotted lines) and \pyt\ (dashed lines)
normalised to the integrated measured cross section (using the
factors indicated in parentheses) are also shown. The bottom part of
each figure shows the ratio of the predictions to the measured cross
section. The inner (outer) error bars represent the statistical
uncertainties (the statistical and systematic uncertainties added in
quadrature) and the filled bands represent the theoretical
uncertainty of the NLO QCD predictions. For most of the points, the inner and outer error
bars are smaller than the marker size and, thus, not visible.
The visible thin error bars represent the outer error bars.
}
\label{fig04}
\end{figure}

\begin{figure}[h]
\setlength{\unitlength}{1.0cm}
\begin{picture} (18.0,17.0)
\put (1.0,7.0){\includegraphics[width=7cm]{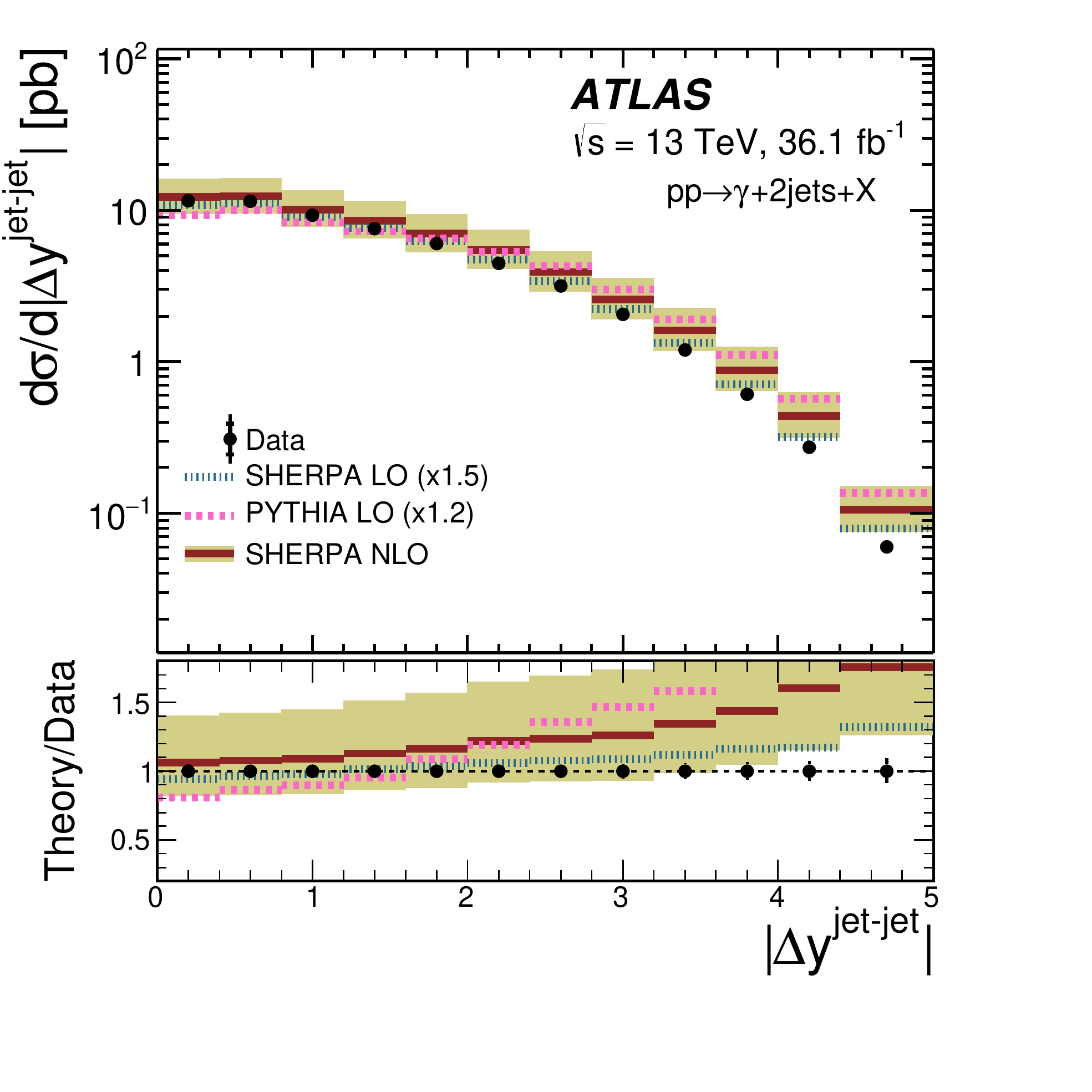}}
\put (8.0,7.0){\includegraphics[width=7cm]{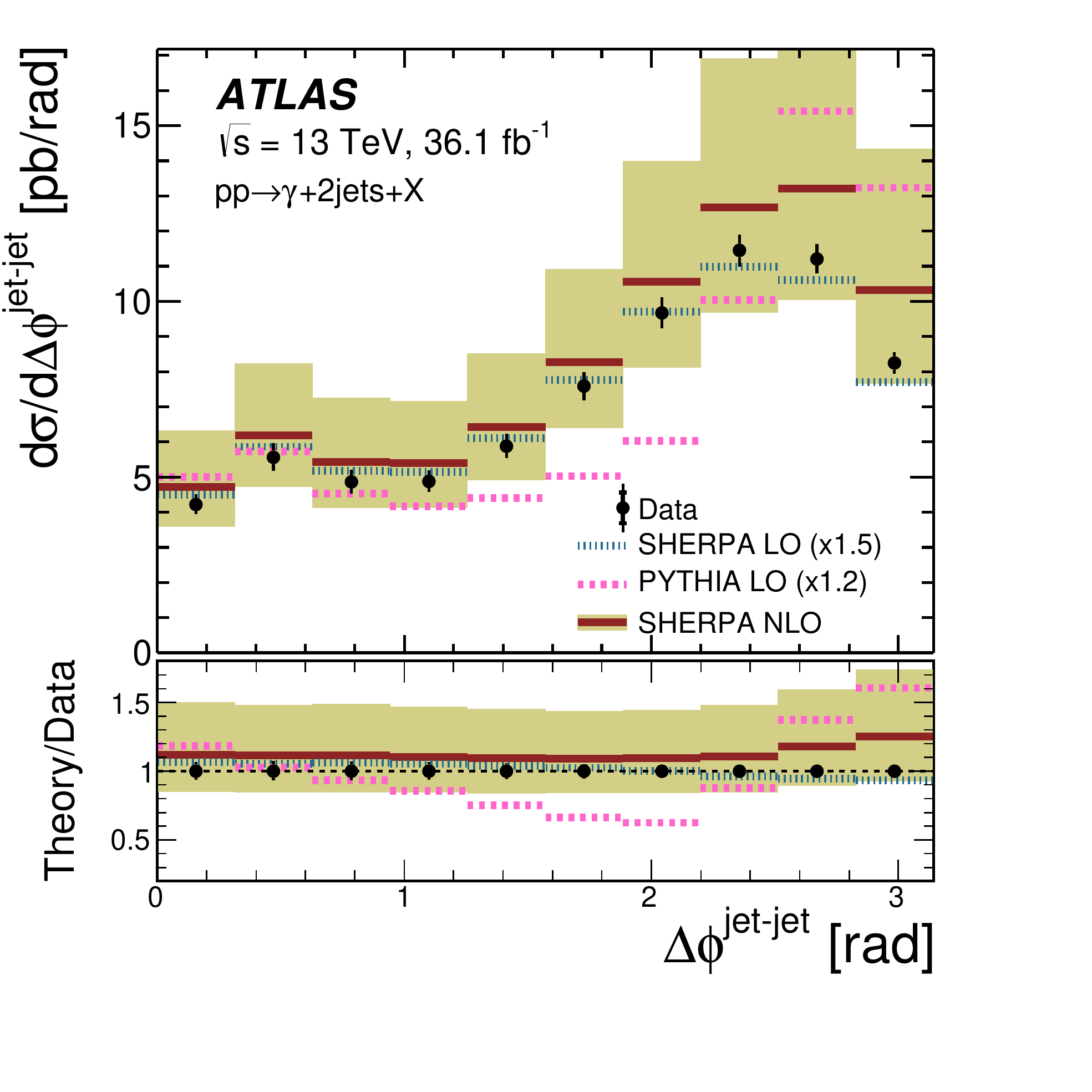}}
\put (1.0,0.0){\includegraphics[width=7cm]{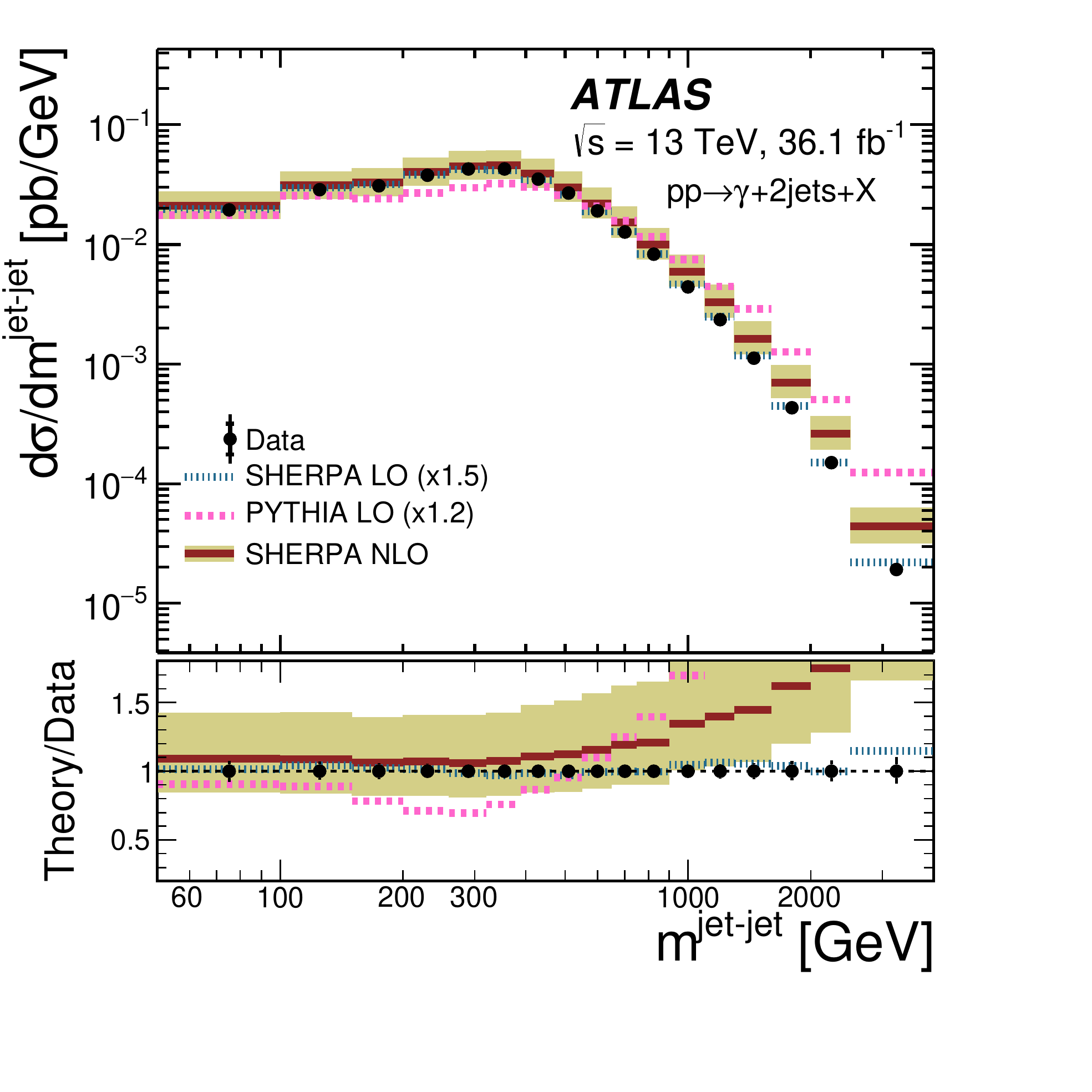}}
\put (8.0,0.0){\includegraphics[width=7cm]{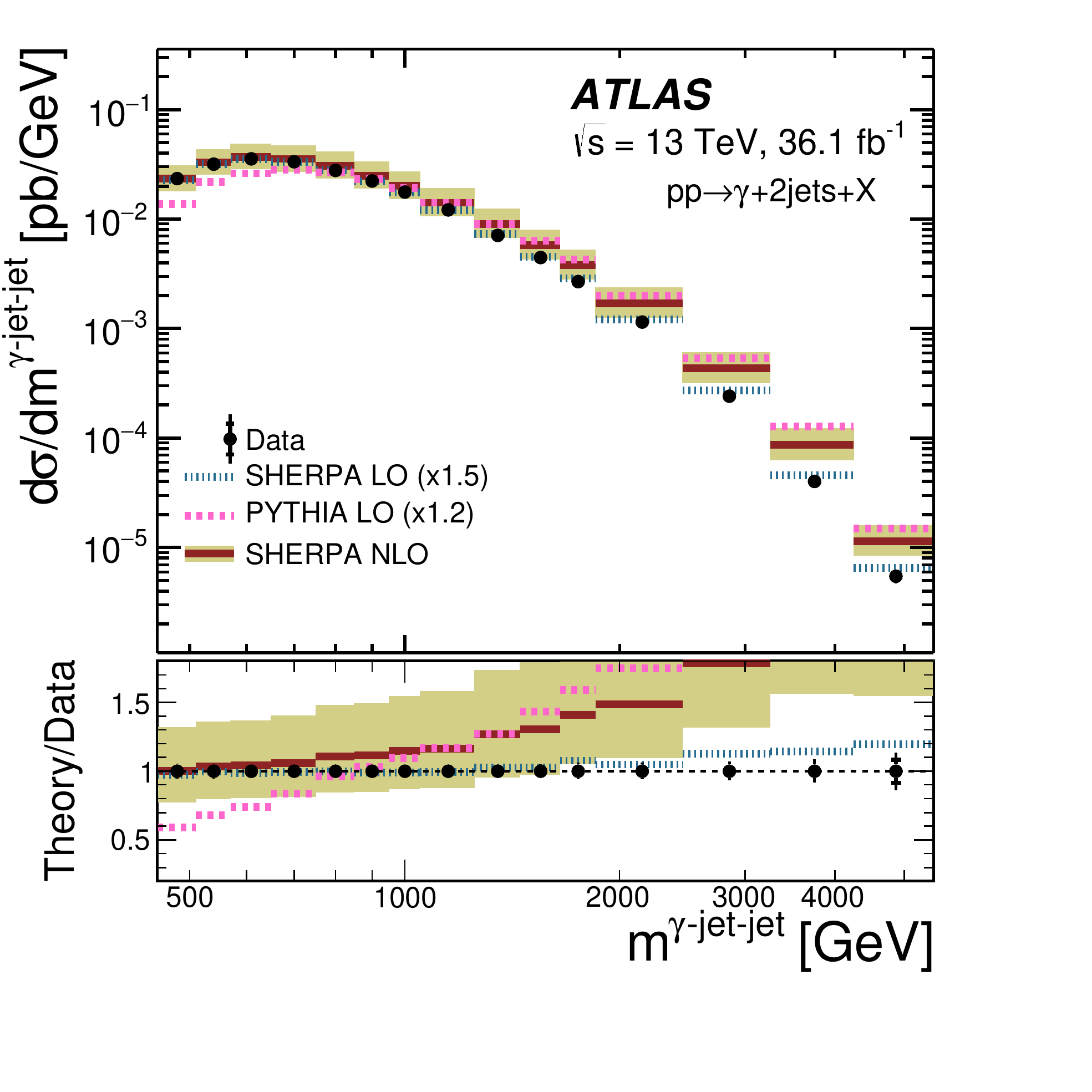}}
\put (4.0,7.5){{\small (a)}}
\put (11.0,7.5){{\small (b)}}
\put (4.0,0.5){{\small (c)}}
\put (11.0,0.5){{\small (d)}}
\end{picture}
\caption
{
Measured cross sections for isolated-photon plus two-jet production
(dots) as functions of (a) $|\deltarapjj|$, (b) $\deltaphijj$, (c)
$\mjj$ and (d) $\mgjj$ for the total phase space. The NLO QCD
predictions from \sher\ (solid lines) based on the NNPDF3.0 PDFs are
also shown. The tree-level plus parton-shower predictions from LO
\sher\ (dotted lines) and \pyt\ (dashed lines) normalised to the
integrated measured cross section (using the factors indicated in
parentheses) are also shown. The bottom part of each figure shows
the ratio of the predictions to the measured cross section. The
inner (outer) error bars represent the statistical uncertainties
(the statistical and systematic uncertainties added in quadrature)
and the filled bands represent the theoretical uncertainty of the NLO QCD predictions.
For most of the points, the inner and outer error
bars are smaller than the marker size and, thus, not visible.
The visible thin error bars represent the outer error bars.
}
\label{fig05}
\end{figure}

\begin{figure}[h]
\setlength{\unitlength}{1.0cm}
\begin{picture} (18.0,20.5)
\put (1.0,14.0){\includegraphics[width=7cm]{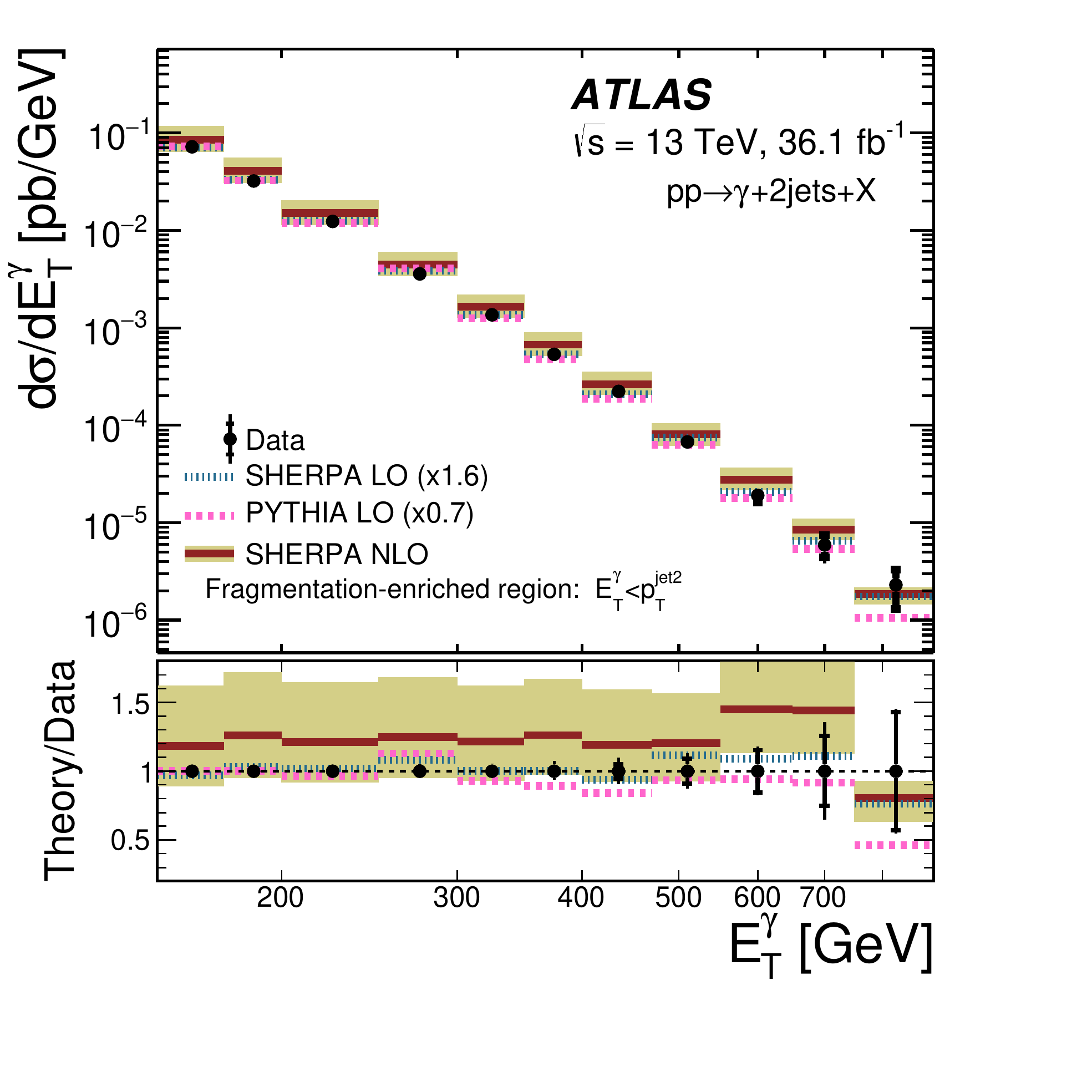}}
\put (8.0,14.0){\includegraphics[width=7cm]{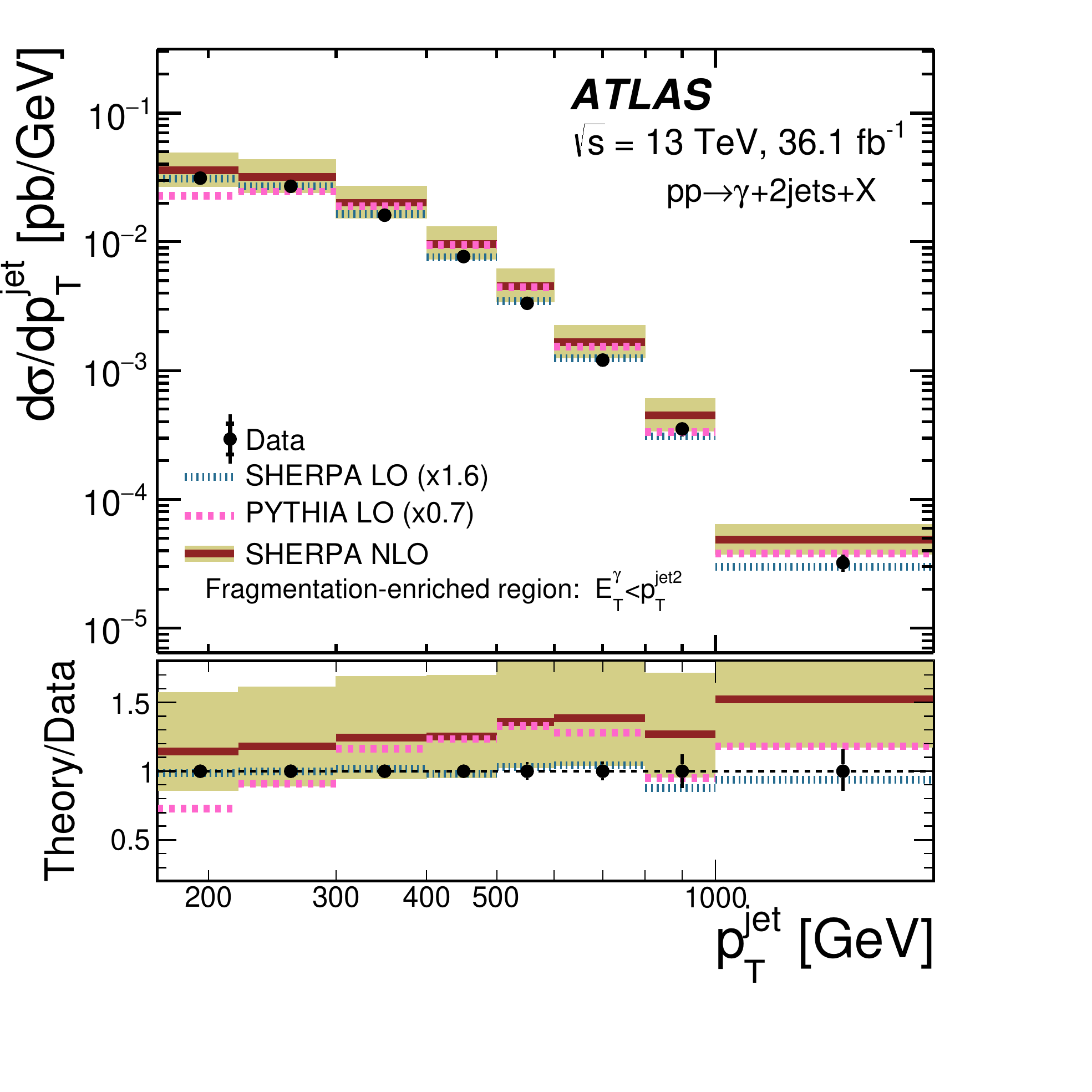}}
\put (1.0,7.5){\includegraphics[width=7cm]{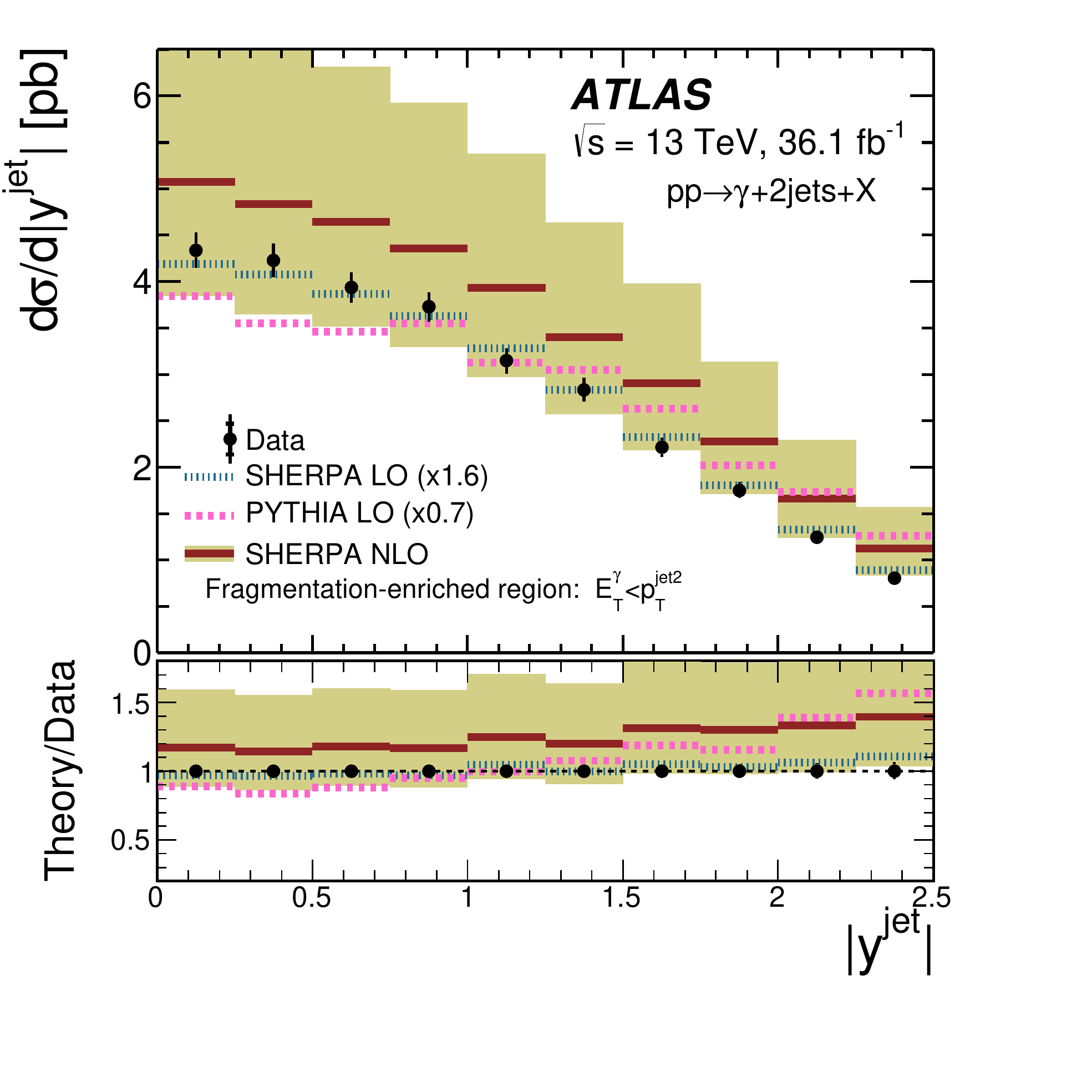}}
\put (1.0,1.0){\includegraphics[width=7cm]{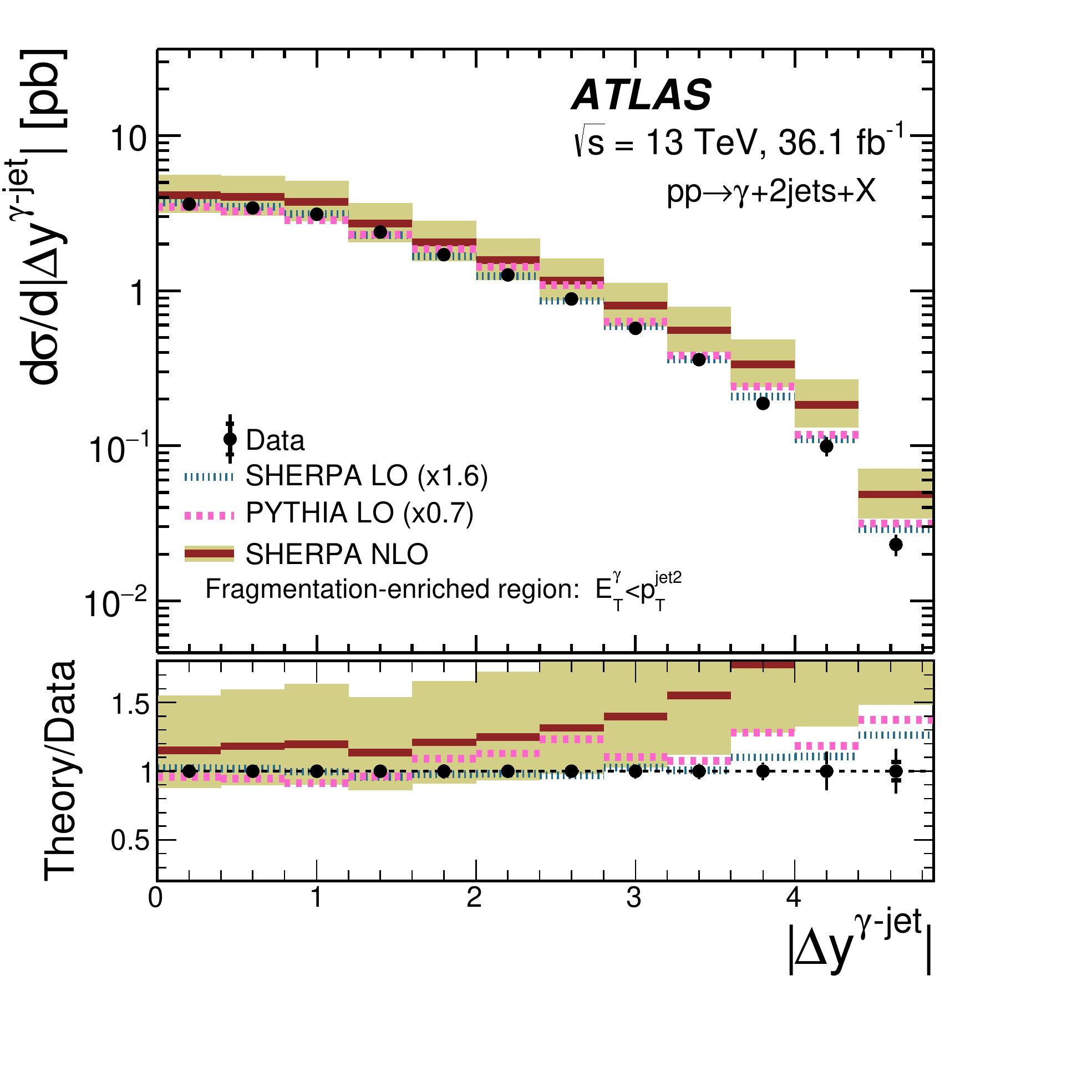}}
\put (8.0,1.0){\includegraphics[width=7cm]{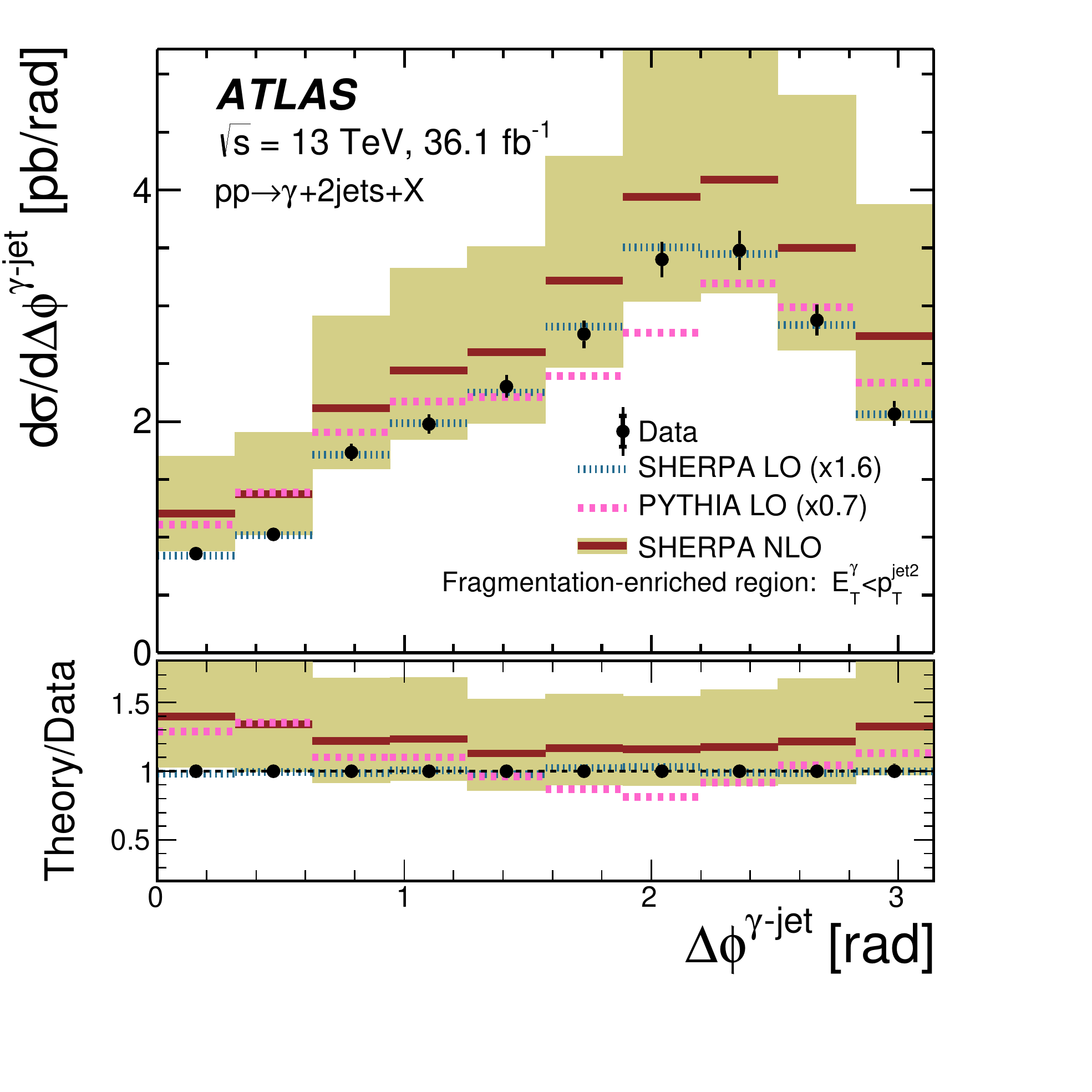}}
\put (4.0,14.5){{\small (a)}}
\put (11.0,14.5){{\small (b)}}
\put (4.0,8.0){{\small (c)}}
\put (4.0,1.5){{\small (d)}}
\put (11.0,1.5){{\small (e)}}
\end{picture}
\vspace{-2.0cm}
\caption
{
Measured cross sections for isolated-photon plus two-jet production
(dots) as functions of (a) $\etg$, (b) $\ptjet$, (c) $|\rapjet|$,
(d) $|\deltarapgj|$ and (e) $\deltaphigj$ for the
fragmentation-enriched phase space. The NLO QCD predictions from
\sher\ (solid lines) based on the NNPDF3.0 PDFs are also shown. The
tree-level plus parton-shower predictions from LO \sher\ (dotted
lines) and \pyt\ (dashed lines) normalised to the integrated
measured cross section (using the factors indicated in parentheses)
are also shown. The bottom part of each figure shows the ratio of
the predictions to the measured cross section. The inner (outer)
error bars represent the statistical uncertainties (the statistical
and systematic uncertainties added in quadrature) and the filled
bands represent the theoretical uncertainty of the NLO QCD predictions. For most of the points,
the inner and outer error bars are smaller than the marker size and,
thus, not visible. The visible thin error bars represent the outer error bars.
}
\label{fig06}
\end{figure}

\begin{figure}[h]
\setlength{\unitlength}{1.0cm}
\begin{picture} (18.0,17.0)
\put (1.0,7.0){\includegraphics[width=7cm]{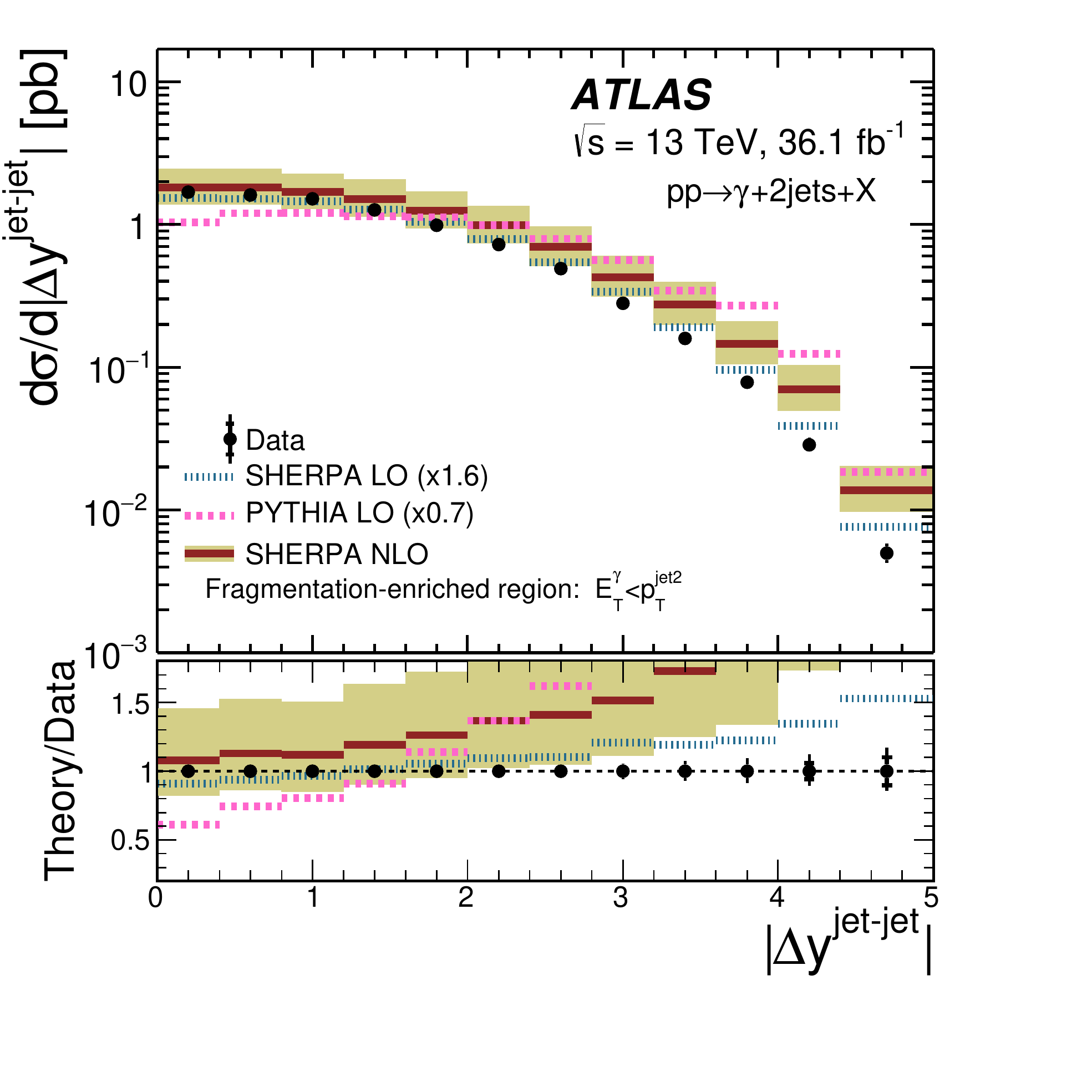}}
\put (8.0,7.0){\includegraphics[width=7cm]{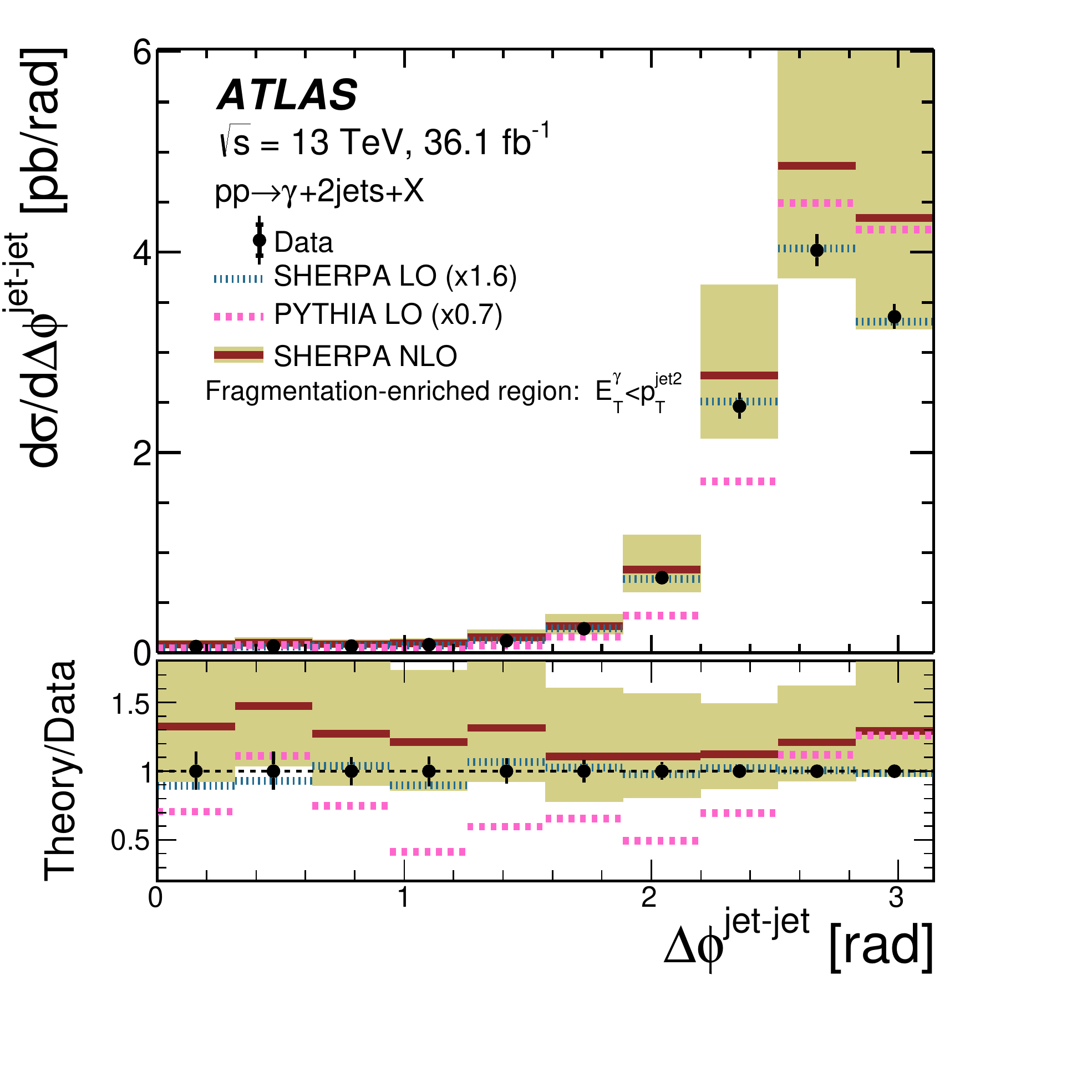}}
\put (1.0,0.0){\includegraphics[width=7cm]{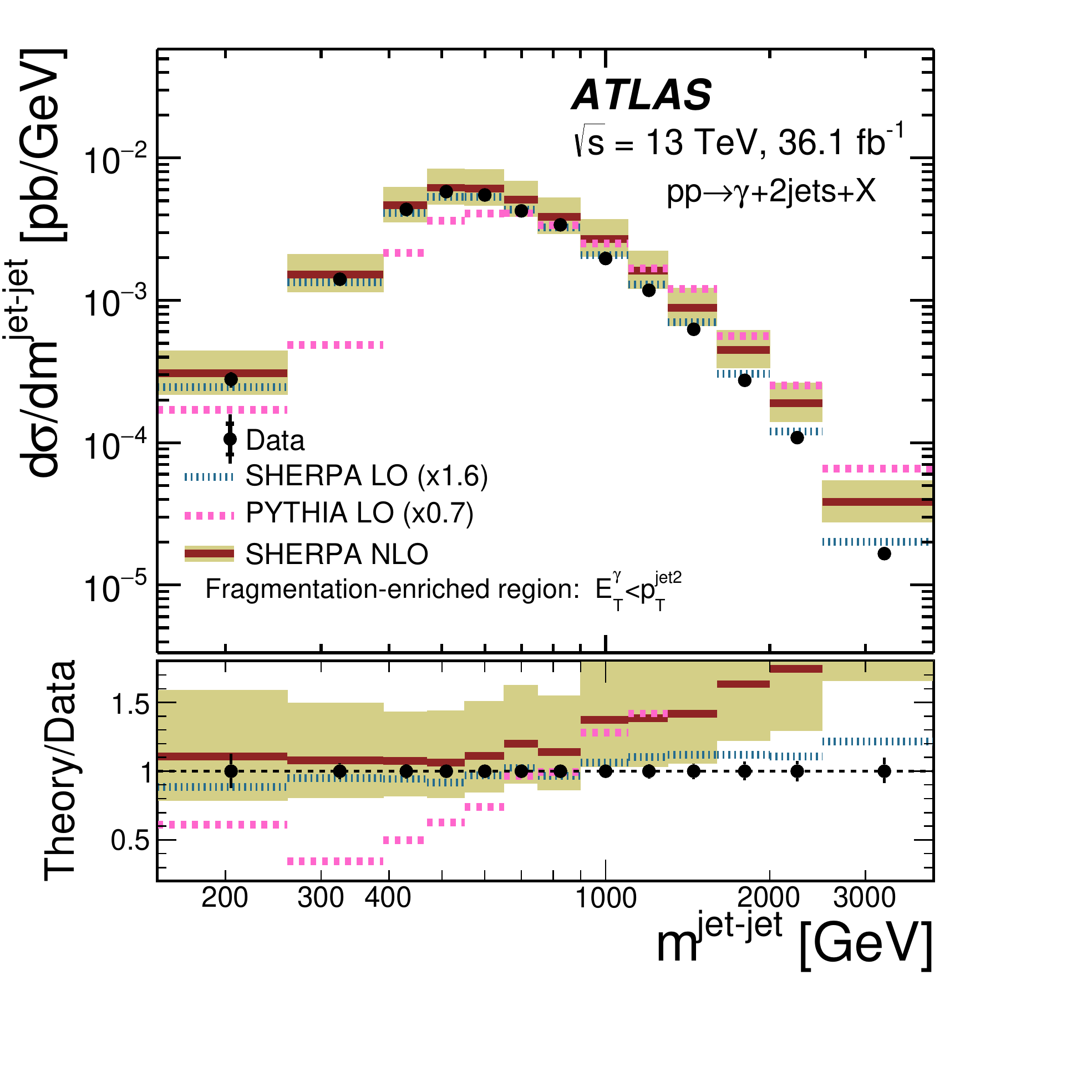}}
\put (8.0,0.0){\includegraphics[width=7cm]{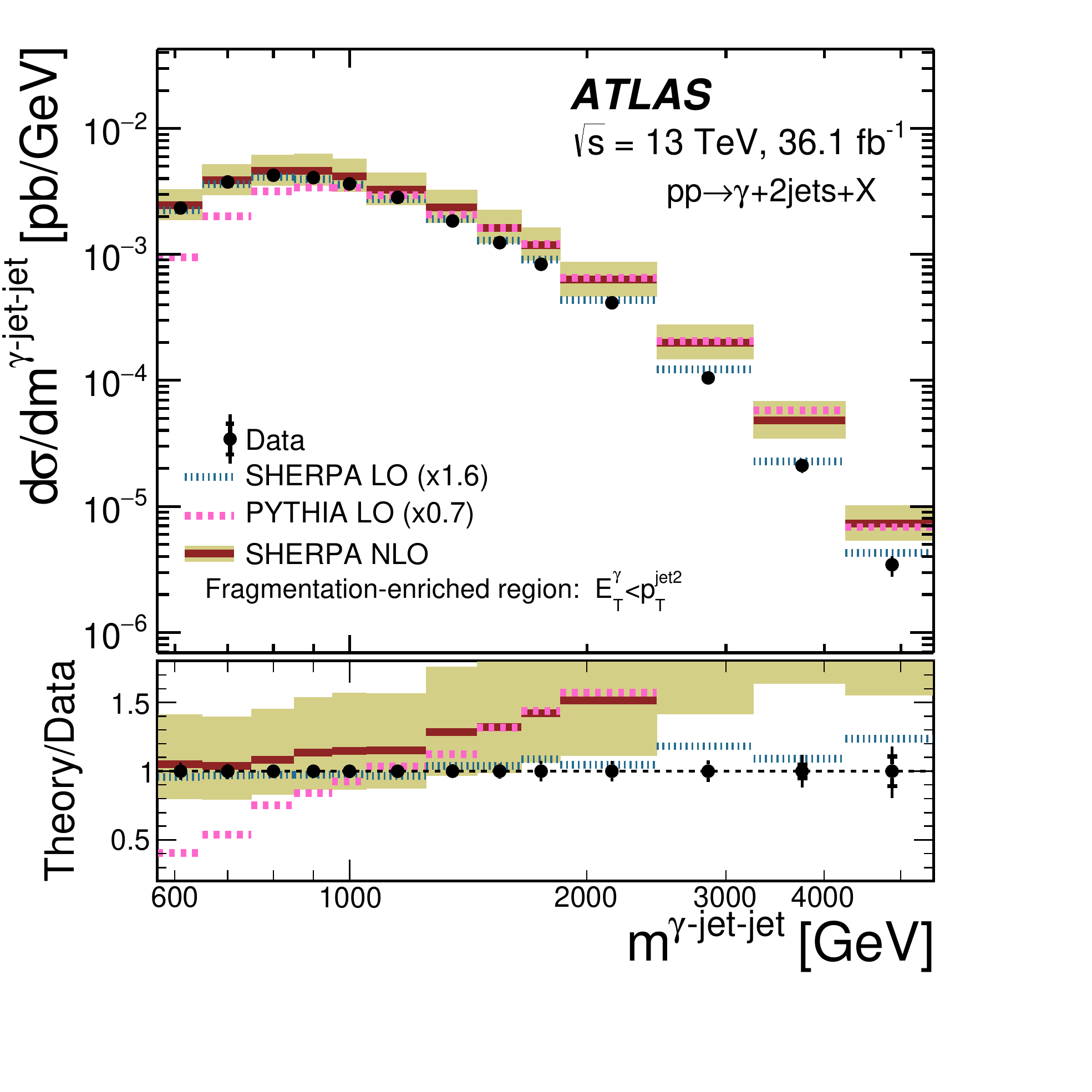}}
\put (4.0,7.5){{\small (a)}}
\put (11.0,7.5){{\small (b)}}
\put (4.0,0.5){{\small (c)}}
\put (11.0,0.5){{\small (d)}}
\end{picture}
\caption
{
Measured cross sections for isolated-photon plus two-jet production
(dots) as functions of (a) $|\deltarapjj|$, (b) $\deltaphijj$, (c)
$\mjj$ and (d) $\mgjj$ for the fragmentation-enriched phase
space. The NLO QCD predictions from \sher\ (solid lines) based on
the NNPDF3.0 PDFs are also shown. The tree-level plus parton-shower
predictions from LO \sher\ (dotted lines) and \pyt\ (dashed lines)
normalised to the integrated measured cross section (using the
factors indicated in parentheses) are also shown. The bottom part of
each figure shows the ratio of the predictions to the measured cross
section. The inner (outer) error bars represent the statistical
uncertainties (the statistical and systematic uncertainties added in
quadrature) and the filled bands represent the theoretical
uncertainty of the NLO QCD predictions. For most of the points, the inner and outer error
bars are smaller than the marker size and, thus, not visible.
The visible thin error bars represent the outer error bars.
}
\label{fig07}
\end{figure}

\begin{figure}[h]
\setlength{\unitlength}{1.0cm}
\begin{picture} (18.0,20.5)
\put (1.0,14.0){\includegraphics[width=7cm]{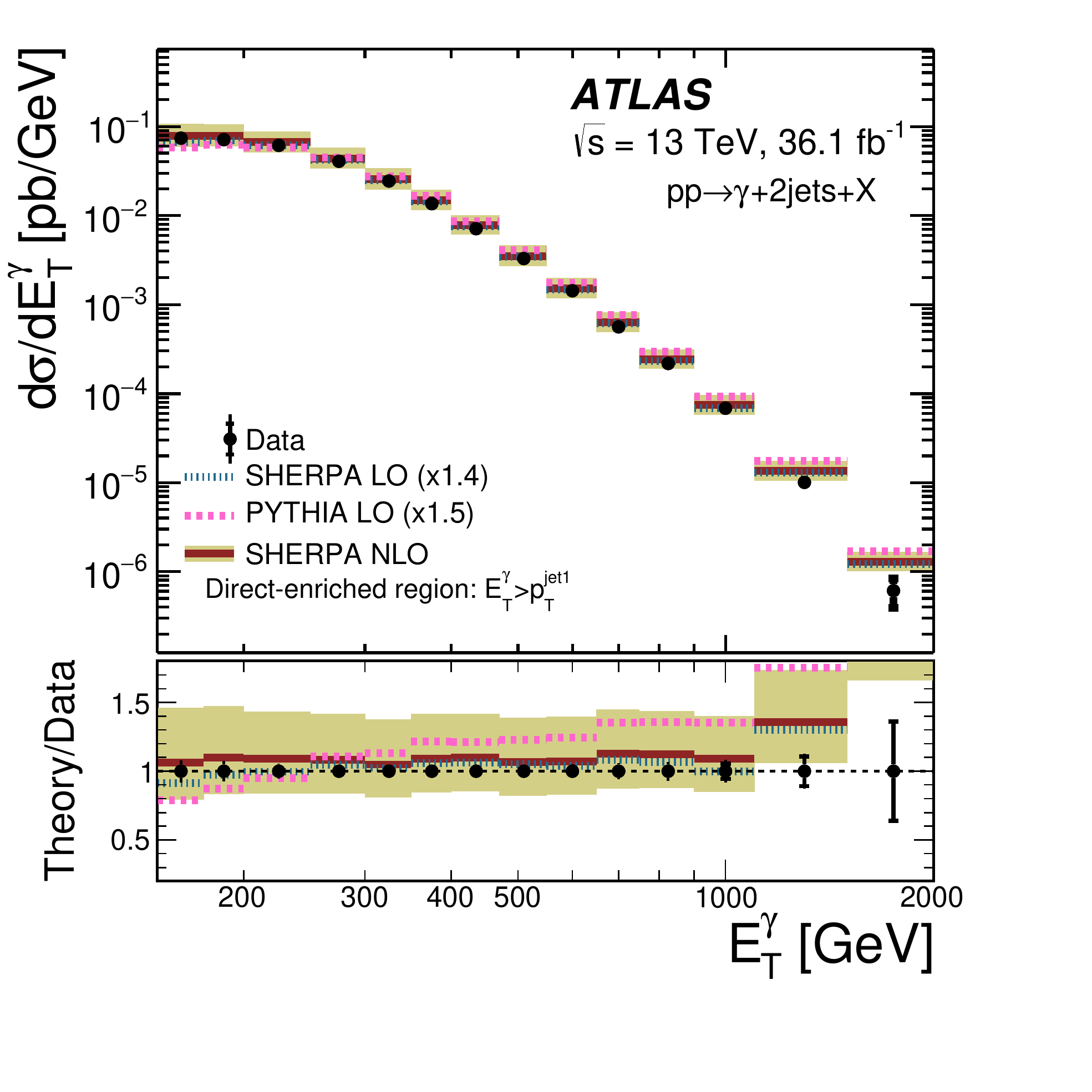}}
\put (8.0,14.0){\includegraphics[width=7cm,height=7cm]{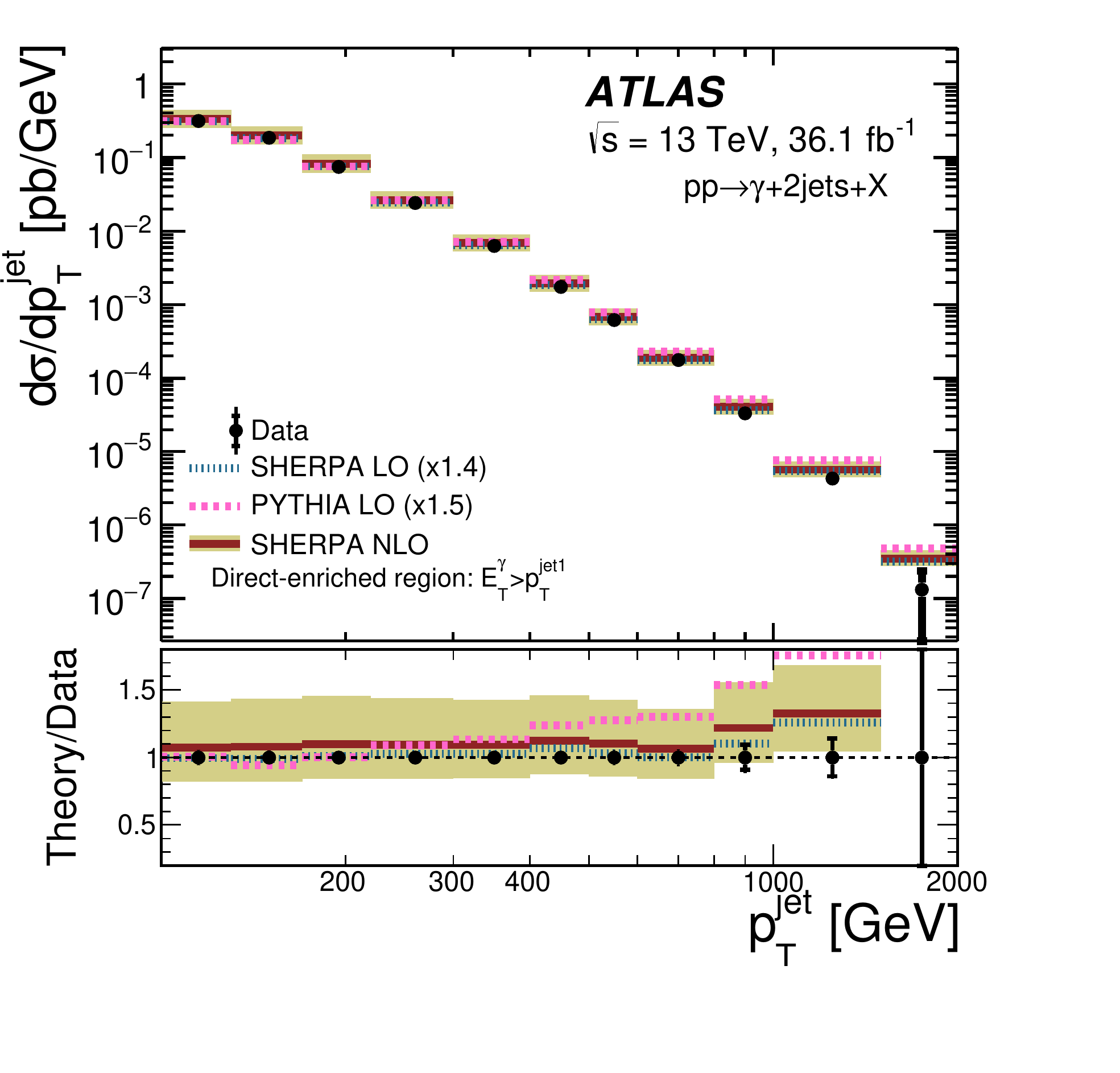}}
\put (1.0,7.5){\includegraphics[width=7cm]{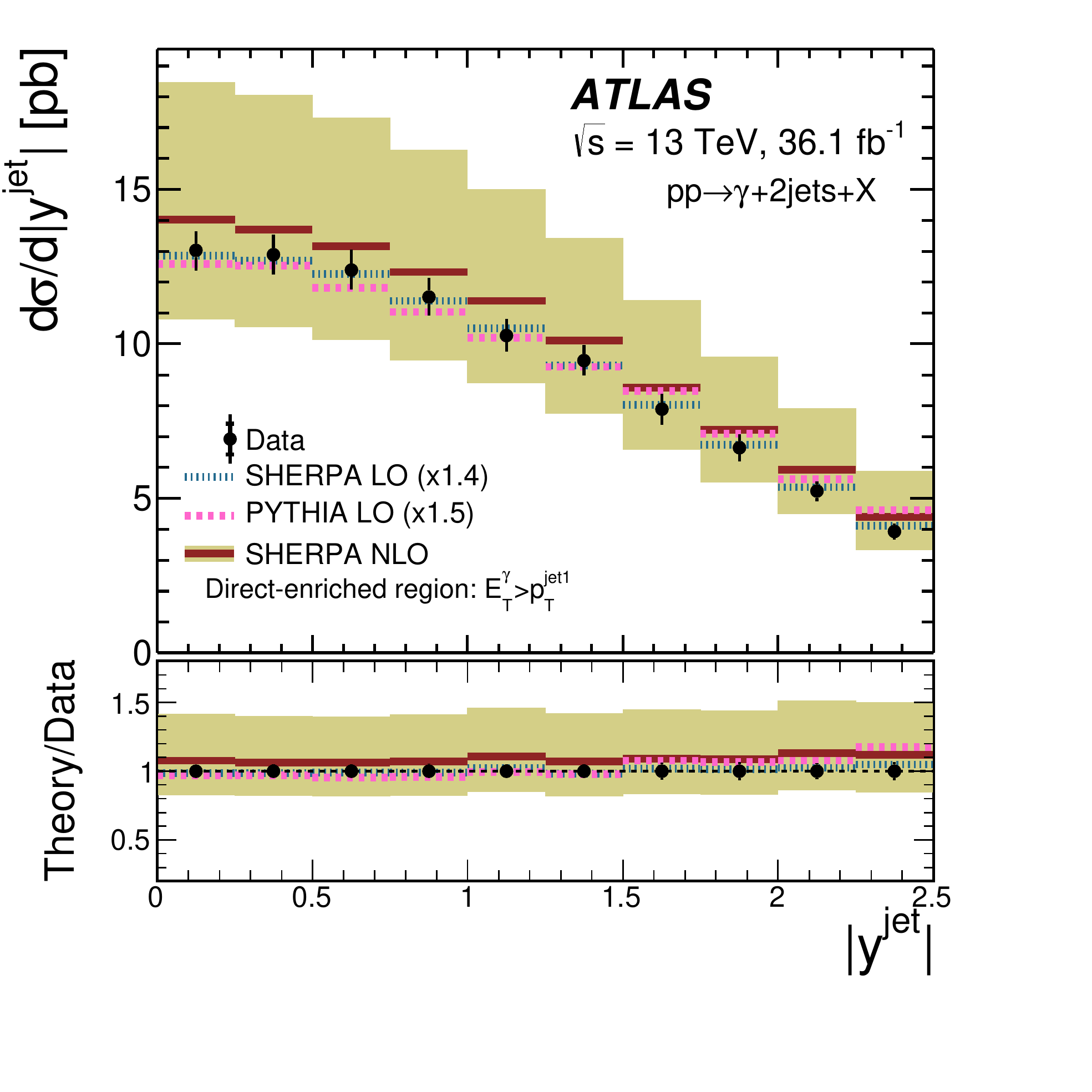}}
\put (1.0,1.0){\includegraphics[width=7cm]{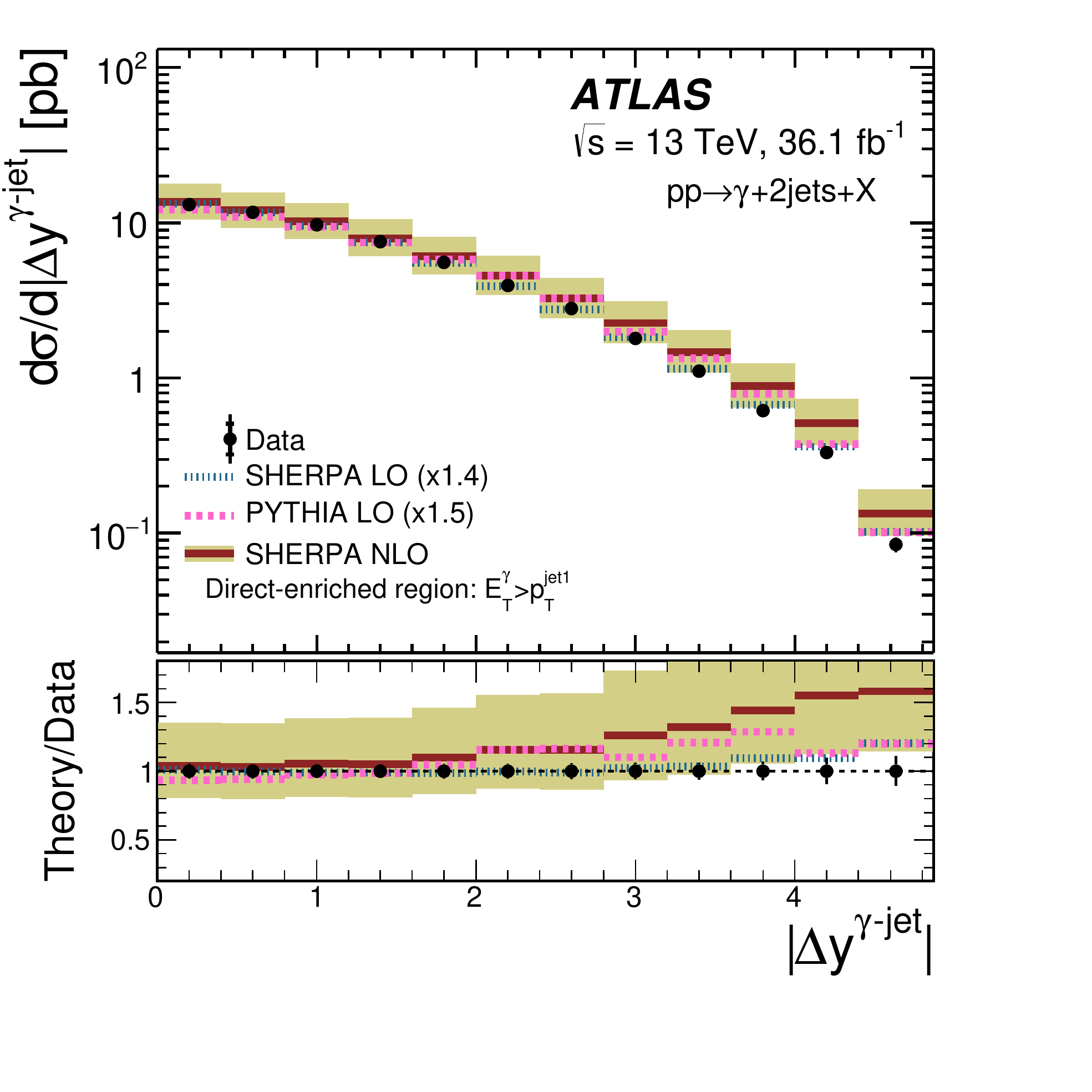}}
\put (8.0,1.0){\includegraphics[width=7cm]{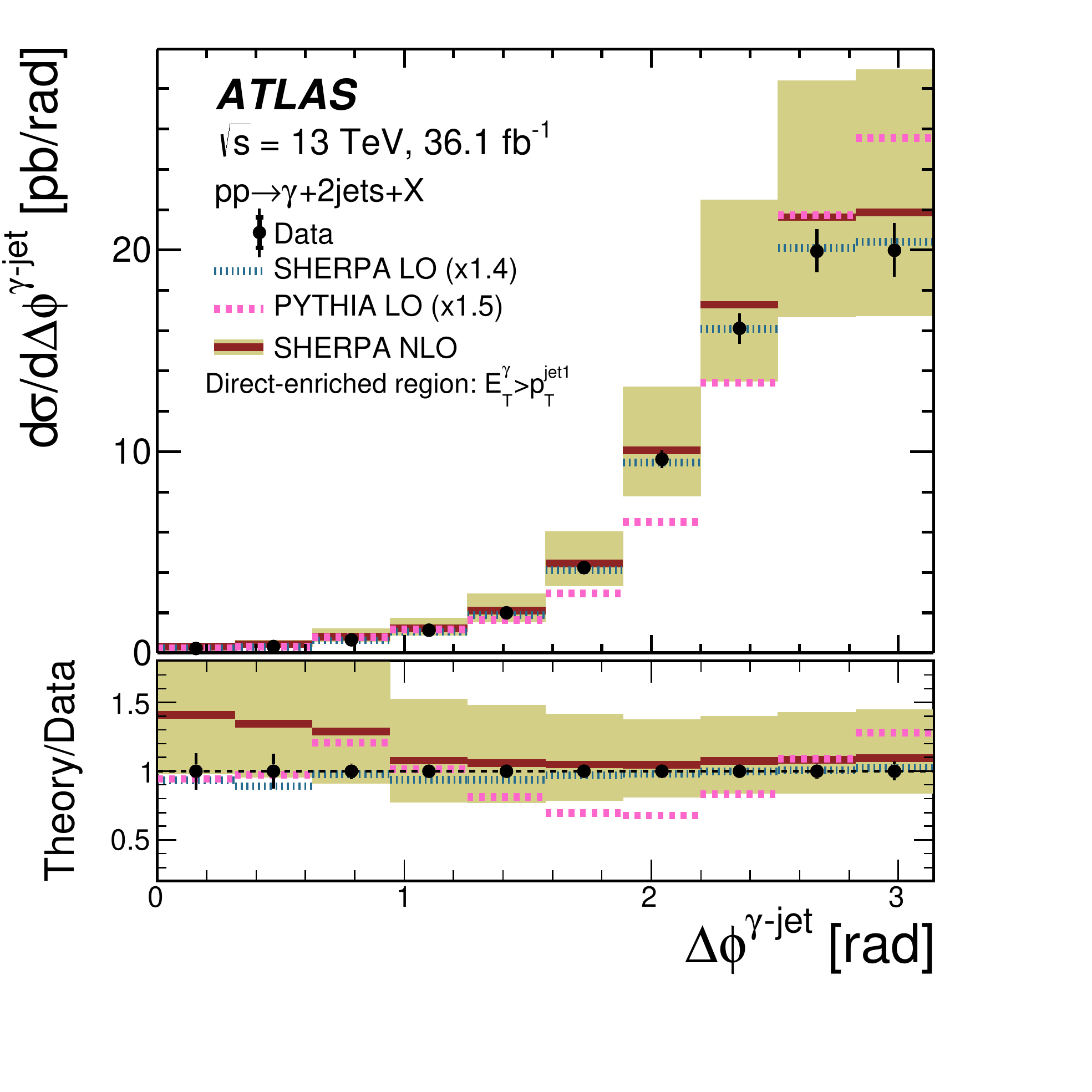}}
\put (4.0,14.5){{\small (a)}}
\put (11.0,14.5){{\small (b)}}
\put (4.0,8.0){{\small (c)}}
\put (4.0,1.5){{\small (d)}}
\put (11.0,1.5){{\small (e)}}
\end{picture}
\vspace{-2cm}
\caption
{
Measured cross sections for isolated-photon plus two-jet production
(dots) as functions of (a) $\etg$, (b) $\ptjet$, (c) $|\rapjet|$,
(d) $|\deltarapgj|$ and (e) $\deltaphigj$ for the direct-enriched
phase space. The NLO QCD predictions from \sher\ (solid lines) based
on the NNPDF3.0 PDFs are also shown. The tree-level plus
parton-shower predictions from LO \sher\ (dotted lines) and \pyt\
(dashed lines) normalised to the integrated measured cross section
(using the factors indicated in parentheses) are also shown. The
bottom part of each figure shows the ratio of the predictions to the
measured cross section. The inner (outer) error bars represent the
statistical uncertainties (the statistical and systematic
uncertainties added in quadrature) and the filled bands represent
the theoretical uncertainty of the NLO QCD predictions. For most of the points, the inner and outer error
bars are smaller than the marker size and, thus, not visible.
The visible thin error bars represent the outer error bars.
}
\label{fig08}
\end{figure}

\begin{figure}[h]
\setlength{\unitlength}{1.0cm}
\begin{picture} (18.0,17.0)
\put (1.0,7.0){\includegraphics[width=7cm]{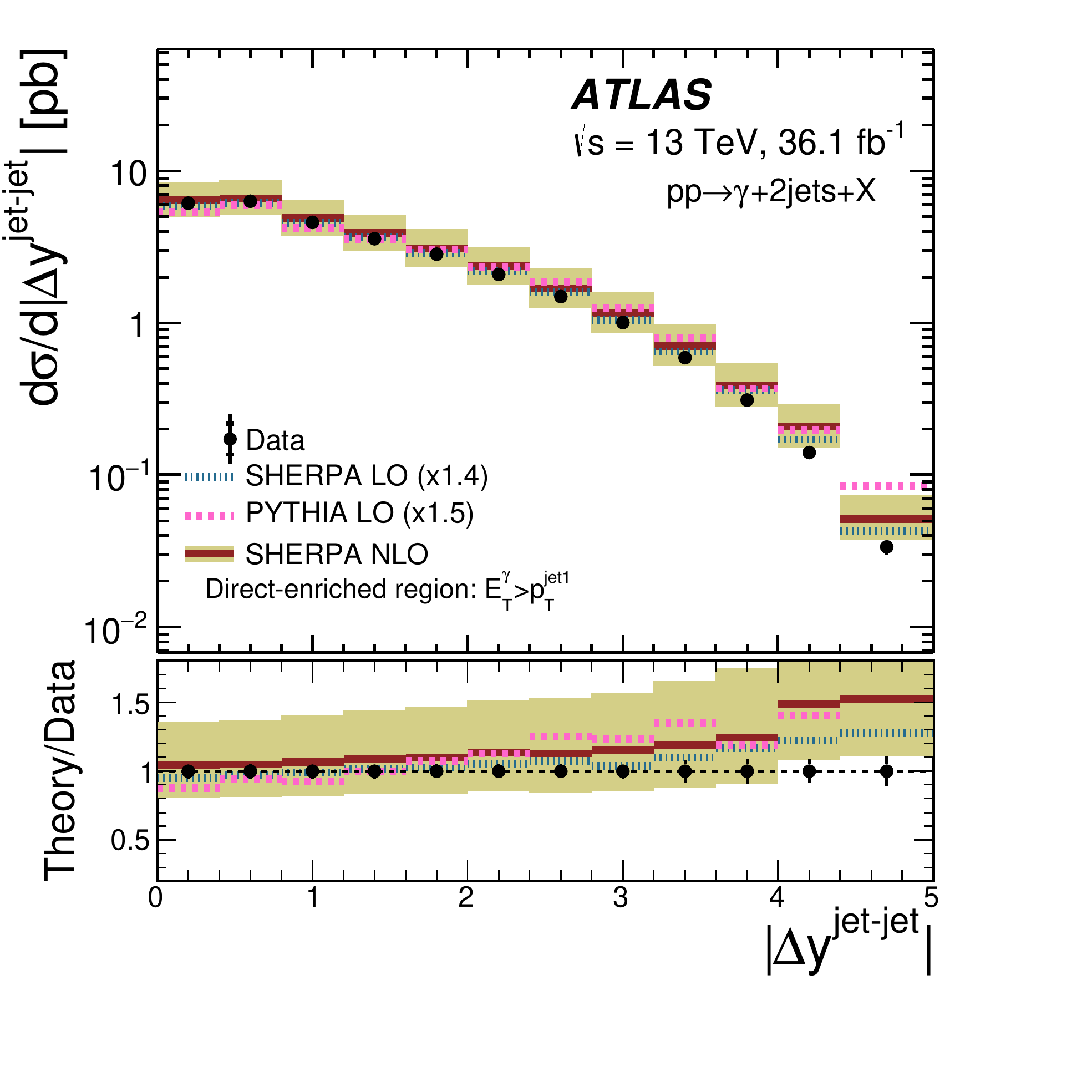}}
\put (8.0,7.0){\includegraphics[width=7cm]{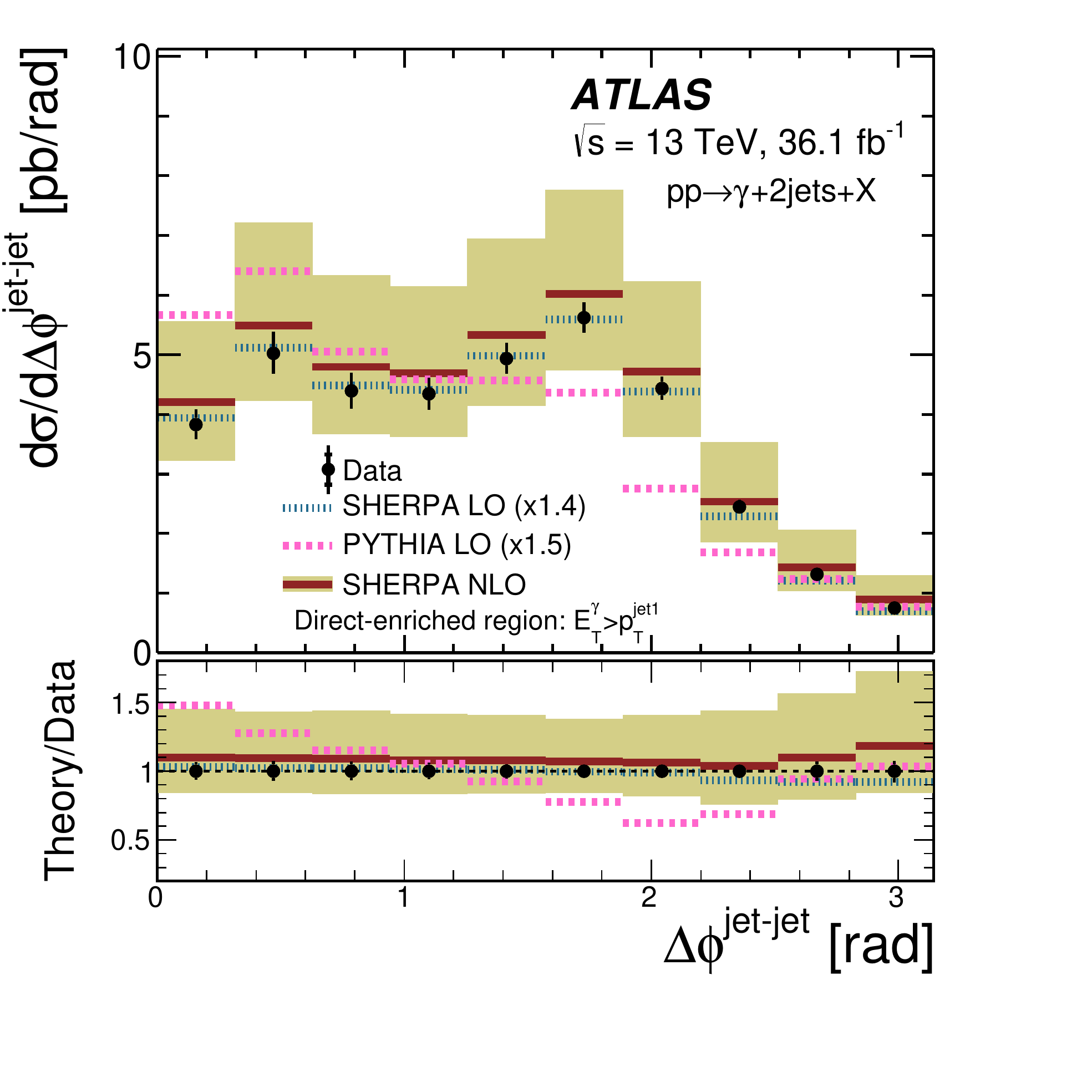}}
\put (1.0,0.0){\includegraphics[width=7cm]{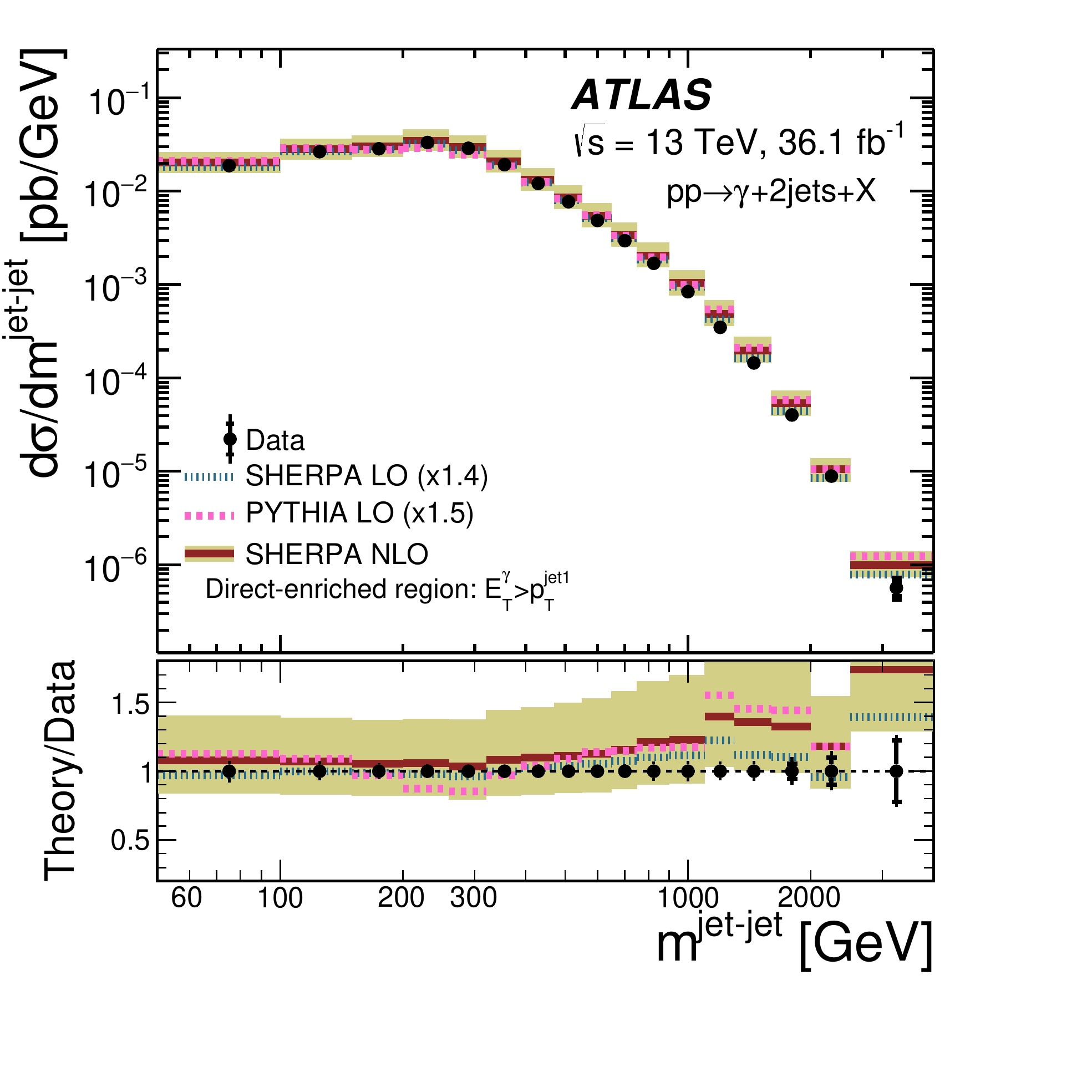}}
\put (8.0,0.0){\includegraphics[width=7cm]{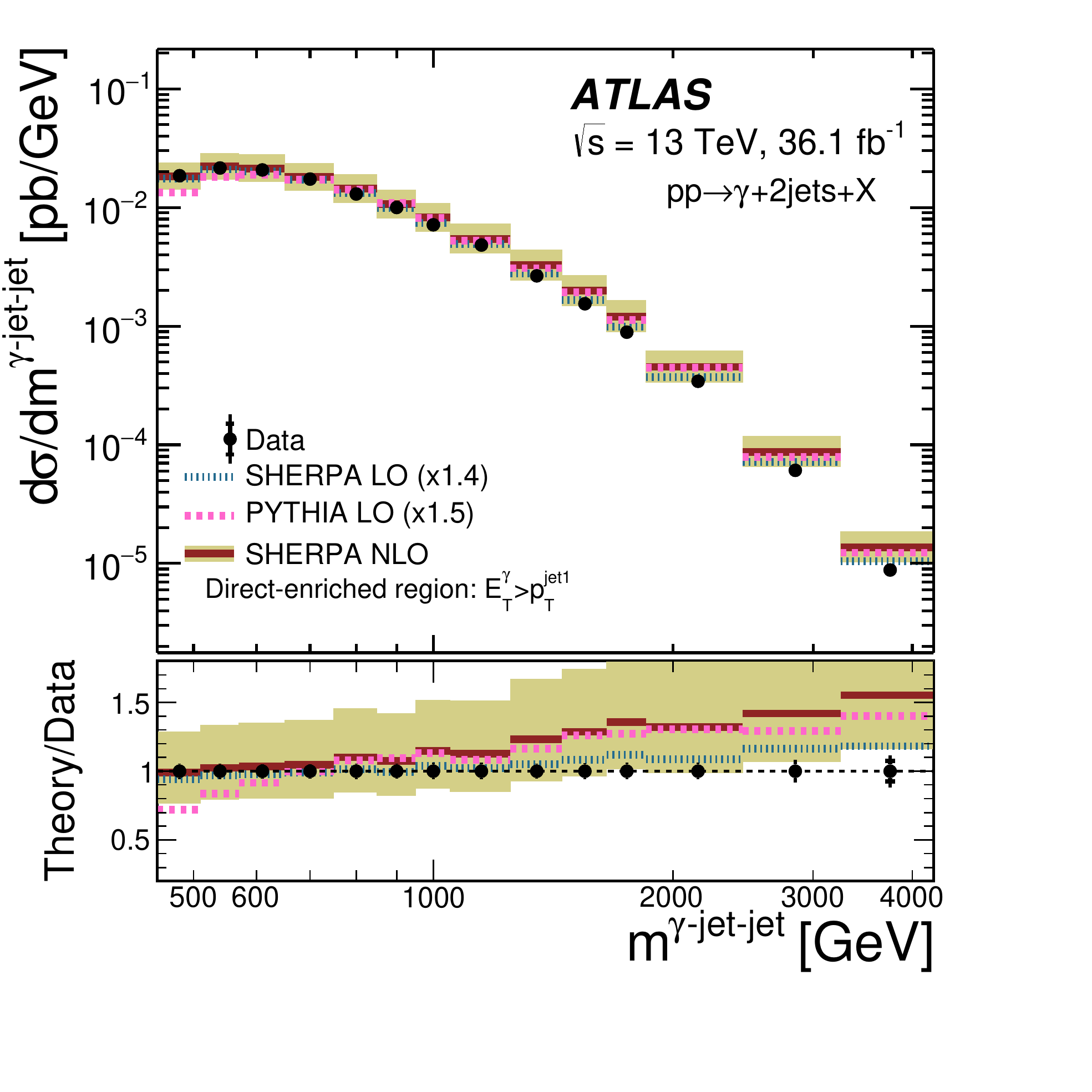}}
\put (4.0,7.5){{\small (a)}}
\put (11.0,7.5){{\small (b)}}
\put (4.0,0.5){{\small (c)}}
\put (11.0,0.5){{\small (d)}}
\end{picture}
\caption
{
Measured cross sections for isolated-photon plus two-jet production
(dots) as functions of (a) $|\deltarapjj|$, (b) $\deltaphijj$, (c)
$\mjj$ and (d) $\mgjj$ for the direct-enriched phase space. The NLO
QCD predictions from \sher\ (solid lines) based on the NNPDF3.0 PDFs
are also shown. The tree-level plus parton-shower predictions from
LO \sher\ (dotted lines) and \pyt\ (dashed lines) normalised to the
integrated measured cross section (using the factors indicated in
parentheses) are also shown. The bottom part of each figure shows
the ratio of the predictions to the measured cross section. The
inner (outer) error bars represent the statistical uncertainties
(the statistical and systematic uncertainties added in quadrature)
and the filled bands represent the theoretical uncertainty of the NLO QCD predictions.
For most of the points, the inner and outer error
bars are smaller than the marker size and, thus, not visible.
The visible thin error bars represent the outer error bars.
}
\label{fig09}
\end{figure}

The predictions of the NLO calculations of \sher\ are compared with the
data in Figures~\ref{fig04} to \ref{fig09}. The uncertainties in the
predictions include those due to missing higher-order terms in the
perturbative expansion, those due to the uncertainties in the PDFs and
those due to the uncertainty in $\asz$, as explained in
Section~\ref{thunc}. The predictions are shown with the theory
normalisation, so the comparison between data and theory is
done in terms of both normalisation and shape, except for the
LO \sher\ predictions (see Section~\ref{compalo}) which are
normalised to the data. The NLO \sher\ predictions
describe the data adequately in shape and normalisation within the
experimental and theoretical uncertainties except for ranges at
high values of $|\deltarapgj|$, $|\deltarapjj|$, $\mjj$ and $\mgjj$,
where the predictions overestimate the data in the total and
fragmentation-enriched phase-space regions. In those ranges, the
overestimation of the predictions relative to the data is larger
in the fragmentation-enriched phase space than when considering the total
phase space. These discrepancies might be attributable to unaccounted higher-order
corrections given the fact that the variation of the renormalisation and
factorisation scales leads to relatively large uncertainties in the
predictions. In particular, the use of $\etg$ as the scale might not
be optimal in the ranges where $\ptjet$ is much larger than $\etg$. A
mismodelling of the high tail in $\mjj$ is also observed for the
measurements of the final state $V+{\mathrm{jet}}+{\mathrm{jet}}$ with
$V=W^{\pm}$ or $Z$
boson~\cite{pl:b775:206,epj:c77:474,jhep:1805:077}. In addition,
decorrelation effects from multiple parton emission might be present
at large values of $|\deltarapgj|$ and $|\deltarapjj|$.
The comparison of the predictions based on different parameterisations of the proton
PDFs shows that the description of the data does not depend
significantly on the specific PDF used. The theoretical uncertainties
are much larger than the experimental ones and, therefore, more
precise calculations are needed, either in the form of resummed
calculations or through the inclusion of higher-order terms. Once the
size of the uncertainties of the predictions is reduced significantly,
a meaningful comparison with predictions including electroweak effects
will be possible.
 
\FloatBarrier
\section{Summary}
\label{conc}
Measurements of the cross sections for the production of an isolated
photon in association with two jets in $pp$ collisions at $\sqrt
s=13$~\TeV, $\ppgjj$, are presented. These measurements are based on
an integrated luminosity of $36.1$~\fb1\ of ATLAS data recorded at the
LHC. The photon is required to have $\etg>150$~\GeV\ and
$|\etag|<2.37$, excluding the region $1.37<|\etag|<1.56$. The jets are
reconstructed using the anti-$k_t$ algorithm with radius parameter
$R=0.4$. The cross sections are measured as functions of $\etg$,
$\ptjet$, $|\rapjet|$, $\deltaphigj$, $|\deltarapgj|$, $\mjj$,
$\deltaphijj$, $|\deltarapjj|$ and $\mgjj$ in three different regions
of phase space, namely the total phase space, the fragmentation-photon enriched
phase space where $\etg<\ptjetsl$ and the direct-photon enriched phase space
where $\etg>\ptjetl$. The measurements extend up to values of $2$~\TeV\
in $\etg$ and $\ptjet$. The dependence of the cross sections on $\mjj$
and $\mgjj$ is measured up to values of $4$~\TeV\ and $5.5$~\TeV,
respectively. The measured cross sections in the fragmentation-photon and
direct-photon enriched phase-space regions exhibit the features expected from
the two underlying production mechanisms, each of which dominates in
one of the two regions, and for some observables, from the effects of the kinematical
constraints that define the two phase-space regions.
 
The predictions of the tree-level plus parton-shower MC model in LO
\sher\ give a good description of the shape of the data distributions,
except at high $\etg$, $|\deltarapjj|$ and $\mgjj$. In contrast, the
predictions from \pyt, for which the sub-leading jet must necessarily
originate from the parton shower, generally fail to describe the shape of
the distributions in the data. The improved description of the data by
the predictions from LO \sher\ is attributed to the inclusion of
tree-level higher-order matrix elements. The precision of the measurements
is significantly better than the differences between the predictions.
The NLO predictions from \sher\ describe the data adequately in shape and
normalisation within the experimental and theoretical uncertainties except
for the regions at high values of $|\deltarapgj|$, $|\deltarapjj|$, $\mjj$ and
$\mgjj$, where the predictions overestimate the data for the total and
fragmentation-photon enriched phase-space regions. There are regions in which
the NLO predictions exhibit trends that differ from those in the data, such as
at high $\etg$ and $\ptjet$, and at low $\deltaphigj$. The theoretical
uncertainties are much larger than those of experimental nature,
preventing a more precise test of the theory. The measurements provide
tests of pQCD at energy scales as high as $2$~\TeV\ for the photon and
jet transverse momenta and scrutinise the NLO QCD description of the
dynamics of isolated-photon plus two-jet production in $pp$ collisions.
 
\section*{Acknowledgements}
 

We thank CERN for the very successful operation of the LHC, as well as the
support staff from our institutions without whom ATLAS could not be
operated efficiently.
 
We acknowledge the support of ANPCyT, Argentina; YerPhI, Armenia; ARC, Australia; FWF, BMWFW, Austria; ANAS, Azerbaijan; SSTC, Belarus; CNPq, FAPESP, Brazil; NSERC, CFI, NRC, Canada; CERN, CERN; CONICYT, Chile; CAS, NSFC, MOST, China; COLCIENCIAS, Colombia; VSC CR, MSMT CR, MPO CR, Czech Republic; DNSRC, DNRF, Denmark; IN2P3-CNRS, CEA-DRF/IRFU, France; SRNSFG, Georgia; MPG, HGF, BMBF, Germany; GSRT, Greece; RGC, Hong Kong SAR, Hong Kong China; Benoziyo Center, ISF, Israel; INFN, Italy; JSPS, MEXT, Japan; JINR, JINR; CNRST, Morocco; NWO, Netherlands; RCN, Norway; MNiSW, NCN, Poland; FCT, Portugal; MNE/IFA, Romania; NRC KI, MES of Russia, Russia Federation; MESTD, Serbia; MSSR, Slovakia; ARRS, MIZ\v{S}, Slovenia; DST/NRF, South Africa; MINECO, Spain; SRC, Wallenberg Foundation, Sweden; Cantons of Bern and Geneva , SNSF, SERI, Switzerland; MOST, Taiwan; TAEK, Turkey; STFC, United Kingdom; DOE, NSF, United states of America. In addition, individual groups and members have received support from CRC, Compute Canada, Canarie, BCKDF, Canada; Marie Sk{\l}odowska-Curie, COST, ERDF, ERC, Horizon 2020, European Union; ANR, Investissements d'Avenir Labex and Idex, France; AvH, DFG, Germany; Herakleitos, Thales and Aristeia programmes co-financed by EU-ESF and the Greek NSRF, Greece; BSF-NSF, GIF, Israel; PROMETEO Programme Generalitat Valenciana, CERCA Generalitat de Catalunya, Spain; Leverhulme Trust, The Royal Society, United Kingdom.
 
The crucial computing support from all WLCG partners is acknowledged gratefully, in particular from CERN, the ATLAS Tier-1 facilities at TRIUMF (Canada), NDGF (Denmark, Norway, Sweden), CC-IN2P3 (France), KIT/GridKA (Germany), INFN-CNAF (Italy), NL-T1 (Netherlands), PIC (Spain), ASGC (Taiwan), RAL (UK) and BNL (USA), the Tier-2 facilities worldwide and large non-WLCG resource providers. Major contributors of computing resources are listed in Ref.~\cite{ATL-GEN-PUB-2016-002}.
 
\printbibliography
 
\newpage
\input{atlas_authlist} 
\end{document}

%% file: atlas_authlist.tex
 
\begin{flushleft}
{\Large The ATLAS Collaboration}

\bigskip

G.~Aad$^\textrm{\scriptsize 101}$,    
B.~Abbott$^\textrm{\scriptsize 128}$,    
D.C.~Abbott$^\textrm{\scriptsize 102}$,    
A.~Abed~Abud$^\textrm{\scriptsize 70a,70b}$,    
K.~Abeling$^\textrm{\scriptsize 53}$,    
D.K.~Abhayasinghe$^\textrm{\scriptsize 93}$,    
S.H.~Abidi$^\textrm{\scriptsize 167}$,    
O.S.~AbouZeid$^\textrm{\scriptsize 40}$,    
N.L.~Abraham$^\textrm{\scriptsize 156}$,    
H.~Abramowicz$^\textrm{\scriptsize 161}$
% ...
\bigskip
\\

% ...
$^{40}$Niels Bohr Institute, University of Copenhagen, Copenhagen; Denmark.\\
% ...
$^{53}$II. Physikalisches Institut, Georg-August-Universit\"{a}t G\"ottingen, G\"ottingen; Germany.\\
% ...
$^{70}$$^{(a)}$INFN Sezione di Pavia;$^{(b)}$Dipartimento di Fisica, Universit\`a di Pavia, Pavia; Italy.\\
% ...
$^{93}$Department of Physics, Royal Holloway University of London, Egham; United Kingdom.\\
% ...
$^{101}$CPPM, Aix-Marseille Universit\'e, CNRS/IN2P3, Marseille; France.\\
$^{102}$Department of Physics, University of Massachusetts, Amherst MA; United States of America.\\
% ...
$^{128}$Homer L. Dodge Department of Physics and Astronomy, University of Oklahoma, Norman OK; United States of America.\\
% ...
$^{156}$Department of Physics and Astronomy, University of Sussex, Brighton; United Kingdom.\\
% ...
$^{161}$Raymond and Beverly Sackler School of Physics and Astronomy, Tel Aviv University, Tel Aviv; Israel.\\
% ...
$^{167}$Department of Physics, University of Toronto, Toronto ON; Canada.\\
% ...
$^{a}$ Also at Borough of Manhattan Community College, City University of New York, New York NY; United States of America.\\
$^{b}$ Also at CERN, Geneva; Switzerland.\\
% ...

\end{flushleft}


%% file: ANA-STDM-2017-32-PAPER.bib
@Article{epj:c76:581,
     author    = "{\colab{ATLAS}}",
      title          = "{Performance of pile-up mitigation techniques for jets in $pp$ collisions at
      $\sqrt s = 8$~TeV using the ATLAS detector}",
      journal        = "Eur. Phys. J. C",
      volume         = "76",
      year           = "2016",
      pages          = "581",
    doi            = "10.1140/epjc/s10052-016-4395-z",
    eprint         = "1510.03823",
    archivePrefix  = "arXiv",
    primaryClass   = "hep-ex",
}

@Booklet{1909.00761,
     author    = "{\colab{ATLAS}}",
     title     = "{Performance of electron and photon triggers in ATLAS during LHC Run 2}",
  howpublished =  "{(2019), }",
    eprint         = "1909.00761",
    archivePrefix  = "arXiv",
    primaryClass   = "hep-ex",
}

@Booklet{1911.04632,
     author    = "{\colab{ATLAS}}",
     title     = "{ATLAS data quality operations and performance for 2015–2018 data-taking}",
  howpublished =  "{(2019), }",
    eprint         = "1911.04632",
    archivePrefix  = "arXiv",
    primaryClass   = "physics.ins-det",
}

@Book{ISBN:978-3-935702-41-6,
     author    = "{G. Bohm and G. Zech}",
     title     = "{Introduction to statistics and data analysis for physicists}",
     edition   = "Deutsches Elektronen-Synchrotron, Hamburg, Germany, ISBN 9783935702881",
     year      = "2014",
    doi            = "10.3204/DESY-BOOK/statistics",
}

@Booklet{ATLAS-CONF-2019-021,
  author    = "{\colab{ATLAS}}",
    title          = "{Luminosity determination in $pp$ collisions at $\sqrt s=13$ TeV 
using the ATLAS detector at the LHC}",
    howpublished   = "{ATLAS-CONF-2019-021}",
    url            = "https://cdsweb.cern.ch/record/2677054",
    year           = "2019",
}

@Article{PERF-2017-03,
    author         = "{\colab{ATLAS}}",
    title          = "{Electron and photon energy calibration with the ATLAS detector using
2015--2016 LHC proton--proton collision data}",
    journal        = "JINST",
    volume         = "14",
    year           = "2019",
    pages          = "P03017",
    doi            = "10.1088/1748-0221/14/03/P03017",
    eprint         = "1812.03848",
    archivePrefix  = "arXiv",
    primaryClass   = "hep-ex",
}

@Article{LUCID2,
    author         = "{G. Avoni \etal}",
    title          = "{The new LUCID-2 detector for luminosity measurement and 
monitoring in ATLAS}",
    journal        = "JINST",
    volume         = "13",
    year           = "2018",
    pages          = "P07017",
    doi            = "10.1088/1748-0221/13/07/P07017",
}

@Article{epj:c63:189,
  author =       "{A.D. Martin, W.J. Stirling, R.S. Thorne and G. Watt}",
  title =        "{Parton distributions for the LHC}",
  journal =      "Eur. Phys. J. C",
  volume =       "63",
  year =         "2009",
  pages =        "189",
    doi            = "10.1140/epjc/s10052-009-1072-5",
    eprint         = "0901.0002",
    archivePrefix  = "arXiv",
    primaryClass   = "hep-ph",
}

@Article{Abbott:2018ikt, 
author = "{B. Abbott \etal}", 
title = "{Production and integration of the ATLAS Insertable B-Layer}", 
journal = "JINST", 
volume = "13", 
year = "2018", 
pages = "T05008", 
doi = "10.1088/1748-0221/13/05/t05008", 
eprint = "1803.00844", 
archivePrefix = "arXiv", 
primaryClass = "physics.ins-det", 
}

@Article{pl:b775:206,
  author    = "{\colab{ATLAS}}",
     title     = "{Measurement of the cross-section for electroweak production of dijets 
in association with a $Z$ boson in $pp$ collisions at $\sqrt s = 13$ TeV with the ATLAS
detector}",
     journal   = "Phys.  Lett. B",
     volume    = "775",
     year      = "2017",
     pages     = "206",
    doi            = "10.1016/j.physletb.2017.10.040",
    eprint         = "1709.10264",
    archivePrefix  = "arXiv",
    primaryClass   = "hep-ex",
}

@Article{epj:c77:474,
     author    = "{\colab{ATLAS}}",
      title          = "{Measurements of electroweak $Wjj$ production and constraints 
on anomalous gauge couplings with the ATLAS detector}",
      journal        = "Eur. Phys. J. C",
      volume         = "77",
      year           = "2017",
      pages          = "474",
    doi            = "10.1140/epjc/s10052-017-5007-2",
    eprint         = "1703.04362",
    archivePrefix  = "arXiv",
    primaryClass   = "hep-ex",
}

@Article{jhep:1805:077,
     author    = "{\colab{ATLAS}}",
     title     = "{Measurement of differential cross sections and $W^+/W^-$
cross-section ratios for $W$ boson production in association with jets at 
$\sqrt s=8$ TeV with the ATLAS detector}",
     journal   = "JHEP",
     volume    = "05",
     year      = "2018",
     pages     = "077",
    doi            = "10.1007/JHEP05(2018)077",
    eprint         = "1711.03296",
    archivePrefix  = "arXiv",
    primaryClass   = "hep-ex",
}

@Article{prl:117:182002,
  author    = "{\colab{ATLAS}}",
     title     = "{Measurement of the Inelastic Proton-Proton Cross Section at
$\sqrt s = 13$ TeV with the ATLAS Detector at the LHC}",
     journal   = "Phys.  Rev. Lett.",
     volume    = "117",
     year      = "2016",
     pages     = "182002",
    doi            = "10.1103/PhysRevLett.117.182002",
    eprint         = "1606.02625",
    archivePrefix  = "arXiv",
    primaryClass   = "hep-ex",
}

@Booklet{ATLAS-CONF-2015-029,
  author    = "{\colab{ATLAS}}",
    title          = "{Selection of jets produced in 13~TeV proton–proton collisions with the 
ATLAS detector}",
    howpublished   = "{ATLAS-CONF-2015-029}",
    url            = "https://cdsweb.cern.ch/record/2037702",
    year           = "2015",
}

@Article{pr:d96:072002,
     author    = "{\colab{ATLAS}}",
     title     = "{Jet energy scale measurements and their systematic uncertainties in 
proton-proton collisions at $\sqrt s = 13$ TeV with the ATLAS detector}",
     journal   = "Phys. Rev. D",
     volume    = "96",
     year      = "2017",
     pages     = "072002",
    doi            = "10.1103/PhysRevD.96.072002",
    eprint         = "1703.09665",
    archivePrefix  = "arXiv",
    primaryClass   = "hep-ex",
}

@Article{epj:c79:20,
     author    = "{\colab{CMS}}",
     title     = "{Measurement of differential cross sections for inclusive isolated-photon
and photon+jet production in proton-proton collisions at $\sqrt s = 13$ TeV}",
     journal   = "Eur. Phys. J. C",
     volume    = "79",
     year      = "2019",
     pages     = "20",
    doi            = "10.1140/epjc/s10052-018-6482-9",
    eprint         = "1807.00782",
    archivePrefix  = "arXiv",
    primaryClass   = "hep-ex",
}

@Article{epj:c79:969,
     author    = "{\colab{CMS}}",
     title     = "{Measurements of triple-differential cross sections for inclusive 
isolated-photon+jet events in $pp$ collisions at $\sqrt s = 8$ TeV}",
     journal   = "Eur. Phys. J. C",
     volume    = "79",
     year      = "2019",
     pages     = "969",
    doi            = "10.1140/epjc/s10052-019-7451-7",
    eprint         = "1907.08155",
    archivePrefix  = "arXiv",
    primaryClass   = "hep-ex",
}

@Article{jhep:1910:203,
     author    = "{\colab{ATLAS}}",
     title     = "{Measurement of the inclusive isolated-photon cross section
in $pp$ collisions at $\sqrt s=13$ TeV using
$36$~\fb1\ of ATLAS data}",
     journal   = "JHEP",
     volume    = "10",
     year      = "2019",
     pages     = "203",
    doi            = "10.1007/JHEP10(2019)203",
    eprint         = "1908.02746",
    archivePrefix  = "arXiv",
    primaryClass   = "hep-ex",
}

@Article{pl:b780:578,
     author    = "{\colab{ATLAS}}",
     title     = "{Measurement of the cross section for isolated-photon plus jet 
production in $pp$ collisions at $\sqrt{s}=13$ TeV using the ATLAS detector}",
     journal   = "Phys. Lett. B",
     volume    = "780",
     year      = "2018",
     pages     = "578",
    doi            = "10.1016/j.physletb.2018.03.035",
    eprint         = "1801.00112",
    archivePrefix  = "arXiv",
    primaryClass   = "hep-ex",
}

@Article{epj:c79:205,
     author    = "{\colab{ATLAS}}",
     title     = "{Measurement of the photon identification efficiencies with the ATL
AS detector using LHC Run 2 data collected in 2015 and 2016}",
    journal        = "Eur. Phys. J. C",
    volume         = "79",
    year           = "2019",
    pages          = "205",
    doi            = "10.1140/epjc/s10052-019-6650-6",
    eprint         = "1810.05087",
    archivePrefix  = "arXiv",
    primaryClass   = "hep-ex",
}

@Article{scipost:7:034,
      author         = "{E. Bothmann \etal}",
      title          = "{Event generation with SHERPA 2.2}",
      journal        = "SciPost Phys.",
      volume         = "7",
      year           = "2019",
      pages          = "034",
    doi            = "10.21468/SciPostPhys.7.3.034",
    eprint         = "1905.09127",
    archivePrefix  = "arXiv",
    primaryClass   = "hep-ph",
}

@Article{PERF-2007-01,
    author         = "{ATLAS Collaboration}",
    title          = "{The ATLAS Experiment at the CERN Large Hadron Collider}",
    journal        = "JINST",
    volume         = "3",
    year           = "2008",
    pages          = "S08003",
    doi            = "10.1088/1748-0221/3/08/S08003",
    primaryClass   = "hep-ex",
}

@Article{epj:c77:490,
     author    = "{\colab{ATLAS}}",
      title          = "{Topological cell clustering in the ATLAS calorimeters and its 
performance in LHC Run 1}",
      journal        = "Eur. Phys. J. C",
      volume         = "77",
      year           = "2017",
      pages          = "490",
    doi            = "10.1140/epjc/s10052-017-5004-5",
    eprint         = "1603.02934",
    archivePrefix  = "arXiv",
    primaryClass   = "hep-ex",
}

@Article{epj:c77:317,
     author    = "{\colab{ATLAS}}",
      title          = "{Performance of the ATLAS trigger system in 2015}",
      journal        = "Eur. Phys. J. C",
      volume         = "77",
      year           = "2017",
      pages          = "317",
    doi            = "10.1140/epjc/s10052-017-4852-3",
    eprint         = "1611.09661",
    archivePrefix  = "arXiv",
    primaryClass   = "hep-ex",
}

@Booklet{1105.1160,
     author    = "{T. Adye}",
     title     = "{Unfolding algorithms and tests using RooUnfold}",
  howpublished =  "{(2011), }",
    eprint         = "1105.1160",
    archivePrefix  = "arXiv",
    primaryClass   = "physics.data-an",
}

@Article{pr:d76:034003,
  author =       "{T. Pietrycki and A. Szczurek}",
  title =        "{Photon-jet correlations in $pp$ and $\pp$ collisions}",
  journal =      "Phys. Rev.  D",
  volume =       "76",
  year =         "2007",
  pages =        "034003",
    doi            = "10.1103/PhysRevD.76.034003",
    eprint         = "0704.2158",
    archivePrefix  = "arXiv",
    primaryClass   = "hep-ph",
}

@Article{pr:d75:013005,
  author =       "{U. Baur}",
  title =        "{Weak boson emission in hadron collider processes}",
  journal =      "Phys. Rev.  D",
  volume =       "75",
  year =         "2007",
  pages =        "013005",
    doi            = "10.1103/PhysRevD.75.013005",
    eprint         = "hep-ph/0611241",
    archivePrefix  = "arXiv",
}

@Article{pr:d79:114024,
  author =       "{Z. Belghobsi \etal}",
  title =        "{Photon-jet correlations and constraints on fragmentation 
functions}",
  journal =      "Phys. Rev. D",
  volume =       "79",
  year =         "2009",
  pages =        "114024",
    doi            = "10.1103/PhysRevD.79.114024",
    eprint         = "0903.4834",
    archivePrefix  = "arXiv",
    primaryClass   = "hep-ph",
}

@Article{pr:d83:052005,
     author    = "{\colab{ATLAS}}",
      title          = "{Measurement of the inclusive isolated prompt photon
                  cross section in $pp$ collisions at $\sqrt s=7$~{TeV}
                  with the ATLAS detector}",
      journal        = "Phys. Rev. D",
      volume         = "83",
      year           = "2011",
      pages          = "052005",
    doi            = "10.1103/PhysRevD.83.052005",
    eprint         = "1012.4389",
    archivePrefix  = "arXiv",
    primaryClass   = "hep-ex",
}

@Article{pl:b706:150,
     author    = "{\colab{ATLAS}}",
     title     = "{Measurement of the inclusive isolated prompt photon
                  cross-section in $pp$ collisions at $\sqrt s=7$~{T}e{V}
                  using 35~\pb1\ of {ATLAS} data}",
     journal   = "Phys. Lett. B",
     volume    = "706",
     year      = "2011",
     pages     = "150",
    doi            = "10.1016/j.physletb.2011.11.010",
    eprint         = "1108.0253",
    archivePrefix  = "arXiv",
    primaryClass   = "hep-ex",
}

@Article{jhep:1608:005,
     author    = "{\colab{ATLAS}}",
     title     = "{Measurement of the inclusive isolated prompt photon cross 
section in $pp$ collisions at $\sqrt s=8$ TeV with the ATLAS detector}",
     journal   = "JHEP",
     volume    = "08",
     year      = "2016",
     pages     = "005",
    doi            = "10.1007/JHEP08(2016)005",
    eprint         = "1605.03495",
    archivePrefix  = "arXiv",
    primaryClass   = "hep-ex",
}

@Article{pl:b770:473,
     author    = "{\colab{ATLAS}}",
     title     = "{Measurement of the cross section for inclusive 
isolated-photon production in $pp$ collisions at $\sqrt{s}=13$ TeV using 
the ATLAS detector}",
     journal   = "Phys. Lett. B",
     volume    = "770",
     year      = "2017",
     pages     = "473",
    doi            = "10.1016/j.physletb.2017.04.072",
    eprint         = "1701.06882",
    archivePrefix  = "arXiv",
    primaryClass   = "hep-ex",
}

@Article{cpc:178:852,
      author         = "{T. Sj\"ostrand, S. Mrenna and P.Z. Skands}",
      title          = "{A brief introduction to {PYTHIA} 8.1}",
      journal        = "Comput. Phys. Commun.",
      volume         = "178",
      year           = "2008",
      pages          = "852",
    doi            = "10.1016/j.cpc.2008.01.036",
    eprint         = "0710.3820",
    archivePrefix  = "arXiv",
    primaryClass   = "hep-ph",
}

@Article{jhep:0902:007,
      author         = "{T. Gleisberg \etal}",
      title          = "{Event generation with {SHERPA} 1.1}",
      journal        = "JHEP",
      volume         = "02",
      year           = "2009",
      pages          = "007",
    doi            = "10.1088/1126-6708/2009/02/007",
    eprint         = "0811.4622",
    archivePrefix  = "arXiv",
    primaryClass   = "hep-ph",
}

@Article{jhep:0603:059,
      author         = "{J. H. K\"uhn, A. Kulesza, S. Pozzorini and M. Schulze}",
      title          = "{Electroweak corrections to hadronic photon production at large 
transverse momenta}",
      journal        = "JHEP",
      volume         = "03",
      year           = "2006",
      pages          = "059",
    doi            = "10.1088/1126-6708/2006/03/059",
    eprint         = "hep-ph/0508253",
    archivePrefix  = "arXiv",
}

@Article{prep:97:31,
      author         = "{B. Andersson, G. Gustafson, G. Ingelman and T. Sj\"ostrand}",
      title          = "{Parton fragmentation and string dynamics}",
      journal        = "Phys. Rept.",
      volume         = "97",
      year           = "1983",
      pages          = "31",
    doi            = "10.1016/0370-1573(83)90080-7",
}

@Article{epj:c36:381,
      author         = "{J.-C. Winter, F. Krauss and G. Soff}",
      title          = "{A modified cluster-hadronisation model}",
      journal        = "Eur. Phys. J. C",
      volume         = "36",
      year           = "2004",
      pages          = "381",
    doi            = "10.1140/epjc/s2004-01960-8",
    eprint         = "hep-ph/0311085",
    archivePrefix  = "arXiv",
    %primaryClass   = "hep-ph",
}

@Article{np:b867:244,
      author         = "{R. D. Ball \etal}",
      title          = "{Parton distributions with LHC data}",
      journal        = "Nucl. Phys. B",
      volume         = "867",
      year           = "2013",
      pages          = "244",
    doi            = "10.1016/j.nuclphysb.2012.10.003",
    eprint         = "1207.1303",
    archivePrefix  = "arXiv",
    primaryClass   = "hep-ph",
}

@Article{pr:d82:074024,
      author         = "{H.-L. Lai \etal}",
      title          = "{New parton distributions for collider physics}",
      journal        = "Phys. Rev. D",
      volume         = "82",
      year           = "2010",
      pages          = "074024",
    doi            = "10.1103/PhysRevD.82.074024",
    eprint         = "1007.2241",
    archivePrefix  = "arXiv",
    primaryClass   = "hep-ph",
}

@Article{nim:a506:250,
      author         = "{S. Agostinelli \etal}",
      title          = "{{GEANT}4--a simulation toolkit}",
      journal        = "Nucl. Instrum. Meth. A",
      volume         = "506",
      year           = "2003",
      pages          = "250",
    doi            = "10.1016/S0168-9002(03)01368-8",
}

@Article{epj:c70:823,
     author    = "{\colab{ATLAS}}",
      title          = "{The {ATLAS} Simulation Infrastructure}",
      journal        = "Eur. Phys. J. C",
      volume         = "70",
      year           = "2010",
      pages          = "823",
    doi            = "10.1140/epjc/s10052-010-1429-9",
    eprint         = "1005.4568",
    archivePrefix  = "arXiv",
    primaryClass   = "physics.ins-det",
}

@Booklet{ATL-PHYS-PUB-2015-026,
  author    = "{\colab{ATLAS}}",
  title     = "{Vertex Reconstruction Performance of the ATLAS Detector at 
$\sqrt s=13$~TeV}",
  howpublished =  "{ATL-PHYS-PUB-2015-026}",
    url            = "https://cds.cern.ch/record/2037717",
    year           = "2015",
}

@Article{jhep:1504:040,
     author    = "{\colab{NNPDF}, R.D. Ball \etal}",
     title     = "{Parton distributions for the LHC run II}",
     journal   = "JHEP",
     volume    = "04",
     year      = "2015",
     pages     = "040",
    doi            = "10.1007/JHEP04(2015)040",
    eprint         = "1410.8849",
    archivePrefix  = "arXiv",
    primaryClass   = "hep-ph",
}

@Article{jhep:1304:027,
     author    = "{S. H\"oche, F. Krauss, M. Sch\"onherr and F. Siegert}",
     title     = "{QCD matrix elements + parton showers. The NLO case}",
     journal   = "JHEP",
     volume    = "04",
     year      = "2013",
     pages     = "027",
    doi            = "10.1007/JHEP04(2013)027",
    eprint         = "1207.5030",
    archivePrefix  = "arXiv",
    primaryClass   = "hep-ph",
}

@Article{jhep:0803:038,
     author    = "{S. Schumann and F. Krauss}",
     title     = "{A parton shower algorithm based on Catani-Seymour dipole factorisation}",
     journal   = "JHEP",
     volume    = "03",
     year      = "2008",
     pages     = "038",
    doi            = "10.1088/1126-6708/2008/03/038",
    eprint         = "0709.1027",
    archivePrefix  = "arXiv",
    primaryClass   = "hep-ph",
}

@Article{jhep:0202:044,
     author    = "{F. Krauss, R. Kuhn and G. Soff}",
     title     = "{AMEGIC++ 1.0, A Matrix Element Generator In C++}",
     journal   = "JHEP",
     volume    = "02",
     year      = "2002",
     pages     = "044",
    doi            = "10.1088/1126-6708/2002/02/044",
    eprint         = "hep-ph/0109036",
    archivePrefix  = "arXiv",
    %primaryClass   = "hep-ph",
}

@Article{prl:108:111601,
     author    = "{F. Cascioli, P. Maierh\"ofer and S. Pozzorini}",
     title     = "{Scattering Amplitudes with Open Loops}",
     journal   = "Phys.  Rev. Lett.",
     volume    = "108",
     year      = "2012",
     pages     = "111601",
    doi            = "10.1103/PhysRevLett.108.111601",
    eprint         = "1111.5206",
    archivePrefix  = "arXiv",
    primaryClass   = "hep-ph",
}

@Article{jp:g44:044007,
     author    = "{F. Siegert}",
     title     = "{A practical guide to event generation for prompt photon production with Sherpa}",
     journal   = "J. Phys. G",
     volume    = "44",
     year      = "2017",
     pages     = "044007",
    doi            = "10.1088/1361-6471/aa5f29",
    eprint         = "1611.07226",
    archivePrefix  = "arXiv",
    primaryClass   = "hep-ph",
}

@Article{epj:c72:1896,
     author    = "{M. Cacciari, G.P. Salam and G. Soyez}",
     title     = "{FastJet user manual}",
     journal   = "Eur. Phys. J. C",
     volume    = "72",
     year      = "2012",
     pages     = "1896",
    doi            = "10.1140/epjc/s10052-012-1896-2",
    eprint         = "1111.6097",
    archivePrefix  = "arXiv",
    primaryClass   = "hep-ph",
}

@Article{pl:b269:445,
     author    = "{S. Keller and J.F. Owens}",
     title     = "{Event structure in photon plus two-jet final states}",
     journal   = "Phys. Lett. B",
     volume    = "269",
     year      = "1991",
     pages     = "445",
    doi            = "10.1016/0370-2693(91)90198-Y",
}

@Article{nim:a362:487,
     author    = "{G. D'Agostini}",
     title     = "{A multidimensional unfolding method based on Bayes' theorem}",
     journal   = "Nucl. Instrum. Meth. A",
     volume    = "362",
     year      = "1995",
     pages     = "487",
    doi            = "10.1016/0168-9002(95)00274-X",
}

@Booklet{ATL-GEN-PUB-2016-002,
    author         = "{\colab{ATLAS}}",
    title          = "{ATLAS Computing Acknowledgements}",
    howpublished   = "{ATL-GEN-PUB-2016-002}",
    url            = "https://cds.cern.ch/record/2202407",
}

@Article{jhep:0804:005,
     author    = "{M. Cacciari, G.P. Salam and G. Soyez}",
     title     = "{The catchment area of jets}",
     journal   = "JHEP",
     volume    = "04",
     year      = "2008",
     pages     = "005",
    doi            = "10.1088/1126-6708/2008/04/005",
    eprint         = "0802.1188",
    archivePrefix  = "arXiv",
    primaryClass   = "hep-ph",
}

@Article{pl:b429:369,
     author    = "{S. Frixione}",
     title     = "{Isolated photons in perturbative QCD}",
     journal   = "Phys. Lett. B",
     volume    = "429",
     year      = "1998",
     pages     = "369",
    doi            = "10.1016/S0370-2693(98)00454-7",
    eprint         = "hep-ph/9801442",
    archivePrefix  = "arXiv",
    %primaryClass   = "hep-ph",
}

@Article{np:b875:483,
     author    = "{\colab{ATLAS}}",
     title     = "{Dynamics of isolated-photon plus jet production in $pp$
  collisions at $\sqrt{s}=7$ TeV with the ATLAS detector}",
     journal   = "Nucl. Phys. B",
     volume    = "875",
     year      = "2013",
     pages     = "483",
    doi            = "10.1016/j.nuclphysb.2013.07.025",
    eprint         = "1307.6795",
    archivePrefix  = "arXiv",
    primaryClass   = "hep-ex",
}

@Article{np:b918:257,
     author    = "{\colab{ATLAS}}",
     title     = "{High-$E_T$ isolated-photon plus jets production in $pp$ collisions at 
                     $\sqrt{s}=8$ TeV with the ATLAS detector}",
     journal   = "Nucl. Phys. B",
     volume    = "918",
     year      = "2017",
     pages     = "257",
    doi            = "10.1016/j.nuclphysb.2017.03.006",
    eprint         = "1611.06586",
    archivePrefix  = "arXiv",
    primaryClass   = "hep-ex",
}

@Article{pr:d88:112009,
     author    = "{\colab{CMS}}",
     title     = "{Rapidity distributions in exclusive $Z+jet$ and $\gamma+jet$ events in $pp$
 collisions at $\sqrt s = 7$ TeV}",
     journal   = "Phys. Rev. D",
     volume    = "88",
     year      = "2013",
     pages     = "112009",
    doi            = "10.1103/PhysRevD.88.112009",
    eprint         = "1310.3082",
    archivePrefix  = "arXiv",
    primaryClass   = "hep-ex",
}

@Article{pr:d88:013009,
     author    = "{T. Becher and X. Garcia i Tormo}",
     title     = "{Electroweak Sudakov effects in $W$, $Z$ and $\gamma$ production at 
large transverse momentum}",
     journal   = "Phys. Rev. D",
     volume    = "88",
     year      = "2013",
     pages     = "013009",
    doi            = "10.1103/PhysRevD.88.013009",
    eprint         = "1305.4202",
    archivePrefix  = "arXiv",
    primaryClass   = "hep-ph",
}

@Article{pr:d92:073011,
     author    = "{T. Becher and X. Garcia i Tormo}",
     title     = "{Addendum to "Electroweak Sudakov effects in $W$, $Z$ and $\gamma$ 
production at large transverse momentum"}",
     journal   = "Phys. Rev. D",
     volume    = "92",
     year      = "2015",
     pages     = "073011",
    doi            = "10.1103/PhysRevD.92.073011",
    eprint         = "1509.01961",
    archivePrefix  = "arXiv",
    primaryClass   = "hep-ph",
}

@Article{jhep:1406:009,
     author    = "{\colab{CMS}}",
     title     = "{Measurement of the triple-differential cross section for photon+jets production 
in proton-proton collisions at $\sqrt s=7$ TeV}",
     journal   = "JHEP",
     volume    = "06",
     year      = "2014",
     pages     = "009",
    doi            = "10.1007/JHEP06(2014)009",
    eprint         = "1311.6141",
    archivePrefix  = "arXiv",
    primaryClass   = "hep-ex",
}

@Article{jhep:1510:128,
     author    = "{\colab{CMS}}",
     title     = "{Comparison of the $Z/\gamma^*+jets$ to $\gamma+jets$ cross sections in
 $pp$ collisions at $\sqrt s=8$ TeV}",
     journal   = "JHEP",
     volume    = "10",
     year      = "2015",
     pages     = "128",
    doi            = "10.1007/JHEP10(2015)128",
    eprint         = "1505.06520",
    archivePrefix  = "arXiv",
    primaryClass   = "hep-ex",
}

@Article{pr:d85:092014,
     author    = "{\colab{ATLAS}}",
     title     = "{Measurement of the production cross section of an isolated 
                  photon associated with jets in proton-proton collisions at
                  $\sqrt s = 7$ TeV with the {ATLAS} detector}",
     journal   = "Phys. Rev. D",
     volume    = "85",
     year      = "2012",
     pages     = "092014",
    doi            = "10.1103/PhysRevD.85.092014",
    eprint         = "1203.3161",
    archivePrefix  = "arXiv",
    primaryClass   = "hep-ex",
}

@Article{jhep:0804:063,
      author         = "{M. Cacciari, G.P. Salam and G. Soyez}",
      title          = "{The anti-$\akt$ jet clustering algorithm}",
      journal        = "JHEP",
      volume         = "04",
      year           = "2008",
      pages          = "063",
    doi            = "10.1088/1126-6708/2008/04/063",
    eprint         = "0802.1189",
    archivePrefix  = "arXiv",
    primaryClass   = "hep-ph",
}

@Booklet{ATL-PHYS-PUB-2012-003,
  author    = "{\colab{ATLAS}}",
    title          = "{Summary of ATLAS Pythia 8 tunes}",
    howpublished   = "{ATL-PHYS-PUB-2012-003}",
    url            = "https://cds.cern.ch/record/1474107",
    year           = "2012",
}

@Article{jhep:0905:053,
      author         = "{S. H{\"o}che, F. Krauss, S. Schumann and F. Siegert}",
      title          = "{QCD matrix elements and truncated showers}",
      journal        = "JHEP",
      volume         = "05",
      pages          = "053",
      doi            = "10.1088/1126-6708/2009/05/053",
      year           = "2009",
      eprint         = "0903.1219",
      archivePrefix  = "arXiv",
      primaryClass   = "hep-ph",
}

@Booklet{ATL-PHYS-PUB-2014-021,
  author    = "{\colab{ATLAS}}",
    title          = "{ATLAS Pythia~8 tunes to 7~\TeV\ data}",
    howpublished   = "{ATL-PHYS-PUB-2014-021}",
    url            = "https://cds.cern.ch/record/1966419",
    year           = "2014",
}

@Booklet{ATLAS-TDR-19,
  author    = "{\colab{ATLAS}}",
    title          = "{ATLAS Insertable B-Layer Technical Design Report}",
    howpublished   = "{CERN-LHCC-2010-013, ATLAS-TDR-19, 2010,\\ {\scriptsize{URL}}:
  \url{https://cds.cern.ch/record/1291633}}",
   note = {{\textit{ATLAS Insertable B-Layer Technical Design Report
  Addendum}}, ATLAS-TDR-19-ADD-1, 2012, {\scriptsize{URL}}:
  \url{https://cds.cern.ch/record/1451888}}
 }
